\documentclass[preprint,3p,12pt]{elsarticle}
\usepackage{mathrsfs}
\usepackage{amsmath}
\usepackage{stmaryrd}
\usepackage{bbding}
\usepackage{dcolumn}
\usepackage{graphicx}
\usepackage{amsfonts}
\usepackage{amssymb}
\usepackage{psfrag}
\usepackage{wrapfig}
\usepackage{subfigure}
\usepackage{makeidx}
\usepackage{bm}
\usepackage{epsf}
\usepackage{color}
\usepackage{epsfig}
\usepackage{setspace}
\usepackage{graphicx}
\usepackage{epstopdf}
\usepackage{psfrag}
\usepackage{subfigure}
\usepackage{multirow}
\usepackage{diagbox}
\usepackage{verbatim}
\newtheorem{rmk}{Remark}

\usepackage{makecell}
\usepackage{float}
\usepackage{esint}
\usepackage{comment}
\usepackage{movie15}
\usepackage{hyperref}

\newcommand{\mathsym}[1]{{}}
\newcommand{\unicode}[1]{{}}
\begin{document}
	
\title{Two-step multi-resolution reconstruction-based compact gas-kinetic scheme on tetrahedral mesh}
	
	\author[HKUST1]{Xing Ji}
	\ead{xjiad@connect.ust.hk}
	
	\author[HKUST2]{Fengxiang Zhao}
	\ead{fzhaoac@connect.ust.hk}
	
	\author[HKUST2]{Wei Shyy}
	\ead{weishyy@ust.hk}
	
	\author[HKUST1,HKUST2,HKUST3]{Kun Xu\corref{cor1}}
	\ead{makxu@ust.hk}

	\address[HKUST1]{Department of Mathematics, Hong Kong University of Science and Technology, Clear Water Bay, Kowloon, Hong Kong}
	\address[HKUST2]{Department of Mechanical and Aerospace Engineering, Hong Kong University of Science and Technology, Clear Water Bay, Kowloon, Hong Kong}
	\address[HKUST3]{Shenzhen Research Institute, Hong Kong University of Science and Technology, Shenzhen, China}
	\cortext[cor1]{Corresponding author}

\begin{abstract}
In this paper, a third-order compact gas-kinetic scheme (GKS) on unstructured tetrahedral mesh is constructed for the compressible Euler and Navier-Stokes solutions. The time-dependent gas distribution function at a cell interface
is used to calculate the fluxes for the updating the cell-averaged flow variables and to evaluate the time accurate cell-averaged flow variables as well for evolving the cell-averaged gradients of flow variables.
With the accurate evolution model for both flow variables and their slopes,
the quality of the scheme depends closely on the accuracy and reliability of the initial reconstruction of flow variables.
The reconstruction scheme becomes more challenge on tetrahedral mesh, where the conventional second-order unlimited least-square reconstruction can make the scheme be linearly unstable when using cell-averaged conservative variables alone with von Neumann neighbors.
Benefiting from the evolved cell-averaged slopes, on tetrahedral mesh the GKS is linearly stable from a compact third-order smooth reconstruction
with a large CFL number.
In order to further increase the robustness of the high-order compact GKS for capturing discontinuous solution,
a new two-step multi-resolution weighted essentially non-oscillatory (WENO) reconstruction will be proposed.
The novelty of the reconstruction includes the following.
Firstly, it releases the stability issue from a second-order compact reconstruction through the introduction of a pre-reconstruction step.
Secondly, in the third-order non-linear reconstruction, only one more large stencil is added beside those in the second-order one, which  significantly simplifies the high-order reconstruction.
At the same time, the high-order wall boundary treatment is carefully designed by combining
the constrained least-square technique and the WENO procedure, where a quadratic element is adopted in the reconstruction for the
curved boundary.
Numerical tests  for both inviscid and viscous flow at low and high speed are presented from the second-order and third-order compact GKS.
The proposed third-order scheme shows good robustness in high speed flow computation and favorable mesh adaptability in cases with complex geometry.

\end{abstract}


\maketitle

\section{Introduction}
The simulation of compressible flow with complex geometry is important in the engineering applications.
The use of unstructured mesh is especially favored from its geometric flexibility.
On such a mesh even for a second-order finite volume method (FVM) it is not easy to achieve
the same performance as that in the structure mesh.
Commonly, slope reconstruction schemes, such as cell-based Green-Gauss method \cite{mavriplis2003revisiting} combined with different types of limiters, are robust and widely used in the commercial or open-source software \cite{zhao2020phenglei}.
However, these methods can easily deteriorate the spatial accuracy for skewed mesh and become over-dissipative for flow simulation with discontinuities.
The least-square reconstruction with cell-averaged variables and von Neumann neighbors only preserves a strictly second-order accuracy,
but suffers from the linear instability on tetrahedral grid \cite{haider2009stability}. Therfore, the reconstruction stencil has to be further extended.
In order to ensure linear stability,  a two-step second-order weighted essentially non-oscillatory (WENO) method
has been proposed \cite{xia2014finite} with the attempt of keeping a compact reconstruction  even with an extended stencil.
The high-order WENO-FVMs have been continuously developed and applied to large-scale aeronautical simulations \cite{antoniadis2017assessment}.
But, the compactness can be hardly kept in the high-order FVMs.
Even though the extended stencils in reconstruction can improve the robustness of the schemes,
 difficulties still exist in the parallel programming and boundary treatment.
The recently proposed multi-resolution reconstruction with only five equivalent sub-stencils greatly releases the above problems \cite{zhu2020new} even with the inclusion of neighbor-to-neighbor cells.

The compact methods with the updates of multiple degrees of freedom (DOFs) for each cell have been developed extensively in the past decades.
Two main representatives are the DG \cite{shu2016weno-dg-review} and the FR/CPR methods \cite{Huynh2007FR,yu2014accuracy}, which hybridize the finite volume framework with the finite element method or the finite difference method.
These methods can achieve arbitrary spatial order of accuracy with only the targeted cell as the reconstruction stencil, and yields great mesh adaptability and high scalability.
Successful examples have been demonstrated in large eddy simulation (LES) \cite{wang2017towards} and RANS simulation \cite{yang2019robust} for subsonic flows.
For the flow simulation with discontinuities, these methods have less robustness against the traditional high-order FVMs.
In addition, these methods have restricted explicit time steps and high memory-consumption \cite{luo2010reconstructed}.
The $P_NP_M$ \cite{dumbser2010arbitrary} and reconstructed-DG (rDG) methods \cite{luo2010reconstructed} were targeting to overcome  the above weakness with the release of the compactness of the DG methods.
Large time step and less memory requirement can be achieved in the rDG methods in comparison with the same order DG ones.

In recent years, a class of high-order compact GKS has been developed.
The GKS is based on a time accurate evolution model in the construction of the gas distribution function at a cell interface \cite{xu2014direct}.
The time-dependent solution provides not only the fluxes across a cell interface, but also the corresponding flow variables.
As a result, both the cell-averaged flow variables and their slopes can be updated simultaneously through the divergence theorem.
For the DG/rDG methods, the similar DOFs are obtained differently with explicit governing equations.
Due to their differences in the updating schemes, the compact GKS can use a larger time step and has better robustness than the corresponding DG methods. For example, a CFL number around 0.5 can be taken for the third-order compact GKS \cite{ji2020hweno}
while it is restricted to be less than 0.33 for the third-order P1P2-rDG scheme.
The P1P2-rDG is claimed to be unstable on tetrahedral mesh with smooth reconstruction.
However, as shown in this paper the third-order compact GKS is stable with a CFL number of 1 with the same compact stencil.
Due to the use of time accurate evolution model, another benefit of GKS is to use the two-stage fourth-order temporal discretization method \cite{li2016twostage} or other multi-stage multi-derivative time marching scheme \cite{multi-derivative}.
Although the gas-kinetic flux function has a high computational cost than the time-independent Riemann solvers,
the HGKS can achieve fourth-order temporal accuracy with only two stages\cite{Pan2016twostage}, instead of four-stages in the fourth-order Runge-Kutta (RK) time discretization.
Overall, the compact GKS  turns out to be more efficient in comparison with Riemann-solver-based RK methods \cite{ji2020three,ji2020hweno}.

In this paper, a compact third-order GKS on tetrahedral mesh will be presented.
The scheme is linearly stable for smooth flow with unlimited constrained-least-square reconstruction on a compact stencil with von Neumann neighbors only.
A direct extension of the HWENO-type reconstruction on the hexahedral mesh \cite{ji2020three} to the current tetrahedron mesh is not successful, which shows poor robustness and mesh adaptability.
The main reasons may be the following.
Firstly, the coefficient matrices for the first-order polynomials based on the cell-averaged conservative variables
on the biased sub-stencils depend too sensitively  on the quality of the tetrahedron and can easily become singular.
Secondly, the central first-order sub-stencil cannot provide a proper measurement of the smoothness of the local flow field.
In the FVM,  the unlimited second-order least-square reconstruction on such central stencil was proven to be linearly unstable \cite{haider2009stability}, same to the second-order GKS in the numerical tests.
Extended stencils have to be used to ensure the stability under finite volume framework.
Recently, a two-step WENO reconstruction with a compact stencil in each step has been proposed \cite{xia2014finite}.
The key idea is to firstly reconstruct the first-order spatial derivatives by using the  unlimited least-square reconstruction
and store them in each cell. Then, at the second step, new first-order spatial derivatives on the targeting cell are obtained by a weighted combination of all the pre-computed spatial derivatives.
At the same time, a multi-resolution reconstruction has been proposed in a hierarchical way,
i.e., the Nth-order of accuracy can be achieved by N central stencils from first-order to Nth-order \cite{zhu2020new}.
Inspired from the above two approaches, a two-step multi-resolution reconstruction is designed in the current scheme.
A linearly stable second-order WENO reconstruction is obtained first in a compact manner.
Then, a third-order compact reconstruction is constructed with only one more large stencil based on the above second-order one.
The reconstruction becomes simple and efficient.
For example, the robustness and mesh adaptability of the scheme have been enhanced due to the extended sub-stencils, and
the memory overhead is even sightly reduced because there is no need for storing the polynomial coefficients of the biased sub-stencils in each cell.
Benefiting from the compact reconstruction in each step, the WENO procedure can be easily extended to the boundary reconstruction.
In order to keep high-order accuracy at boundaries,
 a third-order one-sided reconstruction without ghost cells are designed for the adiabatic and isothermal walls at each Gaussian point.
To keep the high-order spatial reconstruction, a quadratic element is used to recover the curved boundary.
Stringent tests including supersonic flow passing through an air-vehicle validate the robustness of the current compact scheme with complex geometry.

This paper is organized as follows.
The basic framework of the compact high-order GKS on tetrahedron mesh is presented in Section 2.
In Section 3, the basic formulation for the two-stage temporal discretization is given.
In Section 4, the details for the two-step multi-resolution WENO reconstruction on tetrahedral mesh is presented.
Numerical examples from nearly incompressible to highly compressible flows are given in Section 5.
A concluding remark is given in the last section.

\section{Compact finite volume gas-kinetic scheme}

The 3-D gas-kinetic BGK equation \cite{BGK} is
\begin{equation}\label{bgk}
f_t+\textbf{u}\cdot\nabla f=\frac{g-f}{\tau},
\end{equation}
where $f=f(\textbf{x},t,\textbf{u},\xi)$ is the gas distribution function, which is a function of space $\textbf{x}$, time $t$, particle velocity $\textbf{u}$, and internal variable $\xi$.
 $g$ is the equilibrium state approached by $f$
and $\tau$ is the collision time.

The collision term satisfies the compatibility condition
\begin{equation}\label{compatibility}
\int \frac{g-f}{\tau} \pmb{\psi} \text{d}\Xi=0,
\end{equation}
where $\pmb{\psi}=(1,\textbf{u},\displaystyle \frac{1}{2}(\textbf{u}^2+\xi^2))^T$,
$\text{d}\Xi=\text{d}u_1\text{d}u_2\text{d}u_3\text{d}\xi_1...\text{d}\xi_{K}$,
$K$ is the number of internal degrees of freedom, i.e.
$K=(5-3\gamma)/(\gamma-1)$ in 3-d case, and $\gamma$
is the specific heat ratio.

In the continuum flow regime with the smoothness assumption,
based on the Chapman-Enskog expansion of the BGK equation the gas distribution function can be expressed as \cite{xu2014direct},
\begin{align*}
f=g-\tau D_{\textbf{u}}g+\tau D_{\textbf{u}}(\tau
D_{\textbf{u}})g-\tau D_{\textbf{u}}[\tau D_{\textbf{u}}(\tau
D_{\textbf{u}})g]+...,
\end{align*}
where $D_{\textbf{u}}={\partial}/{\partial t}+\textbf{u}\cdot \nabla$.
Different hydrodynamic equations can be derived by truncating on different orders of $\tau$.
With the zeroth-order in truncated distribution function $f=g$,
the Euler equations can be recovered by multiplying $\pmb{\psi}$ on Eq.\eqref{bgk} and integrating it over the phase space,
\begin{equation*}\label{euler-conservation}
\begin{split}
\textbf{W}_t+ \nabla \cdot \textbf{F}(\textbf{W})=0.
\end{split}
\end{equation*}
With the first-order truncation, i.e.,
\begin{align} \label{ce-ns}
f=g-\tau (\textbf{u} \cdot \nabla g + g_t),
\end{align}
the N-S equations can be obtained,
\begin{equation*}\label{ns-conservation}
\begin{split}
\textbf{W}_t+ \nabla \cdot \textbf{F}(\textbf{W},\nabla \textbf{W} )=0,
\end{split}
\end{equation*}
with $\tau = \mu / p$ and $Pr=1$.

The conservative variables and their fluxes are the moments of the gas distribution function
\begin{align}\label{point}
\textbf{W}(\textbf{x},t)=\int \pmb{\psi} f(\textbf{x},t,\textbf{u},\xi)\text{d}\Xi
\end{align}
and
\begin{equation}\label{f-to-flux}
\textbf{F}(\textbf{x},t)=
\int \textbf{u} \pmb{\psi} f(\textbf{x},t,\textbf{u},\xi)\text{d}\Xi.
\end{equation}

\begin{rmk}
	The cell-averaged conservative variables can be updated through the interface fluxes under the finite volume framework.
	Besides the fluxes in Eq.\eqref{f-to-flux}, the gas distribution function also provides the flow variables at the cell interface, such as that in Eq.\eqref{point}. It is the key point for the possibility of constructing compact GKS.
	An obvious prerequisite is the time accurate $\textbf{W}(\textbf{x},t)$ at a cell interface.
   In other words, it depends solely on the high-order gas evolution model in the construction of high-order scheme.
\end{rmk}

\subsection{Compact gas-kinetic scheme on tetrahedral mesh}

For a tetrahedral cell $\Omega_i$ in 3-D case, the boundary can be expressed as
\begin{equation*}
\partial \Omega_i=\bigcup_{p=1}^{N_f}\Gamma_{ip},
\end{equation*}
where $N_f=4$ is the number of cell interfaces for cell $\Omega_i$.

The update of the cell averaged conservative flow variables in a finite control volume i from $t_n$ to $t_{n+1}$  can be expressed as
\begin{equation}\label{fv-3d-general}
\textbf{W}^{n+1}_i \left| \Omega_i \right|=\textbf{W}^n_i \left| \Omega_i \right|-\sum_{p=1}^{N_f}\int_{\Gamma_{ip}}\int_{t_n}^{t_{n+1}}
\textbf{F}(\textbf{x},t)\cdot\textbf{n}_p \text{d}t\text{d}s,
\end{equation}
with
\begin{equation}\label{f-to-flux-in-normal-direction}
\textbf{F}(\textbf{x},t)\cdot \textbf{n}_p=\int\pmb{\psi}  f(\textbf{x},t,\textbf{u},\xi) \textbf{u}\cdot \textbf{n}_p \text{d}\Xi,
\end{equation}
where $\textbf{W}_{i}$ is the cell averaged values over cell $\Omega_i$, $\left|
\Omega_i \right|$ is the volume of $\Omega_i$, $\textbf{F}$ is the interface fluxes, and $\textbf{n}_p=(n_1,n_2,n_3)^T$ is the unit vector representing the outer normal direction of $\Gamma_{ip}$.
The semi-discretized form of finite volume scheme can be written as
\begin{equation}\label{semidiscrete}
\frac{\text{d} \textbf{W}_{i}}{\text{d}t}=\mathcal{L}(\textbf{W}_i)=-\frac{1}{\left| \Omega_i \right|} \sum_{p=1}^{N_f} \int_{\Gamma_{ip}}
\textbf{F}(\textbf{W})\cdot\textbf{n}_p \text{d}s.
\end{equation}
For the interface fluxes $\textbf{F}_{ip}(t)$, the numerical quadrature can be adopted and
Eq.\eqref{fv-3d-general} can be rewritten as
\begin{equation}\label{fv-3d-general-quadrature}
\textbf{W}^{n+1}_i \left| \Omega_i \right|
=\textbf{W}^n_i \left| \Omega_i \right|-\sum_{p=1}^{N_f}  \left|\Gamma_{ip}\right| \sum_{k=1}^{M} \omega_k\int_{t_n}^{t_{n+1}}
\textbf{F}(\textbf{x}_{p,k},t)\cdot\textbf{n}_p \text{d}t.
\end{equation}

\begin{figure}[htbp]	
	\centering
	\subfigure[Quadratic triangluar surface]{
		\label{triangle}
		\includegraphics[width=0.48\textwidth]
		{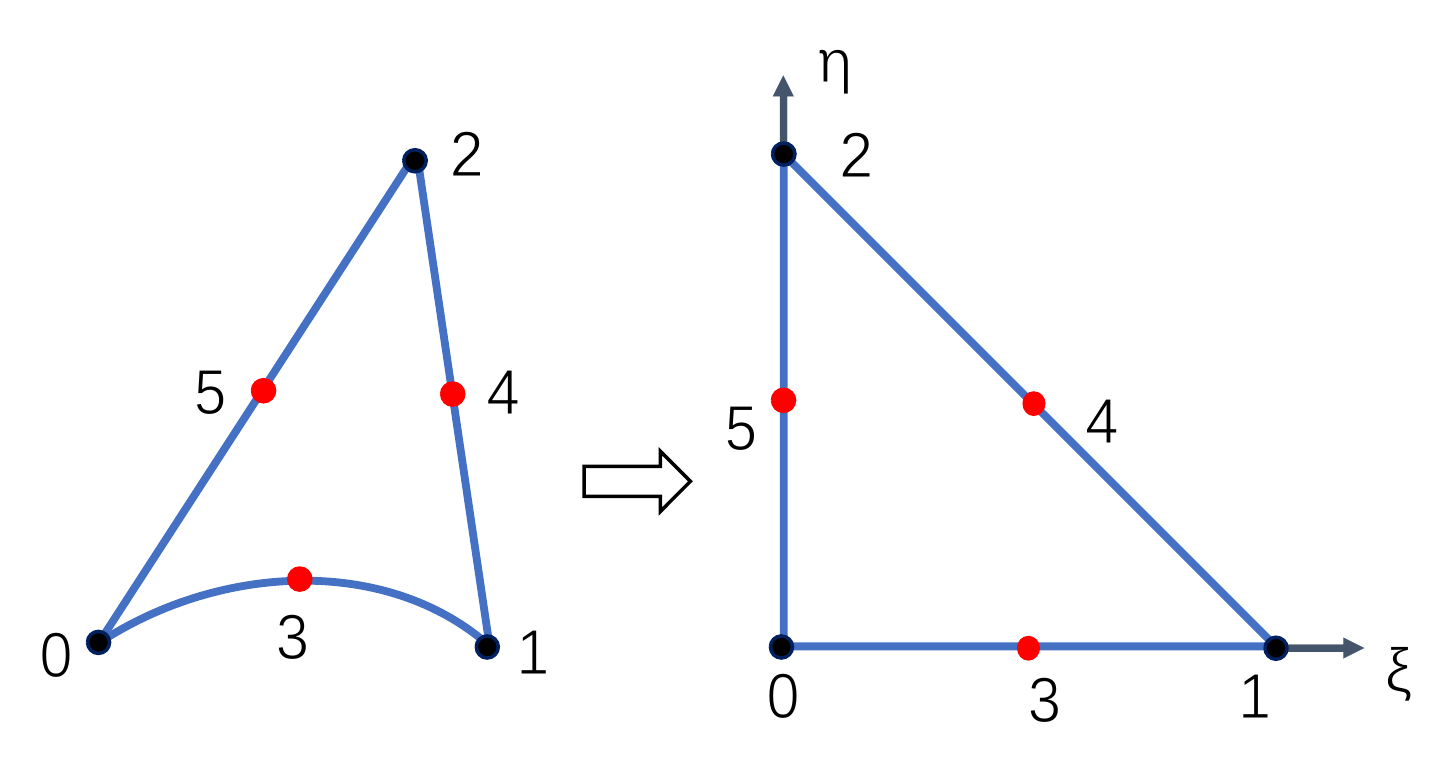}}
	\subfigure[Quadratic tetrahedral volume]{
		\label{tetra}
		\includegraphics[width=0.48\textwidth]
		{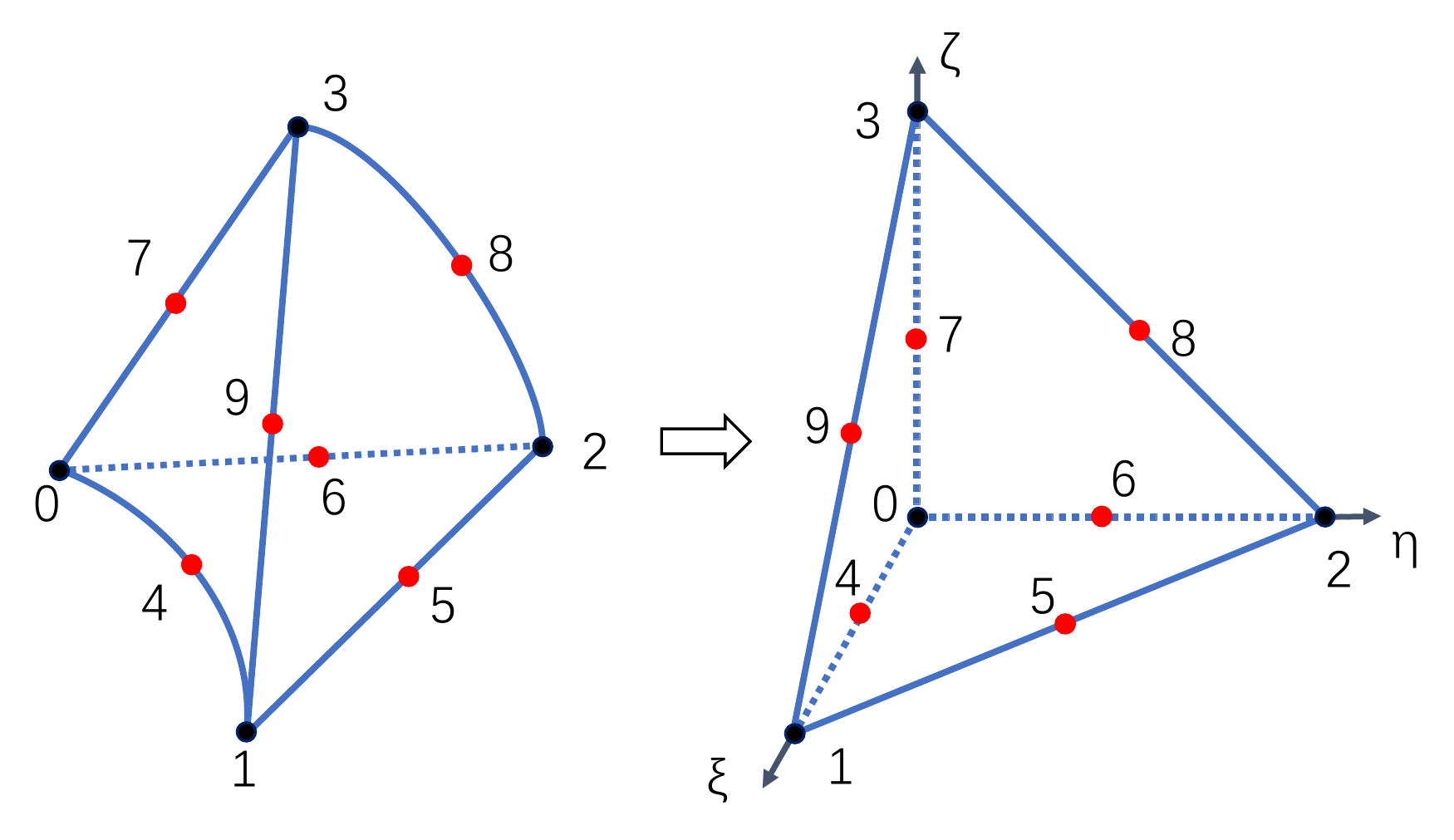}}
	\caption{The controlling points and isoparametric transformation of the quadratic elements.}
	\label{quadratic-element}
\end{figure}

	Nowadays the curved mesh generation has been supported by popular commercial software.
	To be consistent with the spatial accuracy, the quadratic element is applied here to describe the geometry.	
    The controlling points for the quadratic triangle are shown in Fig.~\ref{triangle}.	
	The iso-parametric transformation is used to evaluate the surface integral, which is expressed as
	\begin{align*}
	\textbf{X} (\xi, \eta)= \sum_{l=0}^5 \textbf{x}_l \phi_l (\xi, \eta),
	\end{align*}
	where $\textbf{x}_l$ is the location of the controlling points
	and $\phi_l$ is the base function as follows \cite{wang2017thesis}	
	\begin{equation}
	\begin{array}{l}
	v_{0}=(\xi+\eta-1)(2 \xi+2 \eta-1),~~v_{1}=\xi(2 \xi-1),~~v_{2}=\eta(2 \eta-1), \\
	v_{3}=-4 \xi(\xi+\eta-1),~~v_{4}=4 \xi \eta,~~v_{5}=-4 \eta(\xi+\eta-1).
	\end{array}
	\end{equation}
	
	The flux across $\Gamma_{ip}$ in Eq.\eqref{fv-3d-general} can be transferred to a standard  isosceles right triangle $\tilde{\Gamma}_{ip}$
	\begin{equation*}\label{flux-para-coord}
	\begin{split}
	\textbf{F}_{ip}(t)
	=\int_{\Gamma_{ip}} \textbf{F}(\textbf{x},t)\cdot\textbf{n}_p \text{d}s
	=\int_{\tilde{\Gamma}_{ip}}\textbf{F}(\textbf{W}(\textbf{X}(\xi, \eta))) \cdot\textbf{n}_p \left|\frac{\partial(x, y, z)}{\partial(\xi, \eta)}\right| \text{d}\xi \text{d}\eta.
	\end{split}
	\end{equation*}
	To meet the requirement of a third-order spatial accuracy,
	the above equation can be approximated through Gaussian quadrature as
	\begin{equation*}\label{flux-para-coord-gauss}
	\textbf{F}_{ip}(t)
	= \frac{1}{2} \Delta \xi \Delta \eta \sum_{m=1}^{3} \tilde{\omega}_{m} \textbf{F}_{m}(t) \cdot (\textbf{n}_p )_{m}\left|\frac{\partial(x, y, z)}{\partial(\xi, \eta)}\right|_{m} ,
	\end{equation*}
	where $ \Delta \xi  = \Delta \eta = 1$ and the local normal direction
	$(\textbf{n}_p )_{m}=\left(\boldsymbol{X}_{\xi} \times \boldsymbol{X}_{\eta}\right) /\left\|\boldsymbol{X}_{\xi} \times \boldsymbol{X}_{\eta}\right\|$.
	The standard Gaussian points are
	\begin{equation*}
	\begin{split}
	(\xi,\eta)_{1}=(\frac{1}{6}\Delta \xi ,\frac{1}{6} \Delta \eta),
	~(\xi,\eta)_{2}=(\frac{2}{3}\Delta \xi,\frac{1}{6}\Delta \eta),
	~(\xi,\eta)_{3}=(\frac{1}{6}\Delta \xi,\frac{2}{3}\Delta \eta),
	\end{split}
	\end{equation*}
	with $\tilde{\omega}_{m}=\frac{1}{3},~m=1,2,3$.
	Compared with Eq.\eqref{fv-3d-general-quadrature}, we have
	\begin{equation*}
	\omega_{m} = \frac{1}{2} \tilde{\omega}_{m} \left|\frac{\partial(x, y, z)}{\partial(\xi, \eta)}\right|_{m},~
	\textbf{x}_{p,k} = \textbf{x}((\xi,\eta))_{m},~
	\textbf{n}_{p,k} =(\textbf{n}_{p})_{m},~~m=1,2,3.
	\end{equation*}
    The quadratic element reduces to the linear element when every edge is a straight line.

According to the coordinate transformation, the local coordinate for
the cell interface $\Gamma_{ip}$ is expressed as
$(\widetilde{x}_1,\widetilde{x}_2,\widetilde{x}_3)^T=(0, \widetilde{x}_2,, \widetilde{x}_3)^T$, where
$(\widetilde{x}_2,, \widetilde{x}_3)^T \in \Gamma_{ip}$, and the
velocities in the local coordinate are given by
\begin{equation}\label{rotate-3d}
\begin{split}
\begin{cases}
&\tilde{u_1}=u_1 n_1 + u_2 n_2 + u_3 n_3 , \\
&\tilde{u_2}=-u_1 n_2 + u_2 (n_1+\frac{n_3^2}{1+n_1}) - u_3 \frac{n_2n_3}{1+n_1} , \\[3pt]
&\tilde{u_3}=-u_1 n_3 - u_2 \frac{n_2n_3}{1+n_1}  + u_3 (1-\frac{n_3^2}{1+n_1}),
\end{cases}
~~~ n_1\neq -1.
\end{split}
\end{equation}

The macroscopic conservative flow variables in the local coordinate are expressed as
\begin{align*}
\widetilde{\textbf{W}}(\widetilde{\textbf{x}},t)=\textbf{T}\textbf{W}(\textbf{x},t) ,
\end{align*}
where $\textbf{T}$ is the rotation matrix
\begin{equation}\label{rotate-matrix-3d}
\textbf{T}= \left(
\begin{array}{ccccc}
1 &0 & 0 & 0 & 0 \\
0 & n_1 &n_2 & n_3 &0 \\
0 &-n_2 &n_1+\frac{n_3^2}{1+n_1} & -\frac{n_2n_3}{1+n_1} & 0 \\[3pt]
0 &-n_3 &-\frac{n_2n_3}{1+n_1} & 1-\frac{n_3^2}{1+n_1} &0  \\
0 &0 & 0 & 0 & 1 \\
\end{array}
\right),~~~ n_1\neq -1.
\end{equation}
Note that when $n_1=-1$, Eq.\eqref{rotate-3d} changes to
$(\tilde{u_1},\tilde{u_2},\tilde{u_3})^T=(-u_1,-u_2,u_3)^T$ and
the matrix \eqref{rotate-matrix-3d} is replaced by a diagonal matrix
$\Lambda=\text{diag}(1,-1,-1,1,1)$.

For the gas distribution function in the local coordinate,
$\widetilde{f}(\widetilde{\textbf{x}},t,\widetilde{\textbf{u}},\xi)=f(\textbf{x},t,\textbf{u},\xi)$
and $|\text{d}\textbf{u}|=|\text{d}\widetilde{\textbf{u}}|$, then the numerical fluxes
can be transformed as
\begin{align}\label{flux-global-local}
\textbf{F}(\textbf{x},t)=\int\pmb{\psi} f(\textbf{x},t,\textbf{u},\xi) \textbf{u} \cdot \textbf{n}_p
\text{d} \textbf{u} \text{d} \xi
=\int\pmb{\psi}\widetilde{f}(\widetilde{\textbf{x}},t,\widetilde{\textbf{u}},\xi)
\widetilde{u}_1 \text{d} \widetilde{\textbf{u}} \text{d}\xi .
\end{align}
In the computation, the fluxes are obtained firstly by taking moments of the gas distribution function in the local coordinates
\begin{align}\label{flux-local}
\widetilde{\textbf{F}}(\widetilde{\textbf{x}},t)=\int \widetilde{\pmb{\psi}} \widetilde{f}(\widetilde{\textbf{x}},t,\widetilde{\textbf{u}},\xi)
\widetilde{u}_1 \text{d} \widetilde{\textbf{u}} \text{d}\xi,
\end{align}
where
$\widetilde{\pmb{\psi}}=(1,\widetilde{\textbf{u}},\displaystyle\frac{1}{2}(\widetilde{\textbf{u}}^2+\xi^2))^T$.
According to Eq.\eqref{rotate-3d}, Eq.\eqref{flux-global-local} and Eq.\eqref{flux-local}, the fluxes
in the global coordinate can be expressed as a combination of the
fluxes in the local coordinate
\begin{align}\label{flux-local-to-global}
\textbf{F}(\textbf{W}(\textbf{x},t))\cdot\textbf{n}=
\textbf{T}^{-1}\widetilde{\textbf{F}}(\widetilde{\textbf{W}}(\widetilde{\textbf{x}},t)).
\end{align}

\subsection{Gas-kinetic solver}

In order to construct the numerical
fluxes at $\textbf{x}=(0,0,0)^T$, the integral solution of BGK equation Eq.\eqref{bgk} is used
\begin{equation}\label{integral1}
f(\textbf{x},t,\textbf{u},\xi)=\frac{1}{\tau}\int_0^t g(\textbf{x}',t',\textbf{u},\xi)e^{-(t-t')/\tau}\text{d}t'
+e^{-t/\tau}f_0(\textbf{x}-\textbf{u}t,\textbf{u},\xi),
\end{equation}
where $\textbf{x}=\textbf{x}'+\textbf{u}(t-t')$ is the particle trajectory. $f_0$ is the initial gas distribution function, $g$ is the corresponding
equilibrium state in space and time.
The integral solution basically states a physical process from the particle free transport in $f_0$ in the kinetic scale
to the hydrodynamic flow evolution in the integration of $g$ term.
The flow evolution at the cell interface depends on the ratio of time step
to the  local particle collision time $\Delta t/\tau$.

To construct a time evolving gas distribution function at a cell interface,
the following notations are introduced first
\begin{align*}
a_{x_i} \equiv  (\partial g/\partial x_i)/g=g_{x_i}/g,
A \equiv (\partial g/\partial t)/g=g_t/g,
\end{align*}
where $g$ is the equilibrium state.  The variables $(a_{x_i}, A)$, denoted by $s$,
depend on particle velocity in the form of
\cite{GKS-2001}
\begin{align*}
s=s_j\psi_j =s_{1}+s_{2}u_1+s_{3}u_2+s_{4}u_3
+s_{5}\displaystyle \frac{1}{2}(u_1^2+u_2^2+u_3^2+\xi^2).
\end{align*}
The initial gas distribution function in the solution \eqref{integral1} can be modeled as
\begin{equation*}
f_0=f_0^l(\textbf{x},\textbf{u})\mathbb{H} (x_1)+f_0^r(\textbf{x},\textbf{u})(1- \mathbb{H}(x_1)),
\end{equation*}
where $\mathbb{H}(x_1)$ is the Heaviside function. Here $f_0^l$ and $f_0^r$ are the
initial gas distribution functions on both sides of a cell
interface, which have one to one correspondence with the initially
reconstructed macroscopic variables. The first-order
Taylor expansion for the gas distribution function in space around
$\textbf{x}=\textbf{0}$ can be expressed as
\begin{align}\label{flux-3d-1}
f_0^k(\textbf{x})=f_G^k(\textbf{0})+\frac{\partial f_G^k}{\partial x_i}(\textbf{0})x_i
=f_G^k(\textbf{0})+\frac{\partial f_G^k}{\partial x_1}(\textbf{0})x_1
+\frac{\partial f_G^k}{\partial x_2}(\textbf{0})x_2
+\frac{\partial f_G^k}{\partial x_3}(\textbf{0})x_3,
\end{align}
for $k=l,r$.
According to Eq.\eqref{ce-ns}, $f_{G}^k$ has the form
\begin{align}\label{flux-3d-2}
f_{G}^k(\textbf{0})=g^k(\textbf{0})-\tau(u_ig_{x_i}^{k}(\textbf{0})+g_t^k(\textbf{0})),
\end{align}
where $g^k$ is  the equilibrium state with the form of a Maxwell distribution.
$g^k$ can be fully determined from  the
reconstructed macroscopic variables $\textbf{W}
^l, \textbf{W}
^r$ at the left and right sides of a cell interface
\begin{align}\label{get-glr}
\int\pmb{\psi} g^{l}\text{d}\Xi=\textbf{W}
^l,\int\pmb{\psi} g^{r}\text{d}\Xi=\textbf{W}
^r.
\end{align}
Substituting Eq.\eqref{flux-3d-1} and Eq.\eqref{flux-3d-2} into Eq.\eqref{integral1},
the kinetic part for the integral solution can be written as
\begin{equation}\label{dis1}
\begin{aligned}
e^{-t/\tau}f_0^k(-\textbf{u}t,\textbf{u},\xi)
=e^{-t/\tau}g^k[1-\tau(a_{x_i}^{k}u_i+A^k)-ta^{k}_{x_i}u_i],
\end{aligned}
\end{equation}
where the coefficients $a_{x_1}^{k},...,A^k, k=l,r$ are defined according
to the expansion of $g^{k}$.
After determining the kinetic part
$f_0$, the equilibrium state $g$ in the integral solution
Eq.\eqref{integral1} can be expanded in space and time as follows
\begin{align}\label{equli}
g(\textbf{x},t)= g^{c}(\textbf{0},0)+\frac{\partial  g^{c}}{\partial x_i}(\textbf{0},0)x_i+\frac{\partial  g^{c}}{\partial t}(\textbf{0},0)t,
\end{align}
where $ g^{c}$ is the Maxwellian equilibrium state located at an interface.
Similarly, $\textbf{W}^c$ are the macroscopic flow variables for the determination of the
equilibrium state $ g^{c}$
\begin{align}\label{compatibility2}
\int\pmb{\psi} g^{c}\text{d}\Xi=\textbf{W}^c.
\end{align}
Substituting Eq.\eqref{equli} into Eq.\eqref{integral1}, the hydrodynamic part in the integral solution
can be written as
\begin{equation}\label{dis2}
\begin{aligned}
\frac{1}{\tau}\int_0^t
g&(\textbf{x}',t',\textbf{u},\xi)e^{-(t-t')/\tau}\text{d}t'
=C_1 g^{c}+C_2 a_{x_i}^{c} u_i g^{c} +C_3 A^{c} g^{c} ,
\end{aligned}
\end{equation}
where the coefficients
$a_{x_i}^{c},A^{c}$ are
defined from the expansion of the equilibrium state $ g^{c}$. The
coefficients $C_m, m=1,2,3$ in Eq.\eqref{dis2}
are given by
\begin{align*}
C_1=1-&e^{-t/\tau}, C_2=(t+\tau)e^{-t/\tau}-\tau, C_3=t-\tau+\tau e^{-t/\tau}.
\end{align*}
The coefficients in Eq.\eqref{dis1} and Eq.\eqref{dis2}
can be determined by the spatial derivatives of macroscopic flow
variables and the compatibility condition as follows
\begin{align}\label{co}
&\langle a_{x_1}\rangle =\frac{\partial \textbf{W} }{\partial x_1}=\textbf{W}_{x_1},
\langle a_{x_2}\rangle =\frac{\partial \textbf{W} }{\partial x_2}=\textbf{W}_{x_2},
\langle a_{x_3}\rangle =\frac{\partial \textbf{W} }{\partial x_3}=\textbf{W}_{x_3},\nonumber\\
&\langle A+a_{x_1}u_1+a_{x_2}u_2+a_{x_3}u_3\rangle=0,
\end{align}
where $\left\langle ... \right\rangle$ are the moments of a gas distribution function defined by
\begin{align}\label{co-moment}
\langle (...) \rangle  = \int \pmb{\psi} (...) g \text{d} \Xi .
\end{align}
The details for the evaluation of each term from macroscopic variables can be found in \cite{ji2019high}.

In  smooth flow region, the collision time is determined by $\tau=\mu/p$, where $\mu$ is the dynamic viscosity coefficient and $p$ is the pressure at the cell interface.
In order to properly capture the un-resolved shock structure, additional numerical dissipation is needed.
The physical collision time $\tau$ in the exponential function part can be replaced by a numerical collision time $\tau_n$.
For the
inviscid flow, the collision time $\tau_n$ is modified as
\begin{align*}
\tau_n=\varepsilon \Delta t+C\displaystyle|\frac{p_l-p_r}{p_l+p_r}|\Delta
t,
\end{align*}
where $\varepsilon=0.01$ and $C=1$.
For the viscous flow, the collision time is related to the viscosity coefficient,
\begin{align*}
\tau_n=\frac{\mu}{p}+C \displaystyle|\frac{p_l-p_r}{p_l+p_r}|\Delta t,
\end{align*}
where $p_l$ and $p_r$ denote the pressure on the left and right
sides of the cell interface.
The inclusion of the pressure jump term is to increase  the non-equilibrium transport mechanism in the flux function to mimic the physical process in the shock layer.
Then substitute Eq.\eqref{dis1} and Eq.\eqref{dis2} into Eq.~\eqref{integral1} with $\tau$ and $\tau_n$, the final second-order time dependent gas distribution function becomes
\begin{align}\label{2nd-flux}
f(\textbf{0},t,\textbf{u},\xi)
=&(1-e^{-t/\tau_n}) g^{c}+[(t+\tau)e^{-t/\tau_n}-\tau]a_{x_i}^{c}u_i g^{c}\nonumber
+(t-\tau+\tau e^{-t/\tau_n})A^{c}  g^{c}\nonumber\\
+&e^{-t/\tau_n}g^l[1-(\tau+t)a_{x_i}^{l}u_i-\tau A^l]H(u_1)\nonumber\\
+&e^{-t/\tau_n}g^r[1-(\tau+t)a_{x_i}^{r}u_i-\tau A^r] (1-H(u_1)).
\end{align}

For smooth flow, the time dependent solution in Eq.~\eqref{2nd-flux} can be simplified as \cite{GKS-2001},
\begin{align}\label{2nd-smooth-flux}
f(\textbf{0},t,\textbf{u},\xi)= g^{c}-\tau (a^c_{x_{i}} u_i +A^c)g^{c}+A^c g^{c}t,
\end{align}	
under the assumptions of $g^{l,r}=g^c$, $a^{l,r}_{x_i}=a^c_{x_i}$.
The above gas-kinetic solver for smooth flow has less numerical dissipation than that from the full GKS solver in Eq.~\eqref{2nd-flux}.

\subsection{Direct evolution of the cell averaged first-order spatial derivatives} \label{slope-section}

As shown in Eq.~\eqref{2nd-flux}, a time evolution solution at a cell interface is provided by the gas-kinetic solver,
which is distinguishable from the Riemann solver with a constant solution.
Recall Eq.(\ref{point}), the conservative variables at the Gaussian point  $\textbf{x}_{p,k}$ can be updated by taking moments $\pmb{\psi}$
on the gas distribution function,
\begin{align}\label{point-interface}
\textbf{W}_{p,k}(t^{n+1})=\int \pmb{\psi} f^n(\textbf{x}_{p,k},t^{n+1},\textbf{u},\xi) \text{d}\Xi,~ k=1,...,M.
\end{align}

Then the cell-averaged first-order derivatives within each element at $t^{n+1}$ is given through the Gauss's theorem,
\begin{equation}\label{gauss-formula}
\begin{aligned}
\overline{W}_x^{n+1}
&=\int_{V} \nabla \cdot (\overline{W}(t^{n+1}),0,0) \text{d}V
=\frac{1}{ \Delta V}\int_{\partial V} (1,0,0) \cdot \textbf{n}  \overline{W}(t^{n+1}) \text{d}S \\
&=\frac{1}{ \Delta V}\int_{\partial V} \overline{W}(t^{n+1}) n_1   \text{d}S
=\frac{1}{ \Delta V} \sum_{p=1}^{N_f}\sum_{k=1}^{M} \omega_{p,k} W^{n+1}_{p,k} (n_1)_{p,k} \Delta S_p,
\\
\overline{W}_y^{n+1}
&=\int_{V} \nabla \cdot (0,\overline{W}(t^{n+1}),0) \text{d}V
= \frac{1}{ \Delta V}\int_{\partial V} (0,1,0) \cdot \textbf{n}  \overline{W}(t^{n+1}) \text{d}S \\
& = \frac{1}{ \Delta V}\int_{\partial V} \overline{W}(t^{n+1}) n_2   \text{d}S
=\frac{1}{ \Delta V} \sum_{p=1}^{N_f}\sum_{k=1}^{M} \omega_{p,k} W^{n+1}_{p,k} (n_2)_{p,k} \Delta S_p,
\\
\overline{W}_z^{n+1}
&=\int_{V} \nabla \cdot (0,0,\overline{W}(t^{n+1})) \text{d}V
=\frac{1}{ \Delta V}\int_{\partial V} (0,0,1) \cdot \textbf{n}  \overline{W}(t^{n+1}) \text{d}S \\
&=\frac{1}{ \Delta V}\int_{\partial V} \overline{W}(t^{n+1}) n_3 \text{d}S
=\frac{1}{ \Delta V} \sum_{p=1}^{N_f}\sum_{k=1}^{M} \omega_{p,k} W^{n+1}_{p,k} (n_3)_{p,k} \Delta S_p,
\end{aligned}
\end{equation}
where $\textbf{n}_{p,k}=((n_{1})_{p,k},(n_{2})_{p,k},(n_{3})_{p,k})$ is the outer unit normal direction at each Gaussian point $\textbf{x}_{p,k}$.

\section{Two-stage temporal discretization}

The two-stage fourth-order (S2O4) temporal discretization is
adopted here as that in the previous compact GKSs \cite{ji2020hweno,zhao2020compact,ji2020three}.
Following the definition of Eq.\eqref{semidiscrete},
a fourth-order time-accurate solution for cell-averaged conservative flow variables $\textbf{W}_i$ are updated by
\begin{equation}\label{s2o4}
\begin{aligned}
\textbf{W}_i^*&=\textbf{W}_i^n+\frac{1}{2}\Delta t\mathcal
{L}(\textbf{W}_i^n)+\frac{1}{8}\Delta t^2\frac{\partial}{\partial
	t}\mathcal{L}(\textbf{W}_i^n), \\
\textbf{W}_i^{n+1}&=\textbf{W}_i^n+\Delta t\mathcal
{L}(\textbf{W}_i^n)+\frac{1}{6}\Delta t^2\big(\frac{\partial}{\partial
	t}\mathcal{L}(\textbf{W}_i^n)+2\frac{\partial}{\partial
	t}\mathcal{L}(\textbf{W}_i^*)\big),
\end{aligned}
\end{equation}
where
$\mathcal{L}(\textbf{W}_i^n)$ and $\frac{\partial}{\partial t}\mathcal{L}(\textbf{W}_i^n)$ are given by
\begin{equation} \label{flux-operator}
\begin{aligned}
\mathcal{L}(\textbf{W}_i^n)&= -\frac{1}{\left| \Omega_i \right|} \sum_{p=1}^{N_f}
\sum_{k=1}^{M} \omega_{p,k} \textbf{F}(\textbf{x}_{p,k},t_n)\cdot \textbf{n}_{p,k},\\
\frac{\partial}{\partial t}\mathcal{L}(\textbf{W}_i^n)&= -\frac{1}{\left| \Omega_i \right|} \sum_{p=1}^{N_f}\sum_{k=1}^{M} \omega_{p,k}
\partial_t \textbf{F}(\textbf{x}_{p,k},t_n)\cdot \textbf{n}_{p,k}, \\
\frac{\partial}{\partial t}\mathcal{L}(\textbf{W}_{i}^*)&=-\frac{1}{\left| \Omega_i \right|} \sum_{p=1}^{N_f}\sum_{k=1}^{M} \omega_{p,k}
\partial_t \textbf{F}(\textbf{x}_{p,k},t_*)\cdot \textbf{n}_{p,k}.
\end{aligned}
\end{equation}
The proof for the fourth-order accuracy in time is shown in \cite{li2016twostage}.

In order to obtain the numerical fluxes $\textbf{F}_{p,k}$ and their time derivatives  $\partial_t \textbf{F}_{p,k}$ at $t_n$ and $t_*=t_n + \Delta t/2$,
the time accurate solution in Eq.\eqref{2nd-flux} can be approximated as a linear function of time.
Let's first
introduce the following notation,
\begin{align*}
\mathbb{F}_{p,k}(\textbf{W}^n,\delta)=\int_{t_n}^{t_n+\delta} \textbf{F}_{p,k}(\textbf{W}^n,t)\text{d}t.
\end{align*}
For convenience, assume $t_n=0$,
the flux in the time interval $[t_n, t_n+\Delta t]$ is expanded in
the linear form
\begin{align*}
\textbf{F}_{p,k}(\textbf{W}^n,t)=\textbf{F}_{p,k}^n+ t \partial_t \textbf{F}_{p,k}^n  .
\end{align*}
The coefficients $\textbf{F}_{p,k}^n$ and $\partial_t\textbf{F}_{p,k}^n$ can be
fully determined by
\begin{align*}
\textbf{F}_{p,k}(\textbf{W}^n,t_n)\Delta t&+\frac{1}{2}\partial_t
\textbf{F}_{p,k}(\textbf{W}^n,t_n)\Delta t^2 =\mathbb{F}_{p,k}(\textbf{W}^n,\Delta t) , \\
\frac{1}{2}\textbf{F}_{p,k}(\textbf{W}^n,t_n)\Delta t&+\frac{1}{8}\partial_t
\textbf{F}_{p,k}(\textbf{W}^n,t_n)\Delta t^2 =\mathbb{F}_{p,k}(\textbf{W}^n,\Delta t/2).
\end{align*}
By solving the linear system, we have
\begin{equation}\label{linear-system}
\begin{aligned}
\textbf{F}_{p,k}(\textbf{W}^n,t_n)&=(4\mathbb{F}_{p,k}(\textbf{W}^n,\Delta t/2)-\mathbb{F}_{p,k}(\textbf{W}^n,\Delta t))/\Delta t,\\
\partial_t \textbf{F}_{p,k}(\textbf{W}^n,t_n)&=4(\mathbb{F}_{p,k}(\textbf{W}^n,\Delta t)-2\mathbb{F}_{p,k}(\textbf{W}^n,\Delta t/2))/\Delta
t^2.
\end{aligned}
\end{equation}
Finally, with Eq.\eqref{flux-operator} and \eqref{linear-system},
$W_{i}^{n+1}$ at $t^{n+1}$ can be updated by Eq.\eqref{s2o4}.

The time dependent gas distribution function at a cell interface is updated in a similar way,
\begin{equation}\label{step-du}
\begin{split}
&f^*=f^n+\frac{1}{2}\Delta tf_t^n,\\
&f^{n+1}=f^n+\Delta tf_{t}^*.
\end{split}
\end{equation}

In order to construct the first-order time derivative of the gas distribution function,
the distribution function in Eq.(\ref{2nd-flux}) is approximated by the linear function
\begin{align*}
f(t)=f(\textbf{x}_{p,k},t,\textbf{u},\xi)=f^n+
f_{t}^n(t-t^n).
\end{align*}
According to the gas-distribution function at
$t=0$ and $\Delta t$
\begin{align*}
f^n&=f(0),\\
f^n&+f_{t}^n\Delta t=f(\Delta t),
\end{align*}
the coefficients $f^n, f_{t}^n$ can be
determined by
\begin{align*}
f^n&=f(0),\\
f^n_t&=(f(\Delta t)-f(0))/\Delta t.\\
\end{align*}
Thus, $f^*$ and $f^{n+1}$ are fully determined and the macroscopic flow variables at the cell interface can be obtained by Eq.~\eqref{point}.
Theoretically, a fourth-order temporal accuracy can be achieved for the conservative flow variables on arbitrary mesh. The proof is given  in \cite{zhao2020compact}.

\section{Compact HWENO reconstruction}

In this section, a compact HWENO-type reconstruction is designed to get the piecewise discontinuous flow variables and their first-order derivatives
at each Gaussian point on both sides of a cell interface.
The reconstruction procedure for an inner cell is given first, then the special treatment for the boundary cell is presented in subsequent section  \ref{boundary-recon}.

\subsection {Smooth reconstruction}

As a starting point of WENO reconstruction, a linear reconstruction
will be presented first. For
a piecewise smooth function $Q( \textbf{x} )$ over cell $\Omega_{0}$, a
polynomial $P^r(\textbf{x})$ with degree $r$ can be constructed to
approximate $Q(\textbf{x})$ as follows
\begin{equation*}
P^r(\textbf{x})=Q(\textbf{x})+O(\Delta h^{r+1}),
\end{equation*}
where $\Delta h \sim |\Omega_{0}|^{\frac{1}{3}}$ is the equivalent cell size.
In order to achieve a third-order accuracy and satisfy conservative property,
the following quadratic polynomial over cell $\Omega_{0}$ is obtained
\begin{equation}\label{p2-def}
P^2(\textbf{x})= \overline{Q}_{0}+\sum_{|k|=1}^2a_kp^k(\textbf{x}),
\end{equation}
where $\overline{Q}_{0}$ is the cell averaged value of $Q(\textbf{x})$ over cell $\Omega_{0}$, $k=(k_1,k_2,k_3)$, $|k| = k_1+k_2+k_3$.
The $p^k(\textbf{x})$ are basis functions, which are given by
\begin{align}\label{base}
\displaystyle p^k(\textbf{x})=x_1^{k_1}x_2^{k_2}x_3^{k_3}-\frac{1}{\left| \Omega_{0} \right|}\displaystyle\iiint_{\Omega_{0}}x_1^{k_1}x_2^{k_2}x_3^{k_3} \text{d}V.
\end{align}

The controlling points for a quadratic tetrahedron are shown in Fig.~\ref{tetra}.
The iso-parametric transformation is used to evaluate the volume integral, which can be written as
	\begin{align*}
	\textbf{X} (\xi, \eta, \zeta)= \sum_{i=0}^9 \textbf{x}_i \phi_i (\xi, \eta, \zeta),
	\end{align*}
	where  $\textbf{x}_i$ is the location of the ith controlling point
	and $\phi_i$ is the base function as follows \cite{wang2017thesis},
	\begin{equation}
	\begin{array}{l}
	v_{1}=(-1+\zeta+\eta+\xi)(-1+2 \zeta+2 \eta+2 \xi),\\
	v_{2}=\xi(-1+2 \xi), ~~
	v_{3}=\eta(-1+2 \eta), ~~
	v_{4}=\zeta(-1+2 \zeta), ~~\\
	v_{5}=-4 \xi(-1+\zeta+\eta+\xi),~~
	v_{6}=4 \eta \xi,~~\\
	v_{7}=-4 \eta(-1+\zeta+\eta+\xi),~~
	v_{8}=-4 \zeta(-1+\zeta+\eta+\xi),~~\\
	v_{9}=4 \zeta \xi,~~
	v_{10}=4 \zeta \eta .
	\end{array}
	\end{equation}
	Then, the integration of monomial in Eq.~\eqref{base} becomes
	\begin{equation} \label{base-int-change}
	\begin{array}{l}
	\int_{\Omega} x^{k_1} y^{k_2} z^{k_3} \mathrm{d} x \mathrm{d} y \mathrm{d} z
	=\int_{\tilde{\Omega}} x^{k_1} y^{k_2} z^{k_3} (\xi,\eta,\zeta)
	\left|\frac{\partial(x, y, z)}{\partial(\xi, \eta, \zeta)}\right|
	\mathrm{d} \xi \mathrm{d} \eta \mathrm{d} \zeta.
	\end{array}\end{equation}
	It can be evaluated numerically as
	\begin{equation}\begin{array}{l}
	\iiint_{\Omega} x^{k_1} y^{k_2} z^{k_3} \mathrm{d} x \mathrm{d} y \mathrm{d} z
	=\sum_{m=1}^{M} \omega_{m} x^{k_1} y^{k_2} z^{k_3} \left(\xi, \eta, \zeta\right)_{m} \left|\frac{\partial(x, y, z)}{\partial(\xi, \eta, \zeta)}\right|_{m} \Delta \xi \Delta \eta \Delta \zeta,
	\end{array}\end{equation}
	where $\omega_{m}$ is the quadrature weight at the Gaussian point $\left(\xi, \eta, \zeta\right)_{m}$ and $\Delta \xi = \Delta \eta = \Delta \zeta =1 $.
	A five-point Gaussian quadrature with fourth-order spatial accuracy is used with
	\begin{equation*}
	\begin{split}
	&(\xi,\eta,\zeta)_{1}=(\frac{1}{4} ,\frac{1}{4} ,\frac{1}{4}),
	~(\xi,\eta,\zeta)_{2}=(\frac{1}{2},\frac{1}{6},\frac{1}{6}),
	~(\xi,\eta,\zeta)_{3}=(\frac{1}{6},\frac{1}{6},\frac{1}{6}),\\
	&~(\xi,\eta,\zeta)_{4}=(\frac{1}{6},\frac{1}{6},\frac{1}{2}),
	~(\xi,\eta,\zeta)_{5}=(\frac{1}{6},\frac{1}{2},\frac{1}{6}),
	\end{split}
	\end{equation*}
	with $\omega_{1} = -\frac{1}{5},~ \omega_{m}=\frac{3}{40},~m=2,3,4$.

\subsubsection{Stencil for polynomial $P^2(\textbf{x})$ }
In order to achieve a third-order spatial accuracy,
the quadratic polynomial $P^2(\textbf{x})$ on $\Omega_{0}$ is constructed on
the compact stencil $S_2$ including  $\Omega_{0}$ and its all von Neumann neighbors, $\Omega_{m}, m=1,...,4$,
where the averages of $Q(\textbf{x})$ and averaged derivatives of $Q(\textbf{x})$ over each cell are known.

The following values on $S_2$ are used to obtain $P^2(\textbf{x})$,

\begin{itemize}
	\item cell averages $\overline{Q}$ for cell 0, 1, 2, 3, 4,
	\item cell averages of the $x$-direction partial derivative $\overline{Q}_{x_1}$ for cell 1, 2, 3, 4;
	\item cell averages of the $y$-direction partial derivative $\overline{Q}_{x_2}$ for cell 1, 2, 3, 4;
	\item cell averages of the $z$-direction partial derivative $\overline{Q}_{x_3}$ for cell 1, 2, 3, 4.
\end{itemize}

The polynomial $P^2(\textbf{x})$ is required to exactly satisfy
\begin{align} \label{large-stenci-condition-1}
\iiint_{\Omega_{m}}P^2(\textbf{x})\text{d}V=\overline{Q}_{m}\left| \Omega_{m}\right|,
\end{align}
where $\overline{Q}_m$ is the cell averaged value over $\Omega_{m},~ m=1,...,4$,
with the following condition satisfied in a least-square sense
\begin{equation}\label{large-stenci-condition-2}
\begin{split}
\iiint_{\Omega_{m}}
\frac{\partial}{\partial x_1} P^2(\textbf{x})\text{d}V=(\overline{Q}_{x_1})_m|\Omega_{m}|,\\
\iiint_{\Omega_{m}}
\frac{\partial}{\partial x_2} P^2(\textbf{x})\text{d}V=(\overline{Q}_{x_2})_m|\Omega_{m}|,\\
\iiint_{\Omega_{m}}
\frac{\partial}{\partial x_3} P^2(\textbf{x})\text{d}V=(\overline{Q}_{x_3})_m|\Omega_{m}|,
\end{split}	
\end{equation}
where  $\overline{Q}_{x_i}, i=1,2,3$  are the cell averaged directional derivatives over $\Omega_{m}$ in a global coordinate,
respectively.
On a regular mesh, the system has $16$ independent equations.
The constrained least-square method is used to solve the above linear system \cite{li2014efficient}.
The above reconstruction improves the linear stability of the scheme and reduces the numerical errors.
The left and right states $W^{l,r}$ provided by the reconstructed $P^2(\textbf{x})$ yields a linearly stable third-order compact GKS, as validated through numerical tests in Section \ref{test-case}.

\subsubsection{Stencils for polynomials $P^1(\textbf{x})$ and  $P^0(\textbf{x})$ }

In order to deal with discontinuity, lower-order polynomials from the sub-stencils should be determined.
Following the multi-resolution reconstruction in \cite{zhu2020new},
the first-order polynomial $P^1(\textbf{x})$ is determined from the same central stencil as the $P^2(\textbf{x})$ but only with the cell-averaged conservative variables
\begin{itemize}
	\item $\overline{Q}$ for cell 0, 1, 2, 3, 4.
\end{itemize}
The polynomial $P^1(\textbf{x})$ is required to satisfy
\begin{align} \label{p1-stenci-condition}
\iiint_{\Omega_{m}}P^1(\textbf{x})\text{d}V=\overline{Q}_{m}\left| \Omega_{m}\right|, ~~m=1,2,3,4,
\end{align}
in a least-square sense.

Note that the left and right states $W^{l,r}$ solely determined by the reconstructed $P^1(\textbf{x})$ yield an unstable second-order GKS.
The theoretical proof for such instability on the second-order Riemann solver-based-FVM can be found in \cite{haider2009stability}.
The zeroth-order polynomial $P^0(\textbf{x})$ is simply determined by the cell-averaged conservative variables on the targeted cell $\Omega_{0}$ itself,
i.e. $P^0(\textbf{x}) = \overline{Q}_0$.
The coefficient matrices for the above $P^j(\textbf{x}), j = 0,1$ are always invertible.

\subsection{Multi-resolution WENO procedure}
Define three polynomials
\begin{equation}
\begin{split}
p_{2}(\textbf{x})&=\frac{1}{\gamma_{2,2}} P^2(\textbf{x})-\sum_{\ell=0}^{1} \frac{\gamma_{\ell, 2}}{\gamma_{2,2}} p_{\ell}(\textbf{x}),\\
p_{1}(\textbf{x})&=\frac{1}{\gamma_{1,1}} P^{1}(\textbf{x})-\frac{\gamma_{0,1}}{\gamma_{1,1}} P^{0}(\textbf{x}),\\
p_{0}(\textbf{x})&= P^0(\textbf{x}).
\end{split}
\end{equation}
For a third-order reconstruction, the second-order polynomial $P^{2}(\textbf{x})$ can be rewritten as
\begin{equation} \label{rewrite_expression-3rd}
P^2(\textbf{x})=\gamma_{2,2}p_{2}+\gamma_{1,2}p_{1}+\gamma_{0,2}p_{0}
\end{equation}
with arbitrary positive coefficients $\gamma_{m,n}$ satisfying $\gamma_{0,2}+\gamma_{1,2}+\gamma_{2,2}=1, \gamma_{0,1}+\gamma_{1,1}=1 $.

For a second-order reconstruction, the first-order polynomial $P^{1}(\textbf{x})$ can be rewritten as
\begin{equation} \label{rewrite_expression-2nd}
P^1(\textbf{x})=\gamma_{1,1}p_{1}+\gamma_{0,1}p_{0}
\end{equation}
with arbitrary positive coefficients $\gamma_{m,n}$ satisfying $\gamma_{0,1}+\gamma_{1,1}=1 $.
The coefficients are chosen as $\gamma_{2,2}:\gamma_{1,2}:\gamma_{0,2} = 100:10:1$, and
$\gamma_{1,1}:\gamma_{0,1}=10:1$ as suggested in \cite{zhu2020new}.

The smoothness indicators $\beta_{j}, j=1,2$ are defined as
\begin{equation}\label{smooth-indicator}
\beta_j=\sum_{|\alpha|=1}^{r_j}|\Omega|^{ \frac{2}{3}|\alpha|-1}\iiint_{\Omega}\big(D^{\alpha}P_j(\textbf{x})\big)^2
\text{d} V,
\end{equation}
where $\alpha$ is a multi-index and $D$ is the derivative operator, $r_1=1$, $r_2=2$.
The smoothness indicators in Taylor series at $(x_0,y_0)$ have the order
\begin{align*}
\beta_2&=O\{|\Omega_0|^{\frac{2}{3}}[1+O(|\Omega_0|^{\frac{2}{3}})]\}=O(|\Omega_0|)^{\frac{2}{3}} = O(h^2),\\
\beta_1&=O\{|\Omega_0|^{\frac{2}{3}}[1+O(|\Omega_0|^{\frac{1}{3}})]\}=O(|\Omega_0|)^{\frac{2}{3}} = O(h^2).
\end{align*}
Assuming a suitable $\beta_0$,
\begin{align*}
\beta_0&=O\{|\Omega_0|^{\frac{2}{3}}[1+O(|\Omega_0|^{\frac{1}{3}})]\}=O(|\Omega_0|)^{\frac{2}{3}} = O(h^2),
\end{align*}
 a global smoothness indicator $\sigma$ similar to that in \cite{zhu2020new} can be defined
\begin{equation*}
\sigma^{3rd} = (\frac{1}{2}(|\beta_2-\beta_1|+|\beta_2-\beta_0|))^{\frac{4}{3}} = O(|\Omega_0|^2) = O(h^4),
\end{equation*}
and
\begin{equation*}
\sigma^{2nd} = |\beta_1-\beta_0|^{\frac{4}{3}} = O(|\Omega_0|^2) = O(h^4).
\end{equation*}
Then, the corresponding non-linear weights are given by
\begin{equation}\label{non-linear-weight}
\begin{split}
&\omega_{m,n}=\gamma_{m,n}(1+(\frac{\sigma}{\epsilon+\beta_{m}})^2), \\
&\bar{\omega}_{m,n}=\frac{\bar{\omega}_{m,n}}{\sum \omega_{m,n}} = {\gamma}_{m,n} + O(h^4),
\end{split}
\end{equation}
where $m=0,1,2$ for $n=2$ and $m=0,1$ for $n=1$, and $\epsilon$ takes $10^{-8}$ to avoid zero in the denominator.

Replacing $\gamma_{m,n}$ by the normalized non-linear weights $\bar{\omega}_{m,n}$ in Eq.~\eqref{rewrite_expression-3rd} and Eq.~\eqref{rewrite_expression-2nd}, the final reconstructed polynomials are given by
\begin{equation} \label{final_weno_expression-3rd}
R^{3rd}(\textbf{x})=\bar{\omega}_{2,2}p_{2}+\bar{\omega}_{1,2}p_{1}+\bar{\omega}_{0,2}p_{0}
\end{equation}
for a third-order spatial accuracy,
and
\begin{equation} \label{final_weno_expression-2nd}
R^{2nd}(\textbf{x})=\bar{\omega}_{1,1}p_{1}+\bar{\omega}_{0,1}p_{0}
\end{equation}
for a second-order spatial accuracy.

As a result, the non-linear reconstruction meets the requirement for a third-order accuracy  $R(\textbf{x})=P(\textbf{x})+O(h^3)$.
If any of these values yield negative density or pressure,  the first-order reconstruction is used instead.
The desired non-equilibrium states at Gaussian points can be obtained from the weighted polynomials
\begin{align} \label{final-weno-value}
&Q^{l,r}_{p,k}=R^{l,r}(\textbf{x}_{p,k}), ~(Q^{l,r}_{x_i})_{p,k}= \frac{\partial R^{l,r}}{\partial {x_i}}(\textbf{x}_{p,k}).
\end{align}

\subsection{A two-step reconstruction}
According to the definition in Eq.~\eqref{smooth-indicator}, the smooth indicator of the zeroth-order polynomial $P^{0}(\textbf{x})$ is always $0$.
So, a new smooth indicator for $P^{0}(\textbf{x})$ has to be defined and can be given as a non-linear combination of the first-order biased sub-stencils as suggested in \cite{zhu2020new}.
One of the choices is
\begin{align*}
P_{1}^1 ~&\text{on}~ S_1=\{\bar{Q}_0,\bar{Q}_1,\bar{Q}_2,\bar{Q}_3 \},
~~~P_{2}^1 ~\text{on}~ S_2=\{\bar{Q}_0,\bar{Q}_1,\bar{Q}_2,\bar{Q}_4 \},\\
P_{3}^1 ~&\text{on}~ S_3=\{\bar{Q}_0,\bar{Q}_1,\bar{Q}_3 ,\bar{Q}_4 \},
~~~P_{4}^1 ~\text{on}~ S_4=\{\bar{Q}_0,\bar{Q}_2,\bar{Q}_3,\bar{Q}_4 \}.
\end{align*}
In this plan, a total of $16\times 9+3\times4+3\times3\times5 = 192$ words is required on each cell for reconstruction.
However, this plan has two drawbacks:
(1). It cannot save the second-order reconstruction from the linear instability,
since the stencil is not extended;
(2). It shows poor robustness for the third-order reconstruction since the coefficient matrices for these sub-stencils can be close to singular with poor mesh quality. In other words, the smooth indicators for $P^{0}$ can be greater than those of $P^{1}$ and $P^{2}$ under irregular gird and the WENO will fail to suppress oscillations.

Inspired by the method in \cite{xia2014finite}, a two-step reconstruction is designed as follows to maintain the compact manner of the scheme:
\begin{itemize}
	\item Reconstruction Step 1: Construct the first-order polynomial $P^{1}(\textbf{x})$  in each cell by Eq.~\eqref{p1-stenci-condition}.
	Compute the slopes, e.g., $b_1,b_2,b_3$ for each component, and store them.
	A coefficient matrix with dimension $3\times4$ is stored for $P^{1}(\textbf{x})$ and another matrix with dimension $5\times3$ for the slopes.
	\item Reconstruction Step 2: Conduct the multi-resolution reconstruction for each cell, and the smooth indicator $\beta_0$ for $P^{0}(\textbf{x})$ is given as a non-linear combination of the smooth indicators of  $P^{1}_j(\textbf{x})$
	from the neighbor cells, i.e.,
	\begin{equation}\label{p0-non-linear-weight}
	\begin{split}
	&\beta_{0,j} = |\Delta_0| (b_{1,j}^2 + b_{2,j}^2 + b_{3,j}^2), \\
	& \sigma^{1st} = (\frac{1}{6}(\sum |\beta_{0,j}-\beta_{0,k}|))^{\frac{4}{3}},\\
	&\omega^{1st}_{j}=1+\frac{\sigma^{1st}}{\epsilon+\beta_{j}}, \\
	&\bar{\omega}^{1st}_{j}=\frac{\omega_{j}}{\sum \omega_{j}},	\\
	&\beta_{0} =\sum \bar{\omega} ^{1st}_{j} \beta _{0,j},
	\end{split}
	\end{equation}	
where $j,k=1,2,3,4$ and $j>k$. Then, the second-order reconstruction is complete.
\end{itemize}
For the third-order reconstruction, only one extra beta $\beta_2$ in Eq.~\eqref{smooth-indicator} is needed.
A coefficient matrix with dimension $9\times16$ is stored for $P^{2}(\textbf{x})$.

Through the reconstruction step 1, the sub-stencils are extended to neighboring cells of neighbors.
Compared with the first plan, the robustness and mesh adaptability is significantly improved.
$16\times 9+3\times4+3\times5 =171$ words are required on each cell for the two-step reconstruction, which is even less than the first plan.
However, one more communication is needed if the code is parallelized on different nodes.
In this paper, only the numerical results based on the two-step reconstruction
 are presented.

\subsection{Reconstruction for the boundary cells} \label{boundary-recon}

The strategy of the two-step and multi-resolution reconstruction is extended to the boundary condition treatment.
The one-sided reconstruction without ghost cell is adopted here with special care on Dirichlet boundary condition, i.e. the non-slip adiabatic wall and the non-slip isothermal one.
For the non-slip adiabatic boundary, the velocities are constrained.
For the non-slip isothermal boundary condition, both the velocities and the temperature are constrained.

\begin{itemize}
\item Reconstruction Step 1:

For ith (i=0,...,4) conservative variables:
\begin{itemize}	
	\item  If there is no constraint for all the boundary faces:
	\begin{itemize}
		\item If the neighboring cell number is no less than 3,
		construct the first-order polynomial $P^1(\textbf{x})$.
		If the coefficient matrix is found to be nearly singular, which suggests a poor mesh quality, set $P^1(\textbf{x}) = P^0(\textbf{x})$.
		\item If the neighboring cell number is less than 3, using the cell-averaged slopes as the slopes of the first-order polynomial instead.
	\end{itemize}
	\item If there exists at least one constraint for all the boundary faces on the targeted cell, the weighted constrained least square reconstruction involving all the neighbor cells and boundary faces are conducted.
	The weights for those boundary faces that do not have constraint are set to be zero.	
	In this step, each constrained boundary face has one constraint, which is located at the geometric center of the face.
	\begin{itemize}	
		\item If the sum of the neighboring cell number and the constraint number is no less than 3,
    and the constraint number is no greater than 3,
		construct the first-order polynomial $P^1(\textbf{x})$ by using constrained least-square method.
		\item If the constraint number is greater than 3 (which is impossible for tetrahedron mesh), construct the first-order polynomial $P^1(\textbf{x})$ by using the least-square method.
		\item If the sum of the neighboring cell number and the constraint number is less than 3, use the cell-averaged slopes as the slopes of the first-order polynomial instead.
	\end{itemize}
 For non-slip adiabatic wall, each component is reconstructed in the following order.
\begin{itemize}
	\item Step 1. One-sided reconstruction for density.
	\item Step 2. One-sided constrained reconstruction for momentum $\rho \textbf{U} = \rho \textbf{U}_{wall}$ where the reconstructed density is used.	
	\item Step 3. One-sided reconstruction for energy.
\end{itemize}	
For non-slip isothermal wall, each component is reconstructed in the following order.
	\begin{itemize}
		\item Step 1. One-sided reconstruction for density.
		\item Step 2. One-sided constrained reconstruction for momentum $\rho \textbf{U} = \rho \textbf{U}_{wall}$ where the reconstructed density is used.	
		\item Step 3. One-sided constrained reconstruction for energy
	    	$\rho E = \frac{1}{2} \rho \textbf{U}_{wall}^2+\rho T_{wall}/(r-1)$, where the reconstructed density is used.
     \end{itemize}	
\end{itemize}
\item Reconstruction Step 2.

For a second-order reconstruction, the WENO procedure in Eq.~\eqref{final_weno_expression-2nd} is complete.
If a third-order reconstruction is adopted, the second-order polynomial $P^2(\textbf{x})$ is needed.
For the ith (i=0,...,4) conservative variables:
\begin{itemize}	
	\item  If there is no constraint for all the boundary faces:
	\begin{itemize}
		\item If the neighboring cell number is no less than 3 and the coefficient matrix is not singular,
		construct the second-order polynomial $P^2(\textbf{x})$, by constraining the cell-averaged values.
		Otherwise, the first-order polynomial from the $P^2(\textbf{x})$ stencils will be constructed in a least-squares sense.
		\item For the smooth reconstruction, the first-order polynomial using the $P^2(\textbf{x})$ stencils is reconstructed instead.
	\end{itemize}
	\item If there exists at least one constrained face on the targeted cell, the weighted constrained least square reconstruction involving all the neighbor cells and boundary faces are conducted.
	The weights for those boundary faces that do not have constraint are set to be zero.	
	Each constrained triangular face has three constraints, which are located at the corresponding Gaussian points.
		\begin{itemize}	
		\item If the sum of neighboring cell-averaged data and the constraint number is no less than 9, and the constraint number is no greater than 9,
		construct the first-order polynomial $P^1(\textbf{x})$ by using constrained least-square method.
		\item If the constraint number is greater than 9 (which never happens in the tests of this paper), construct the first-order polynomial using the $P^2(\textbf{x})$ stencils by the weighted least-square instead.
		\item If the sum of neighboring cell-averaged data and the constraint number is less than 9, construct the first-order polynomial using the $P^2(\textbf{x})$ stencil by the weighted least-square instead.
	\end{itemize}
    The constrained quantities and reconstruction order for the non-slip wall boundaries are the same as those in the reconstruction step	1.
\end{itemize}
\end{itemize}

It should be emphasized that the above criteria are general for other types of mesh and hybrid mesh.
After obtaining the inner state (assume as $\tilde{\textbf{W}}^r$) at a boundary Gaussian point, a ghost state (assume as $\tilde{\textbf{W}}^l$) can be assigned according to boundary condition under local coordinates.
 There is possible discontinuity between $\tilde{\textbf{W}}^l$ and $\tilde{\textbf{W}}^r$ if the WENO reconstruction is used.
The ghost state setting at the solid wall boundary is given as follows (the tilde is omitted).
\begin{itemize}
	\item Slip wall.
	The conservative variables under local coordinate
	$(\rho, \rho {U}_1, \rho {U}_2, \rho {U}_3, \rho E)^l = (\rho, -\rho {U}_1, \rho {U}_2, \rho {U}_3, \rho E)^r$.
	The normal derivatives  $(\rho {U}_1)_{{x}_1}^l = (\rho {U}_1)_{{x}_1}^r$ while the normal derivatives for other components are ${W_i}_{{x}_1}^l=-{W_i}_{{x}_1}^r, ~~i=0,2,3,4$.
	The tangential derivatives $(\rho {U}_1)_{{x}_j}^l = - (\rho {U}_1)_{{x}_j}^r, ~~j=2,3$ while the tangential derivatives for other components are ${W_i}_{{x}_j}^l={W_i}_{{x}_j}^r, ~~i=0,2,3,4, ~~j=2,3$.	
	\item Non-slip adiabatic wall.
	The conservative variables under local coordinate are given as
	$(\rho, \rho {U}_1, \rho {U}_2, \rho {U}_3, \rho E)^l = (\rho, -\rho {U}_1, -\rho {U}_2, -\rho {U}_3, \rho E)^r$.
	The derivatives for all momenta  $(\rho {U}_i)_{{x}_j}^l=(\rho {U}_i)_{{x}_j}^r, ~~i=1,2,3, ~~j=1,2,3$,  while the normal derivatives for  other components are ${W_i}_{{x}_j}^l=-{W_i}_{{x}_j}^r, ~~i=0,4, ~~j=1,2,3$.
	\item Non-slip isothermal wall.
	\begin{itemize}
		\item Assume the same pressure $p^l=p^r$. The velocity is opposite ${U}_{i}^l=-{U}_{i}^r,~~i=1,2,3$. The temperature is set as $T^l=2T^0-T^r$, where $T_0=T_{wall}$. Then, $\rho ^ l = p^l / R T^l= p^r / R T^l = p^r / R (2T^0-T^r) $.
		\item Use primitive variables $(\rho, \textbf{U}, p)^l$ to get $(\rho, \textbf{U}, \rho E)^l$.
		\item From the chain rule,
		$\partial U_i = \frac{\partial (\rho U_i)- \partial \rho U_i}{\rho} ~~i=1,2,3$.
		Denote
		$Q=\frac{1}{2}\sum U_i^2$, $\partial Q = \sum \partial U_i U_i$.
		Then,
		$\partial \rho E = \partial \rho Q + \partial Q \rho + \frac{1}{\gamma-1}\partial p$ and $\partial p = (\gamma-1)(\partial \rho E - \partial \rho Q - \partial Q \rho) $.
		From $p=\rho RT$, $\partial T = \frac{\partial p - R \partial \rho T}{R\rho}$ is obtained.
      And $\partial U_i ^ l, i=1,2,3,$ $\partial p ^ l$, and  $\partial T ^ l$	are determined.
		\item The derivatives of the primitive variables for the ghost states are set as $ \partial  U_i ^l = \partial  U_i^r,~~i=1,2,3$.
		$\partial T ^ l = \partial T ^ r$.
		$\partial p ^ l =- \partial p ^ r$.
		\item Then, obtain $\partial \rho ^l$ by $ \partial \rho = \frac{\partial p - R \partial T \rho }{RT}$ and
		${\partial (\rho  U_i)} ^l$ by $\partial (\rho U_i) =\partial \rho U_i+\partial U_i \rho$.
		\item Finally, get $\partial (\rho E) ^l $ by $\partial \rho E = \partial \rho Q + \partial Q \rho + \frac{1}{\gamma-1}\partial p$.
	\end{itemize}
\end{itemize}

\begin{figure}[hbt!]	
	\centering
	\includegraphics[width=0.96\textwidth]
	{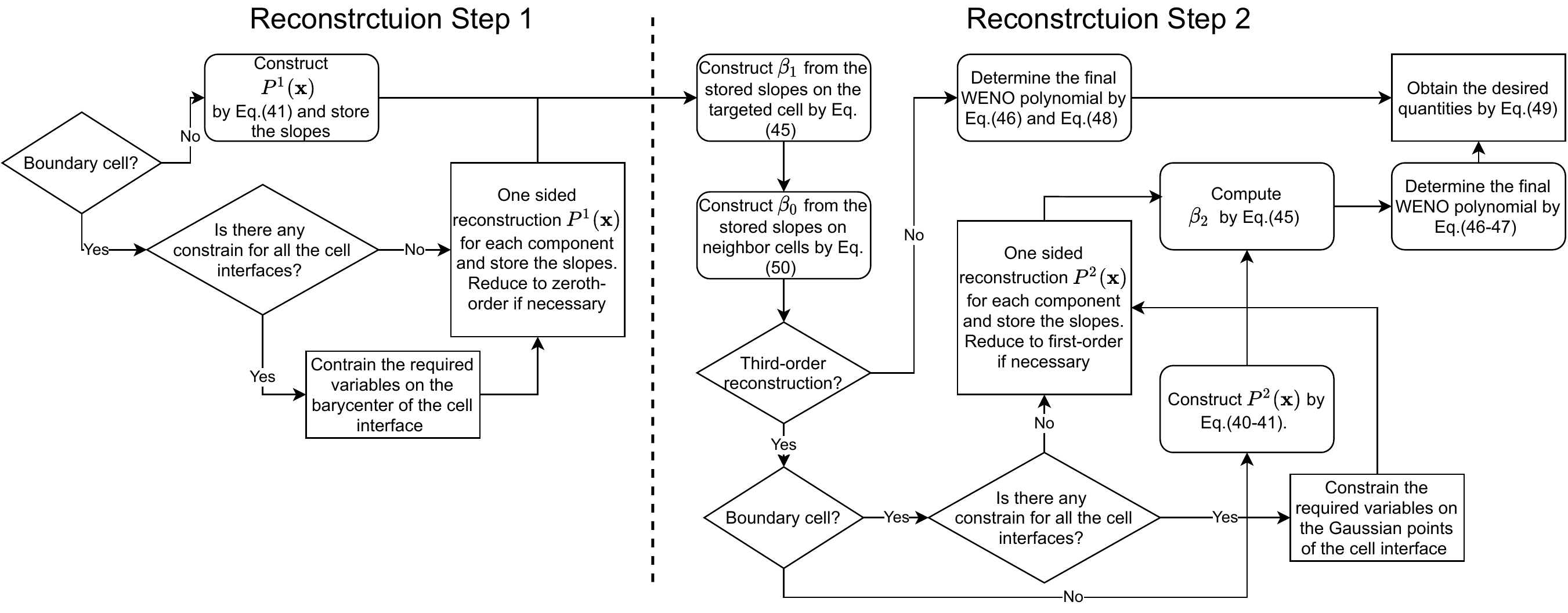}
	\caption{ The process of the compact two-step multi-resolution reconstruction.}
	\label{reconstruction-flowchart}
\end{figure}

A summary for the reconstruction procedure is shown in Fig.~\ref{reconstruction-flowchart}.

\subsection{Reconstruction of the equilibrium state}
The reconstructions for the non-equilibrium states  have the same spatial order and can be used to get the equilibrium state
$ g^{c},g_{x_i}^{c}$ directly by a suitable average of $g^{l,r},g_{x_i}^{l,r}$.
To be consistent with the construction of  $ g^{c}$, we make an analogy of the kinetic-based weighting method for $g_{x_i}^{c}$, which are given by
\begin{align}\label{new-equ-part}
&\int\pmb{\psi} g^{c}\text{d}\Xi=\textbf{W}^c=\int_{u>0}\pmb{\psi}
g^{l}\text{d}\Xi+\int_{u<0}\pmb{\psi} g^{r}\text{d}\Xi, \nonumber \\
&\int\pmb{\psi} g^{c}_{x_i}\text{d}\Xi=\textbf{W}_{x_i}^c=\int_{u>0}\pmb{\psi}
g_{x_i}^{l}\text{d}\Xi+\int_{u<0}\pmb{\psi} g_{x_i}^{r}\text{d}\Xi.
\end{align}
The data for this method has compact support.
In programming, this procedure is included inside the subroutine of the gas distribution function, since it is performed at the local coordinate. Thus, it is also cache-friendly.
This method has been validated in the non-compact WENO5-GKS \cite{ji2020performance}.
In this way, all components of the microscopic slopes in Eq.\eqref{2nd-flux} have been fully obtained.
It is worth to remark that the above reconstruction procedure can be directly implemented to arbitrary mesh.

\section{Numerical examples} \label{test-case}

\begin{figure}[hbt!]	
	\centering
	\includegraphics[width=0.96\textwidth]
	{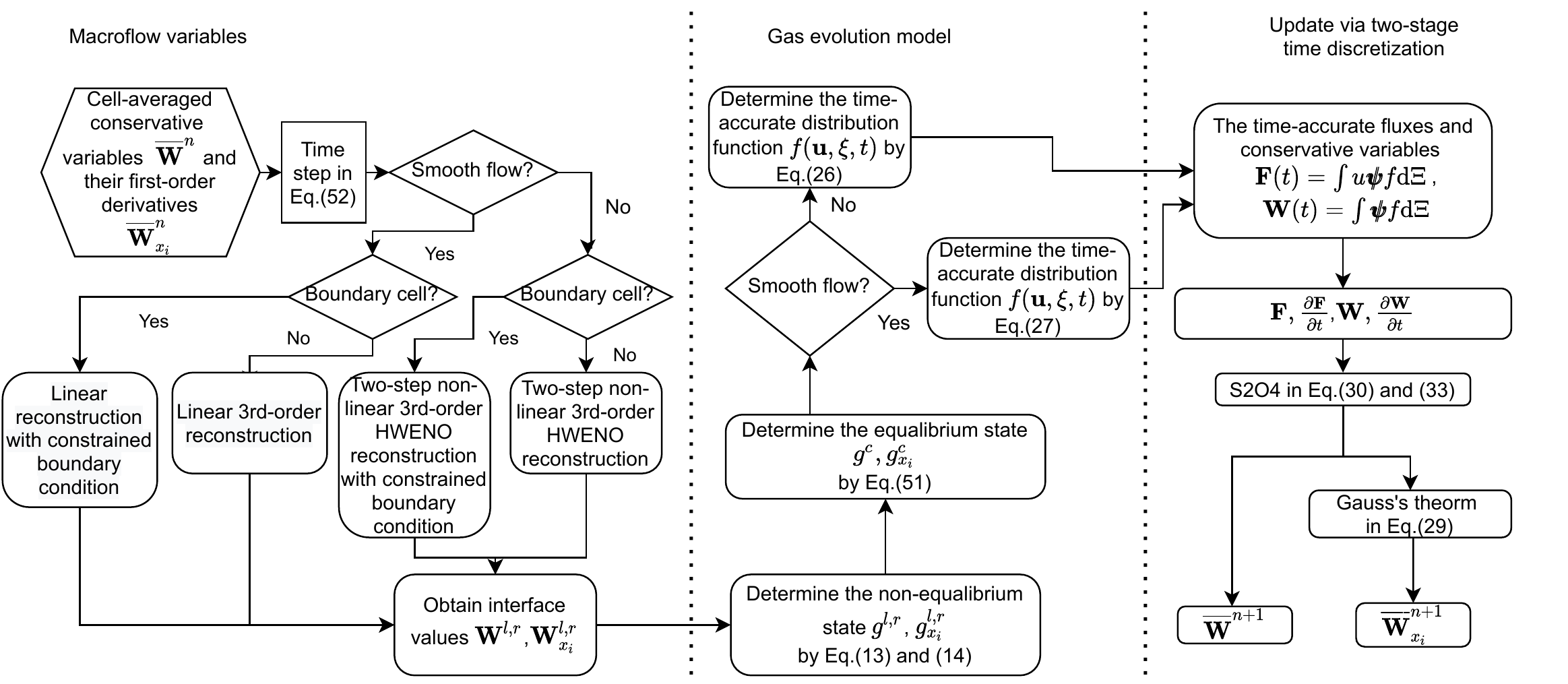}
	\caption{ The brief algorithm of the compact GKS.}
	\label{flowchart}
\end{figure}

In this section, numerical tests will be presented to validate the proposed scheme.
The time step is determined by
\begin{align}\label{time-step}
\Delta t = C_{CFL} \mbox{Min} ( \frac{ \Delta r_i}{|\textbf{U}_i|+(a_s)_i}, \frac{ (\Delta r_i)^2}{3\nu _i}),
\end{align}
where $C_{CFL}$ is the CFL number, and $|\textbf{U}_i|$, $(a_s)_i$, and $\nu _i= (\mu /\rho) _i$ are the magnitude of velocities, sound speed, and kinematic viscosity coefficient for cell i. The $\Delta r_i$ is taken as the approximated inscribed sphere radius of a tetrahedron,
\begin{align*}
\Delta r_i = \frac{3|\Omega_i|}{\sum|\Gamma_{ip}|}.
\end{align*}
All reconstructions will be performed on the conservative variables.
Quadratic elements and a $CFL=1$ are used if no specified.
An algorithm flowchart of the compact GKS is given in Fig.~\ref{flowchart}.

\subsection{3-D sinusoidal wave propagation}
The advection of density perturbation is tested with the initial
condition 
\begin{align*}
\rho(x,y,z)&=1+0.2\sin(\pi (x+y+z)),\\
 \textbf{U}(x,y,z)&=(1,1,1),  \ \ \  p(x,y,z)=1,
\end{align*}
within a cubic domain $[0, 2]\times[0, 2]\times[0, 2]$.
In the computation, a series of uniform meshes with $6 \times N^3$ cells are used.
With the periodic boundary condition in all directions, the analytic
solution is
\begin{align*}
\rho(x,y,z,t)&=1+0.2\sin(\pi(x+y+z-t)),\\
\textbf{U}(x,y,z)&=(1,1,1),\ \ \  p(x,y,z,t)=1.
\end{align*}
The collision time $\tau=0$ is set since the flow is smooth and inviscid.
The $L^1$, $L^2$ and $L^{\infty}$ errors and the corresponding orders with linear and non-linear Z-type weights at $t=2$ for the third-order compact GKS are given in Table~\ref{3d-accuracy-linear-3rd} and Table~\ref{3d-accuracy-weno-3rd}.
The result with non-linear Z-type weights for the second-order scheme is also presented in Table~\ref{3d-accuracy-weno-2nd}.
Expected accuracy is achieved for all cases.

\begin{table}[htp]
	\small
	\begin{center}
		\def\temptablewidth{1\textwidth}
		{\rule{\temptablewidth}{1pt}}
		\begin{tabular*}{\temptablewidth}{@{\extracolsep{\fill}}c|cc|cc|cc}		
			Mesh number & $L^1$ error & Order & $L^2$ error & Order& $L^{\infty}$ error & Order  \\
			\hline
			  $6 \times 5^3$& 2.220404e-02 & ~ & 2.458004e-02 & ~ & 3.674171e-02 & ~ \\
			 $6 \times 10^3$ & 2.714856e-03 & 3.03 & 3.035792e-03 & 3.02 & 4.794437e-03 & 2.94 \\
			 $6 \times 20^3$& 3.285843e-04 & 3.05 & 3.666555e-04 & 3.05 & 6.093576e-04 & 2.98 \\
			 $6 \times 40^3$& 4.360713e-05 & 2.92 & 4.862997e-05 & 2.91 & 8.411243e-05 & 2.87 \\
		\end{tabular*}
		{\rule{\temptablewidth}{0.1pt}}
	\end{center}
	\vspace{-4mm} \caption{\label{3d-accuracy-linear-3rd}
	Accuracy test for the 3D sin-wave
	propagation by the linear third-order compact reconstruction. CFL=1.0.  }
\end{table}

\begin{table}[htp]
	\small
	\begin{center}
		\def\temptablewidth{1\textwidth}
		{\rule{\temptablewidth}{1pt}}
		\begin{tabular*}{\temptablewidth}{@{\extracolsep{\fill}}c|cc|cc|cc}		
			Mesh number & $L^1$ error & Order & $L^2$ error & Order& $L^{\infty}$ error & Order  \\
			\hline
			$6 \times 5^3$&  4.119490e-02 & ~ & 4.675556e-02 & ~ & 7.211452e-02 & ~ \\
			$6 \times 10^3$ & 6.593180e-03 & 2.64 & 8.682551e-03 & 2.43 & 2.501214e-02 & 1.53 \\
			$6 \times 20^3$& 4.217035e-04 & 3.97 & 5.481270e-04 & 3.99 & 1.251195e-03 & 4.32 \\
			$6 \times 40^3$& 4.287225e-05 & 3.30 & 4.947759e-05 & 3.47 & 1.138919e-04 & 3.46 \\
		\end{tabular*}
		{\rule{\temptablewidth}{0.1pt}}
	\end{center}
	\vspace{-4mm} \caption{\label{3d-accuracy-weno-3rd}
		Accuracy test for the 3D sin-wave
		propagation by the third-order compact HWENO reconstruction with $d_0:d_1:d_2=100:10:1$. CFL=1.0.  }
\end{table}

\begin{table}[hbt!]
	\small
	\begin{center}
		\def\temptablewidth{1\textwidth}
		{\rule{\temptablewidth}{1pt}}
		\begin{tabular*}{\temptablewidth}{@{\extracolsep{\fill}}c|cc|cc|cc}
			
			mesh number & $L^1$ error & Order & $L^2$ error & Order& $L^{\infty}$ error & Order  \\
			\hline
			$6 \times 5^3$ & 2.705626e-02 & ~ & 3.377431e-02 & ~ & 6.219542e-02 & ~ \\
			$6 \times 10^3$  & 6.963215e-03 & 1.96 & 7.839272e-03 & 2.10 & 1.311175e-02 &  1.37 \\
			$6 \times 20^3$  & 2.370280e-03 & 1.55 & 2.640328e-03 & 1.57 & 4.165732e-03 & 1.65 \\
			$6 \times 40^3$  &  6.351481e-04 & 1.90 & 7.069711e-04 & 1.90 & 1.100743e-03 & 1.92 \\
		\end{tabular*}
		{\rule{\temptablewidth}{0.1pt}}
	\end{center}
	\vspace{-4mm} \caption{\label{3d-accuracy-weno-2nd} Accuracy test for the 3D sin-wave
		propagation by the second-order WENO reconstruction with $d_0:d_1=10:1$. CFL=1.0.  }
\end{table}

\subsection{One dimensional Riemann problems}

\noindent{\sl{(a) Shu-Osher problem}}

This is the Shu-Osher problem \cite{shu1989efficient} with the initial condition
\begin{align*}
(\rho,U,p) =\begin{cases}
(3.857134, 2.629369, 10.33333), &  0<x \leq 1,\\
(1 + 0.2\sin (5x), 0, 1),  &  1 <x<10.
\end{cases}
\end{align*}

\begin{figure}[hbt!]	
	\centering
	\includegraphics[width=0.44\textwidth]
	{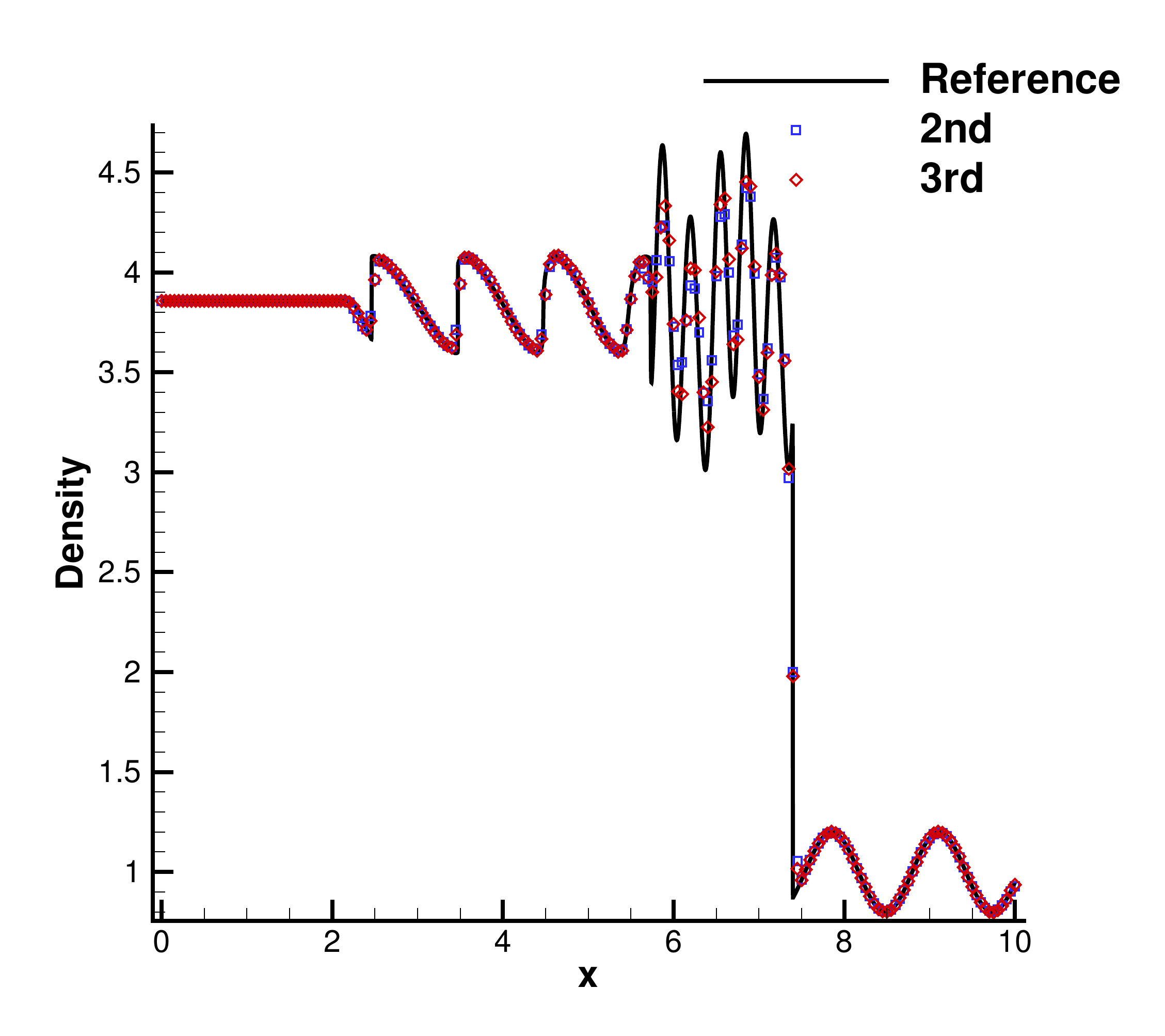}
	\includegraphics[width=0.44\textwidth]
	{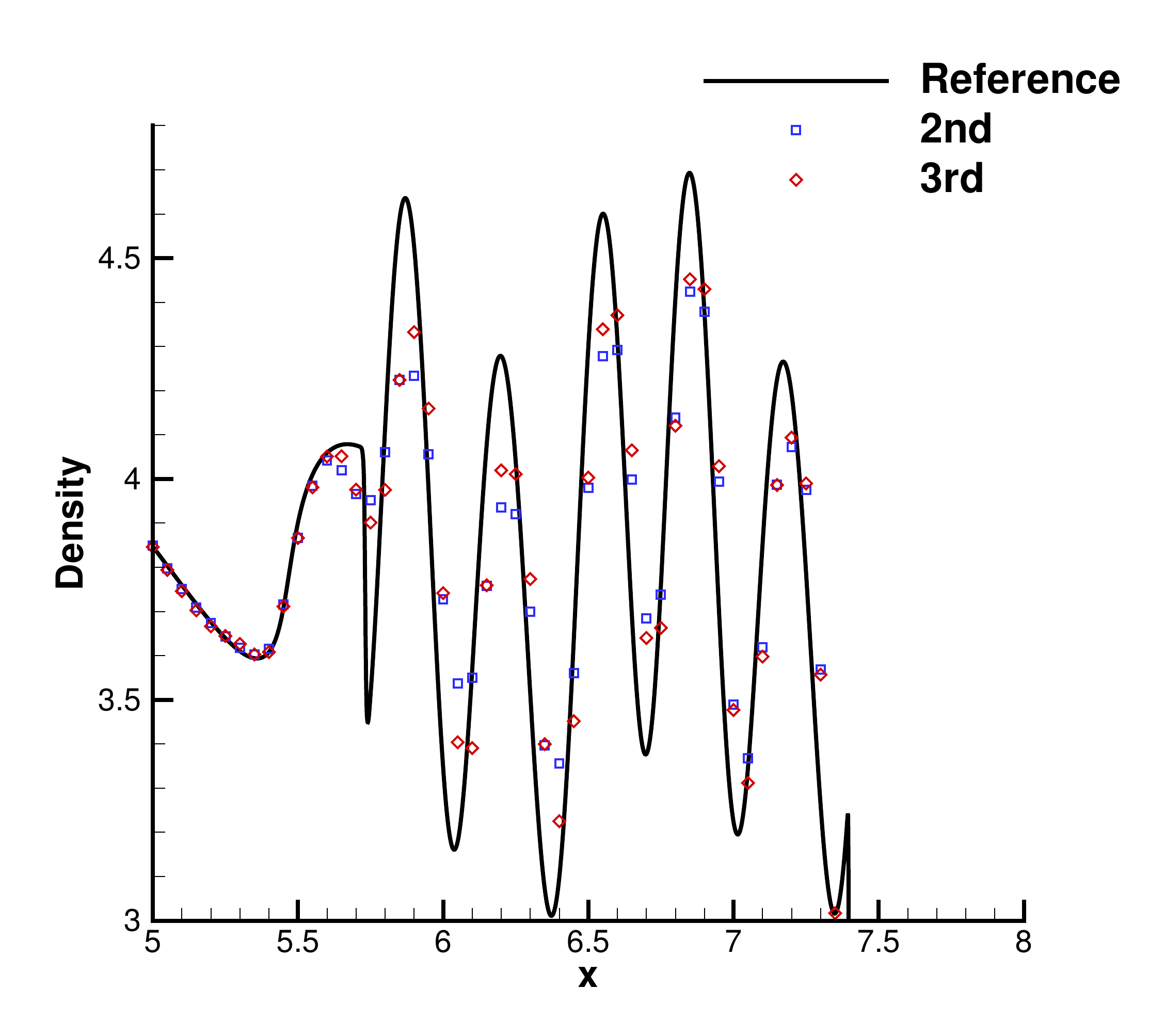}	
	\caption{Shu-Osher problem. Mesh number: $6\times 200 \times 2 \times 2$. }
	\label{shuosher-3d-mesh200}
\end{figure}

\begin{figure}[hbt!]	
	\centering
	\includegraphics[width=0.44\textwidth]
	{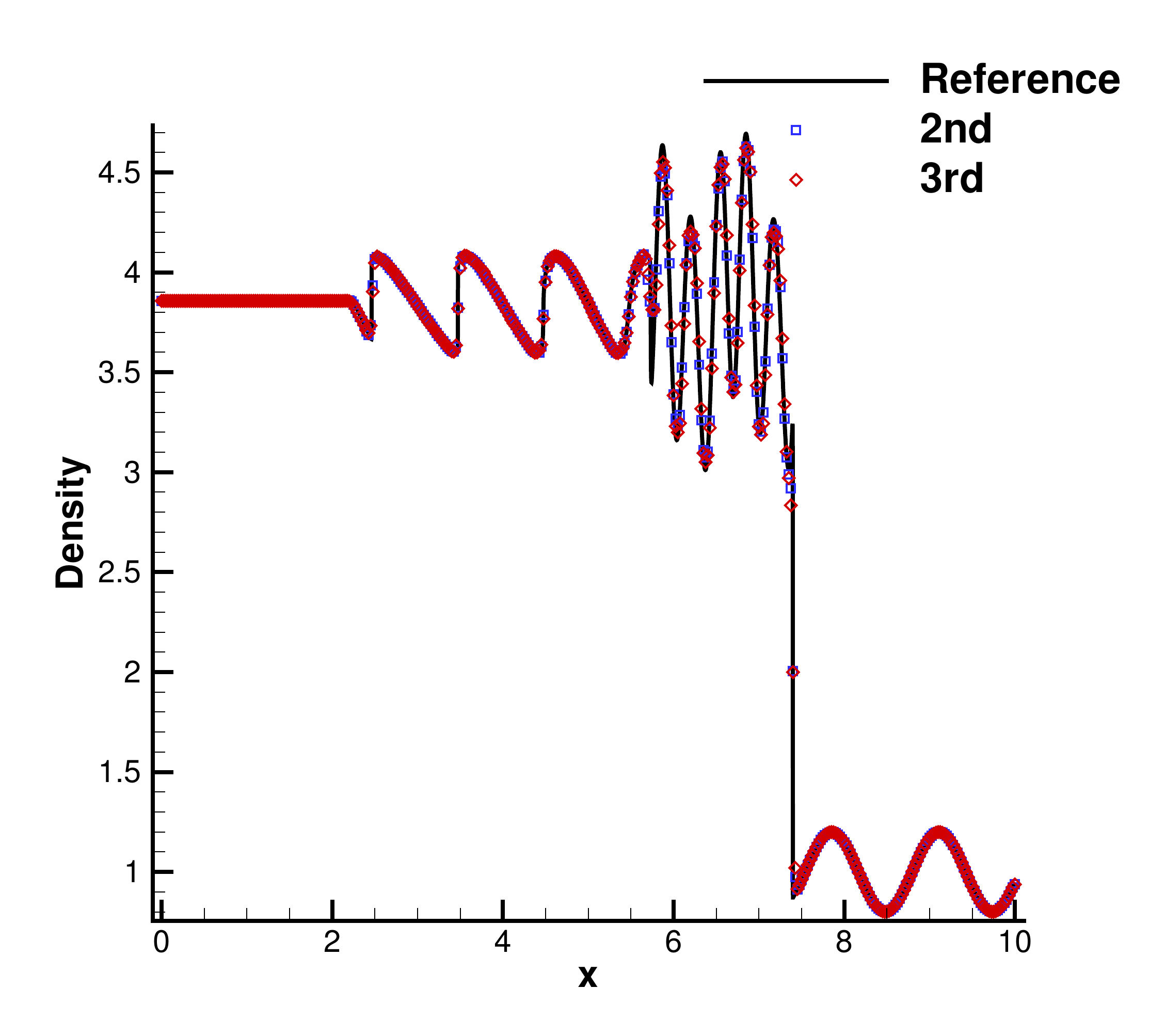}
	\includegraphics[width=0.44\textwidth]
	{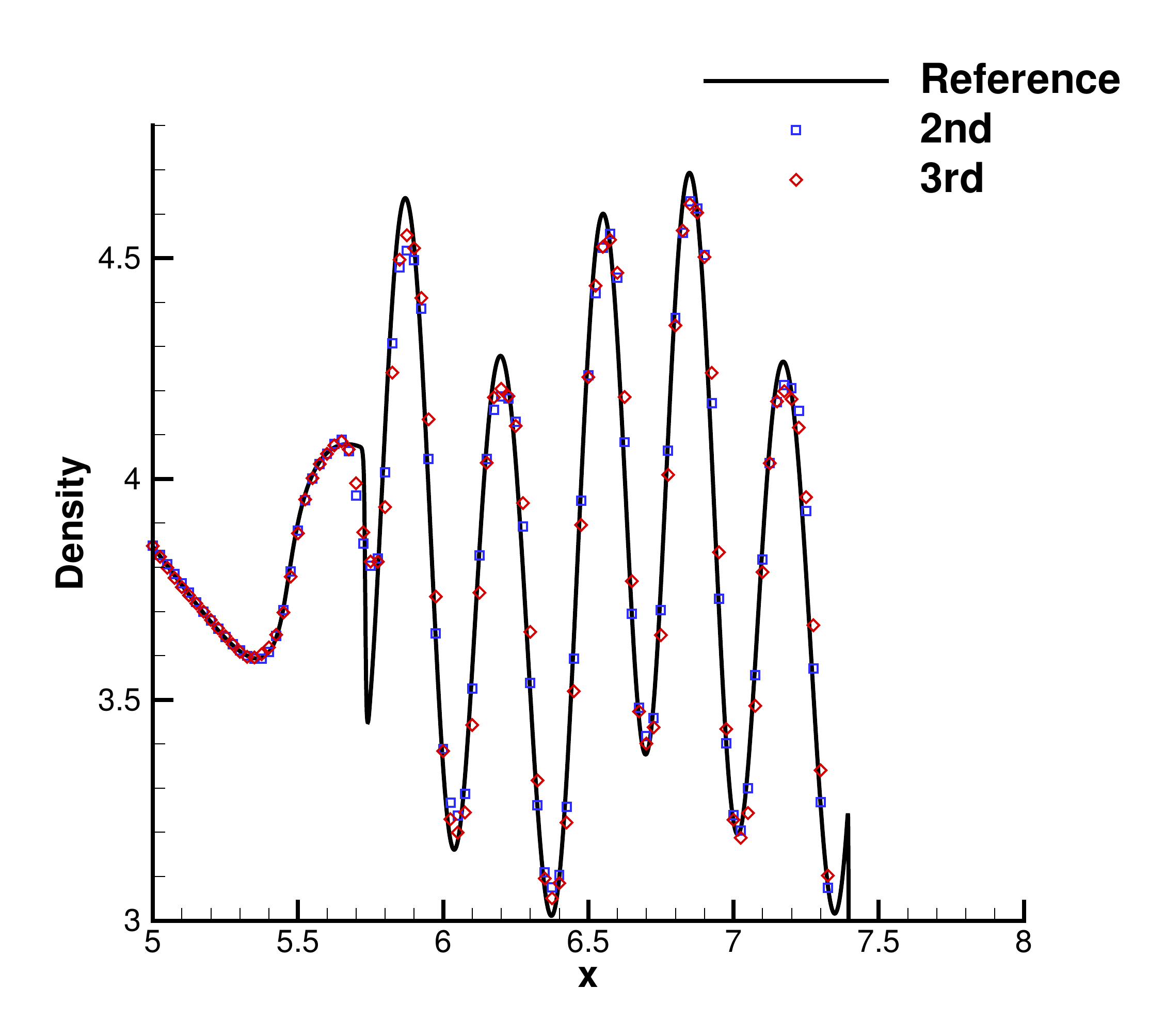}		
	\caption{Shu-Osher problem. Mesh number: $6\times 400 \times 2 \times 2$. }
	\label{shuosher-3d-mesh400}
\end{figure}

The computational domain is $[0, 10]$.
The non-reflecting boundary condition is given on the left boundary, and the
fixed wave profile is extended on the right boundary.
The computed density profiles and local enlargements at $t = 1.8$ with mesh size $1/200$ and $1/400$ are plotted in Fig.~\ref{shuosher-3d-mesh200} and Fiq.~\ref{shuosher-3d-mesh400}.
The third-order compact GKS shows a better resolution in resolving the sinusoidal wave than the second order method on the coarse mesh.
Both schemes resolve the waves nicely with the fine mesh.

\noindent{\sl{(b) Blast wave problem}}

\begin{figure}[hbt!]	
	\centering
	\includegraphics[width=0.44\textwidth]
	{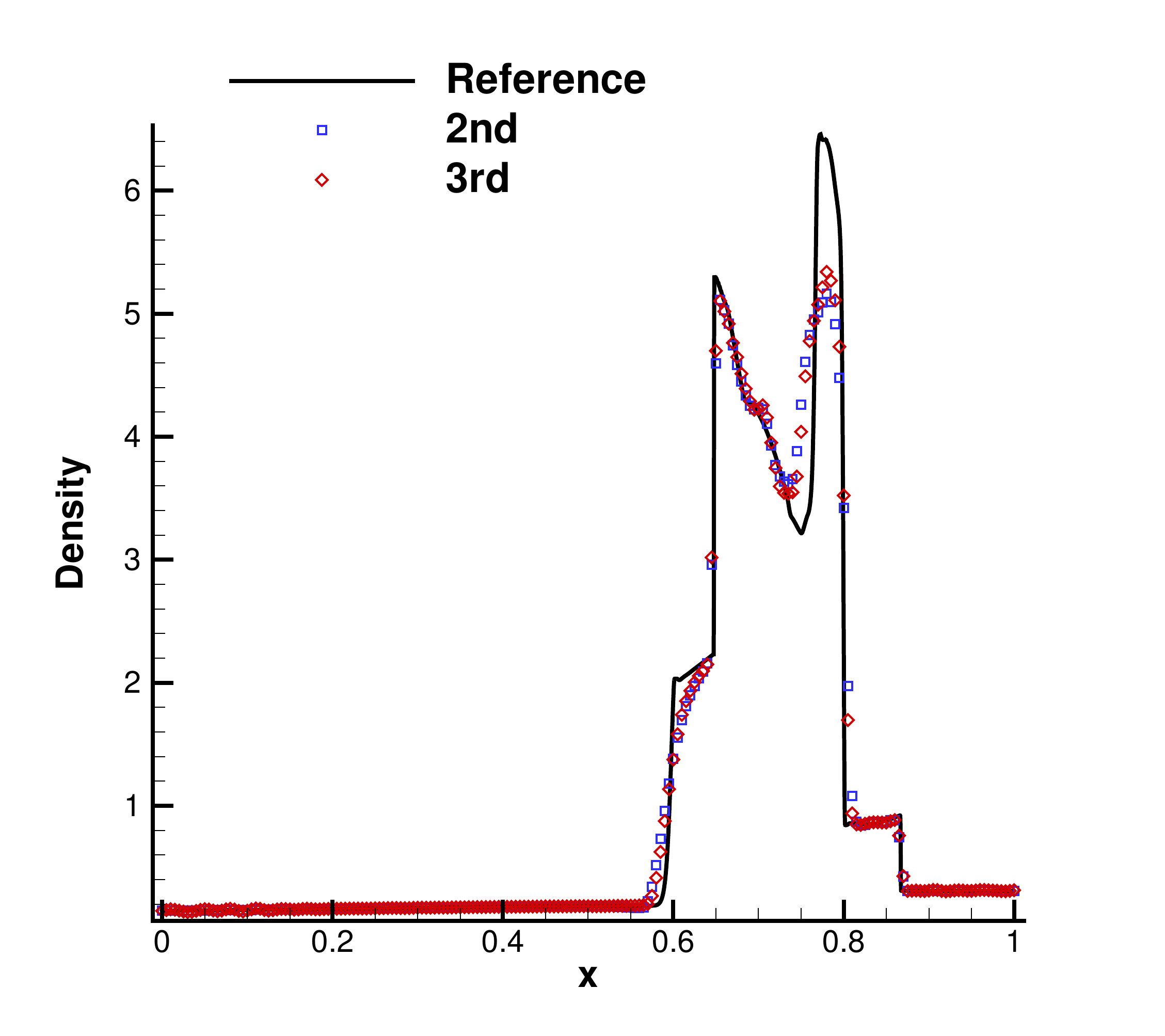}
	\includegraphics[width=0.44\textwidth]
	{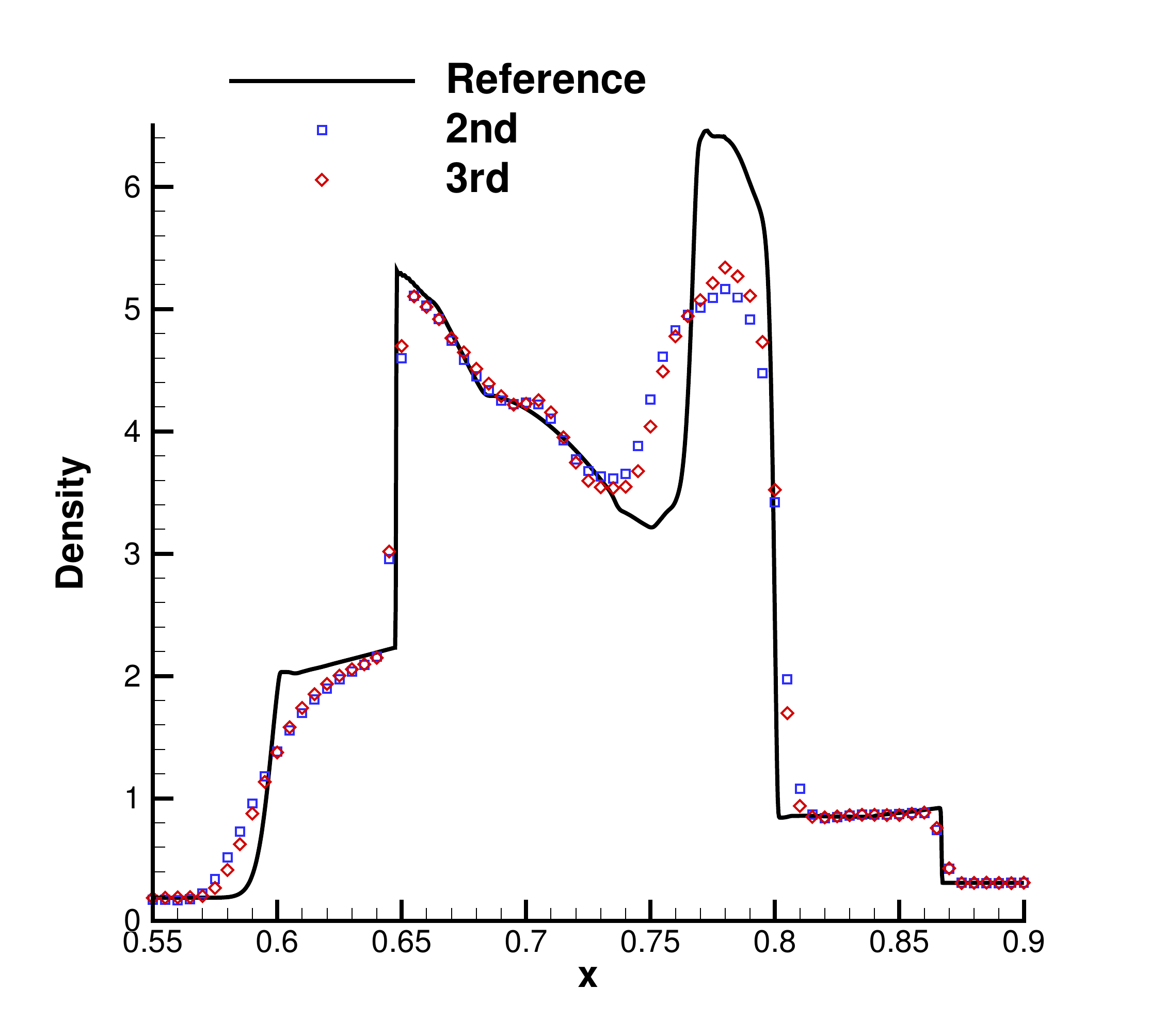}		
	\caption{Blastwave problem. Mesh number: $6\times 200 \times 2 \times 2$. }
	\label{blast-3d-mesh200}
\end{figure}

\begin{figure}[hbt!]	
	\centering
	\includegraphics[width=0.44\textwidth]
	{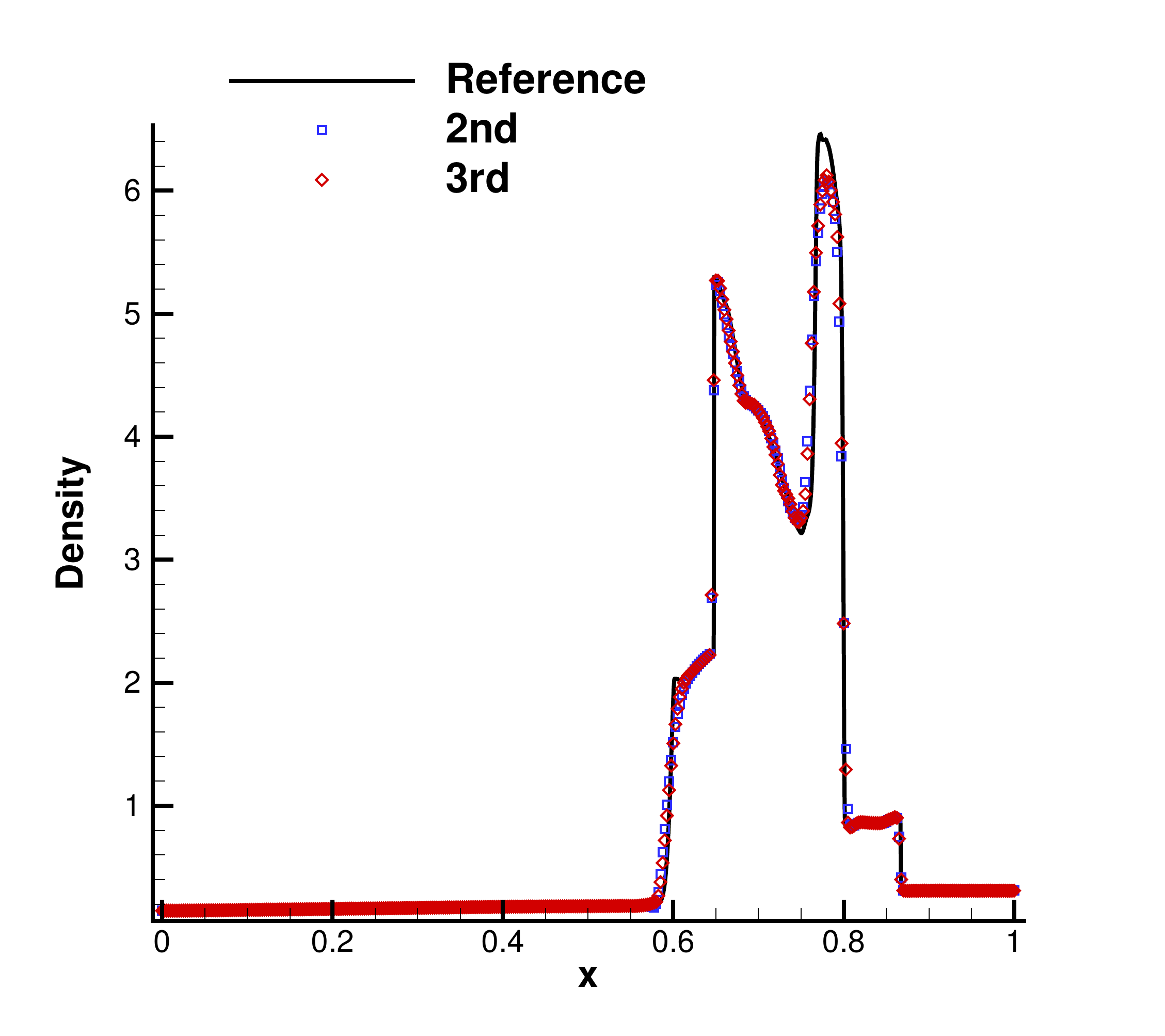}
	\includegraphics[width=0.44\textwidth]
	{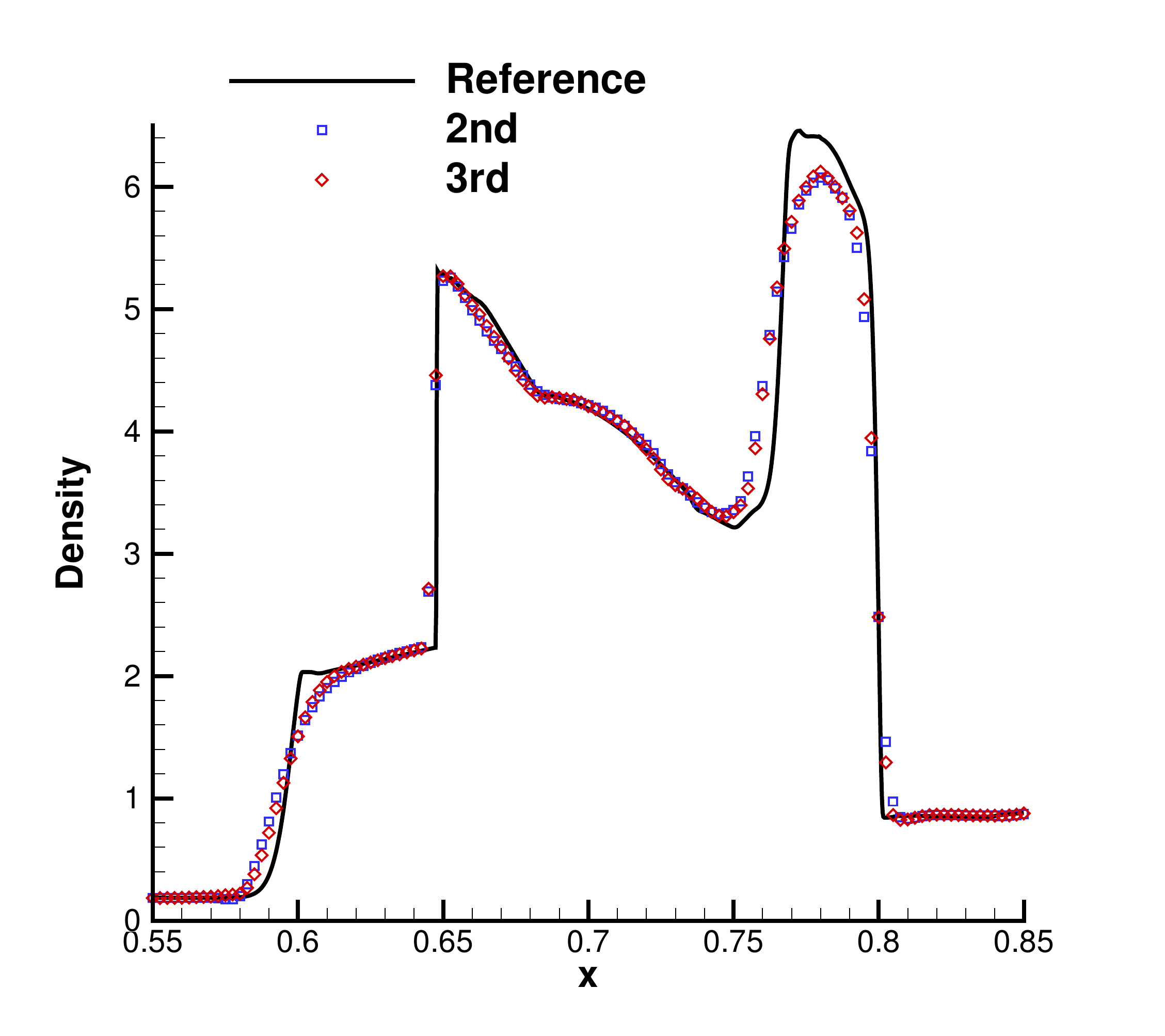}		
	\caption{Blastwave problem. Mesh number: $6\times 400 \times 2 \times 2$. }
	\label{blast-3d-mesh400}
\end{figure}

The initial conditions for the blast wave problem \cite{woodward1984numerical} are given as follows
\begin{equation*}
(\rho,u,p)=\left\{\begin{aligned}
&(1, 0, 1000), 0\leq x<0.1,\\
&(1, 0, 0.01), 0.1\leq x<0.9,\\
&(1, 0, 100),  0.9\leq x\leq 1.
\end{aligned} \right.
\end{equation*}
 In the computational domain,   $6\times 200 \times 2 \times 2$ and $6\times 400 \times 2 \times 2$ mesh points are used.
  Reflection boundary conditions are applied at both ends.
The density distributions at $t=0.038$ are presented in Fig.~\ref{blast-3d-mesh200} and Fig.~\ref{blast-3d-mesh400}.
Both the second-order and third-order schemes show good robustness for such a strong shock-shock interaction.

\subsection{3-D lid-driven cavity flow}

\begin{figure}[htp]	
	\centering	
	\includegraphics[width=0.4\textwidth]{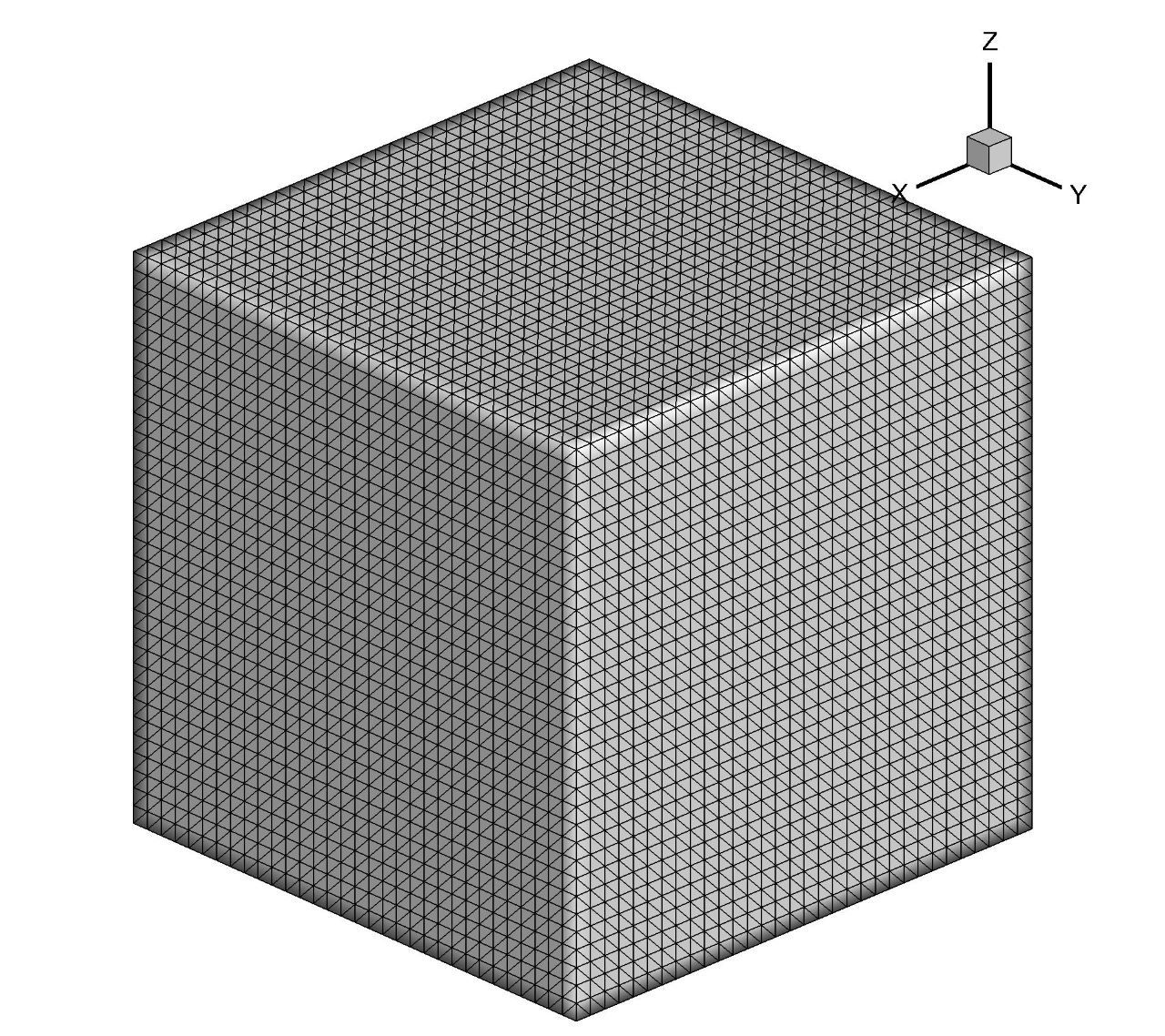}
	\includegraphics[width=0.4\textwidth]{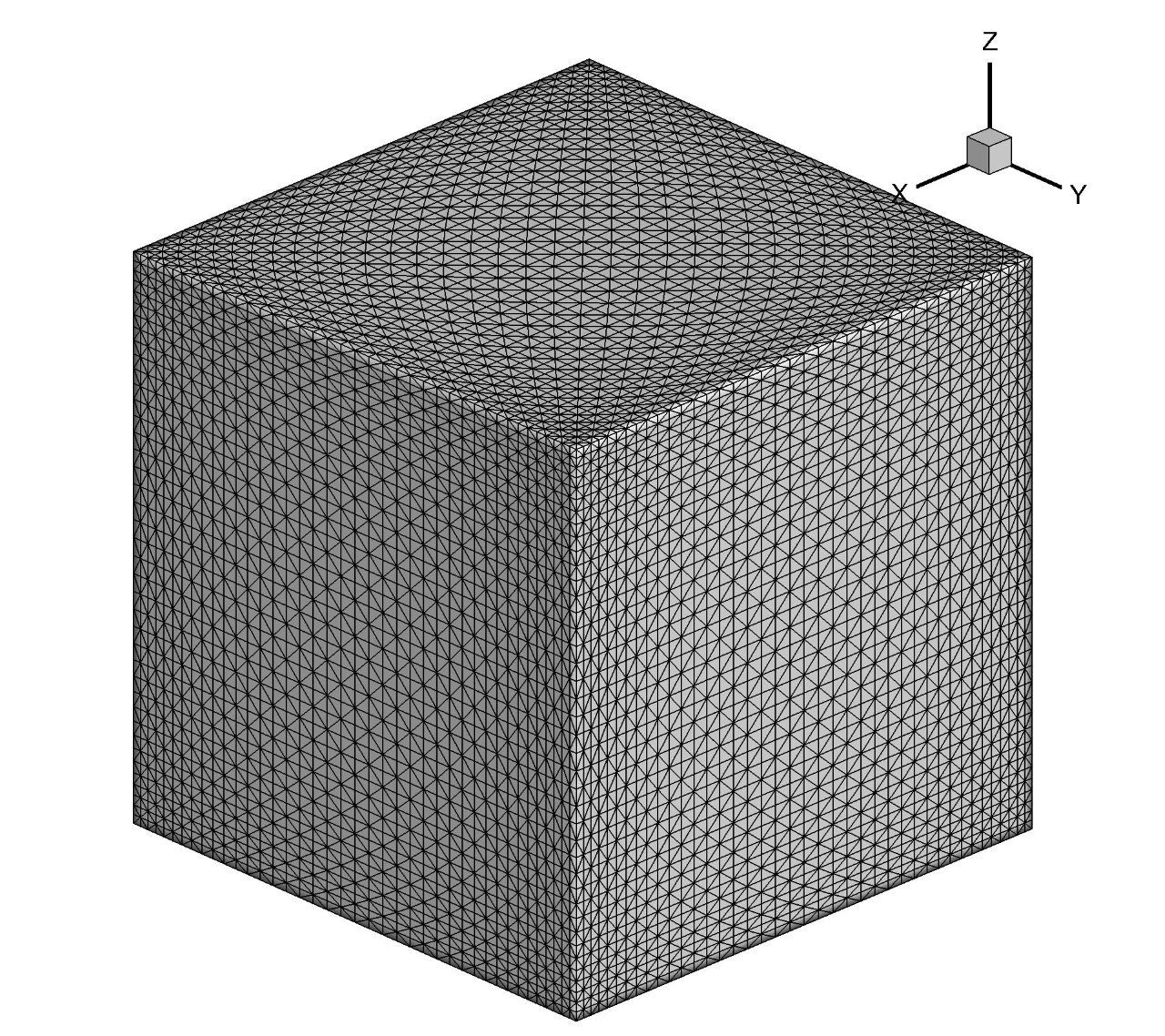}
	\vspace{-4mm} \caption{\label{cavity3d-mesh}
		Lid-driven cavity flow. Left: uniform mesh with near wall size $h=1/32$. Right: non-uniform mesh with near wall size $h=1/64$.}
\end{figure}

A 3-D cavity is bounded in a unit cube and is driven by a
uniform translation of the top boundary. In this case, the flow is
simulated with Mach number $Ma=0.15$ and $\gamma = 5/3$.
All boundaries are isothermal and nonslip.
The computational domain $[-0.5, 0.5]\times[-0.5, 0.5]\times[-0.5, 0.5]$
is covered by a uniform mesh with $6\times 32 \times 32 \times 32$ points and a refined uniform mesh with $5\times 40 \times 40 \times 40$ points, as shown in Fig.~\ref{cavity3d-mesh}.
A CFL number of 0.5 is used.
The flow is initialized with $\rho = 1$, $U_1=0.15$, $U_2=U_3=0$, and $p=1/\gamma$.
Since the flow is nearly incompressible and mesh is regular, the smooth reconstruction and the simplified solver in Eq.~\eqref{2nd-flux} are adopted in the computations. 

\noindent{\sl{(a) Re=1,000}}

\begin{figure}[htp]	
	\centering
	\includegraphics[width=0.4\textwidth]
	{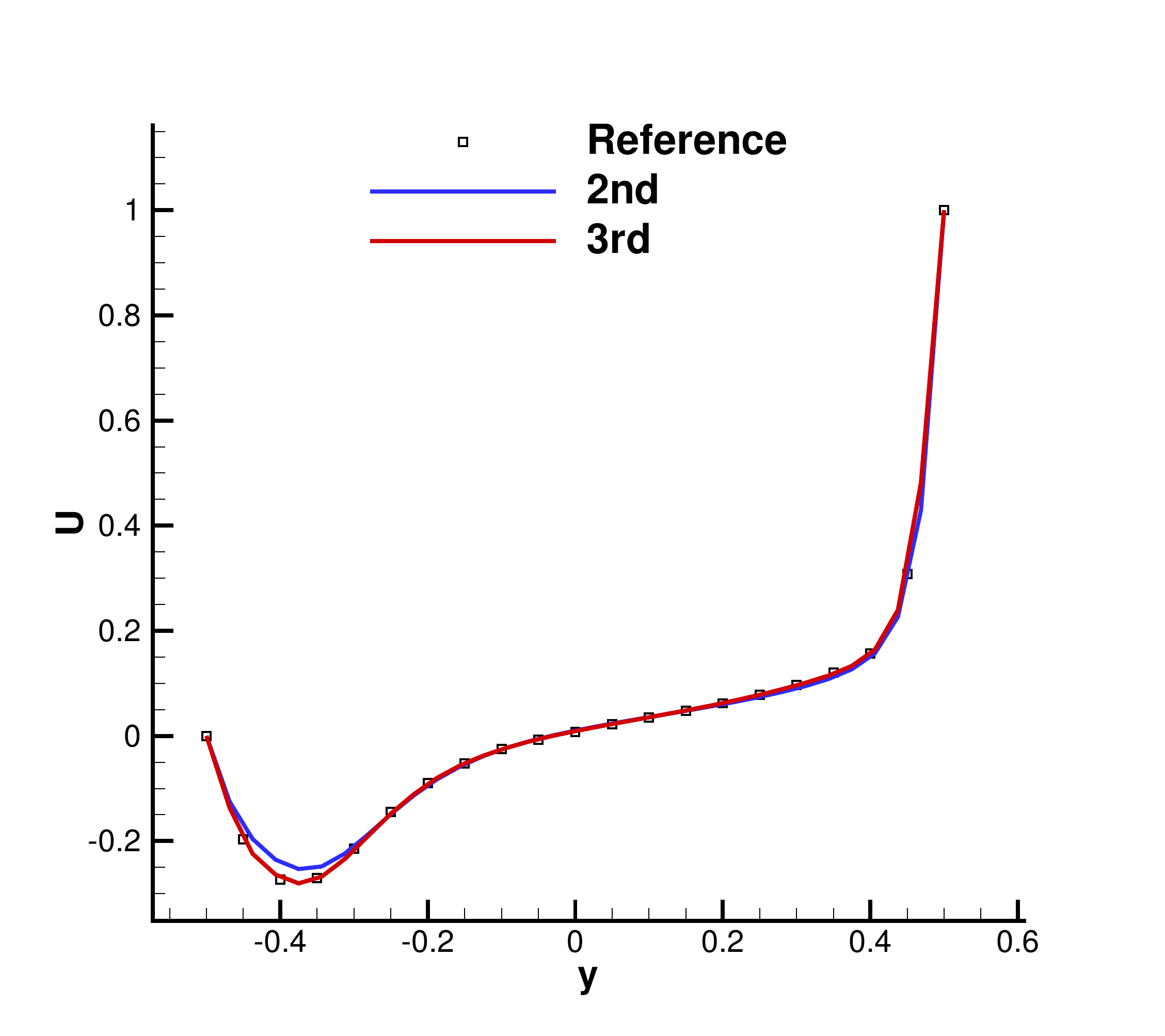}	
	\includegraphics[width=0.4\textwidth]
	{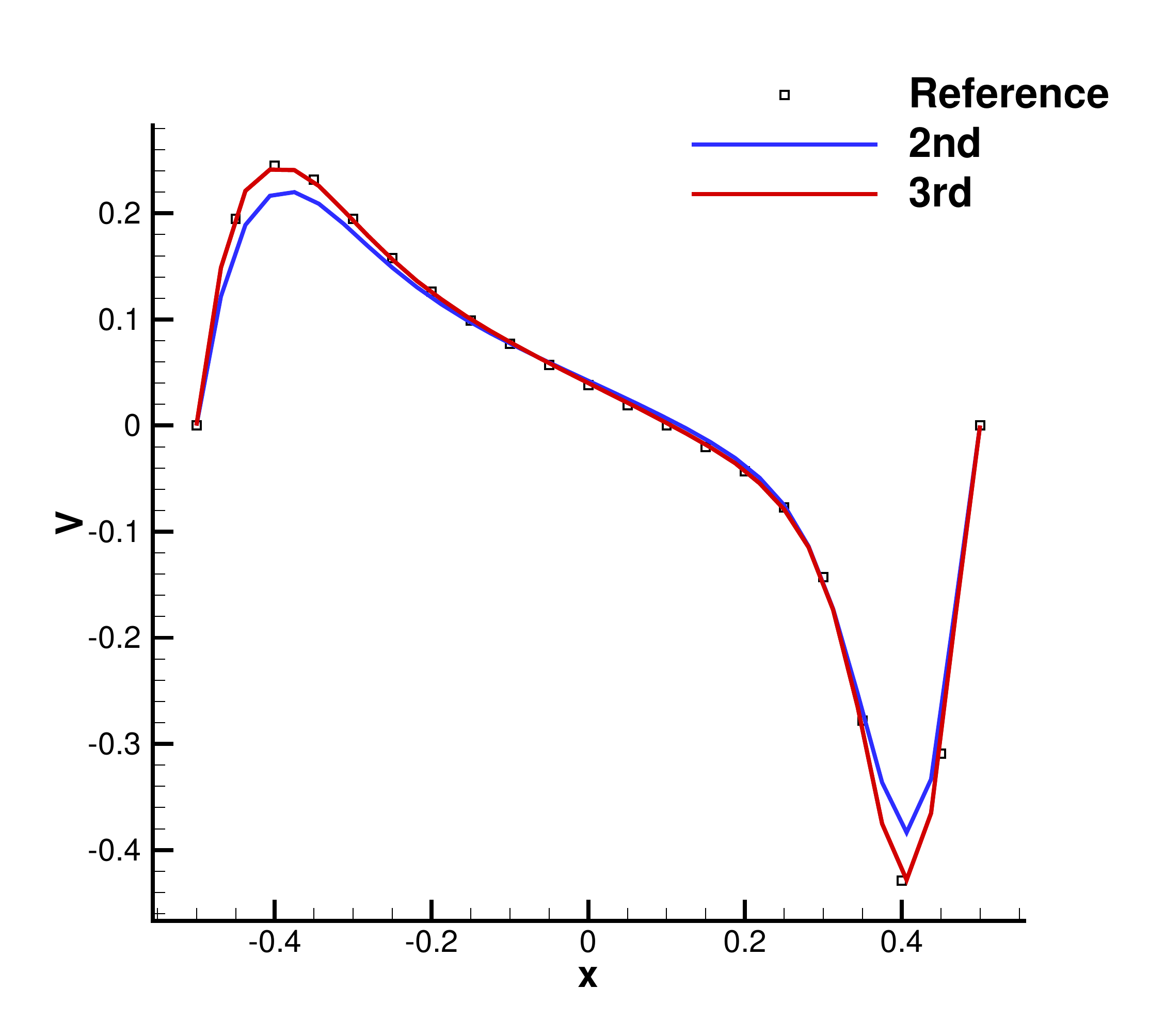}
	\vspace{-4mm} \caption{\label{cavity3d-1000-1}
		Lid-driven cavity flow: Re=1,000. The velocities profiles compared with the reference data in \cite{shu2003numerical}.}
\end{figure}

\begin{figure}[htp]	
	\centering
	\includegraphics[width=0.4\textwidth]
	{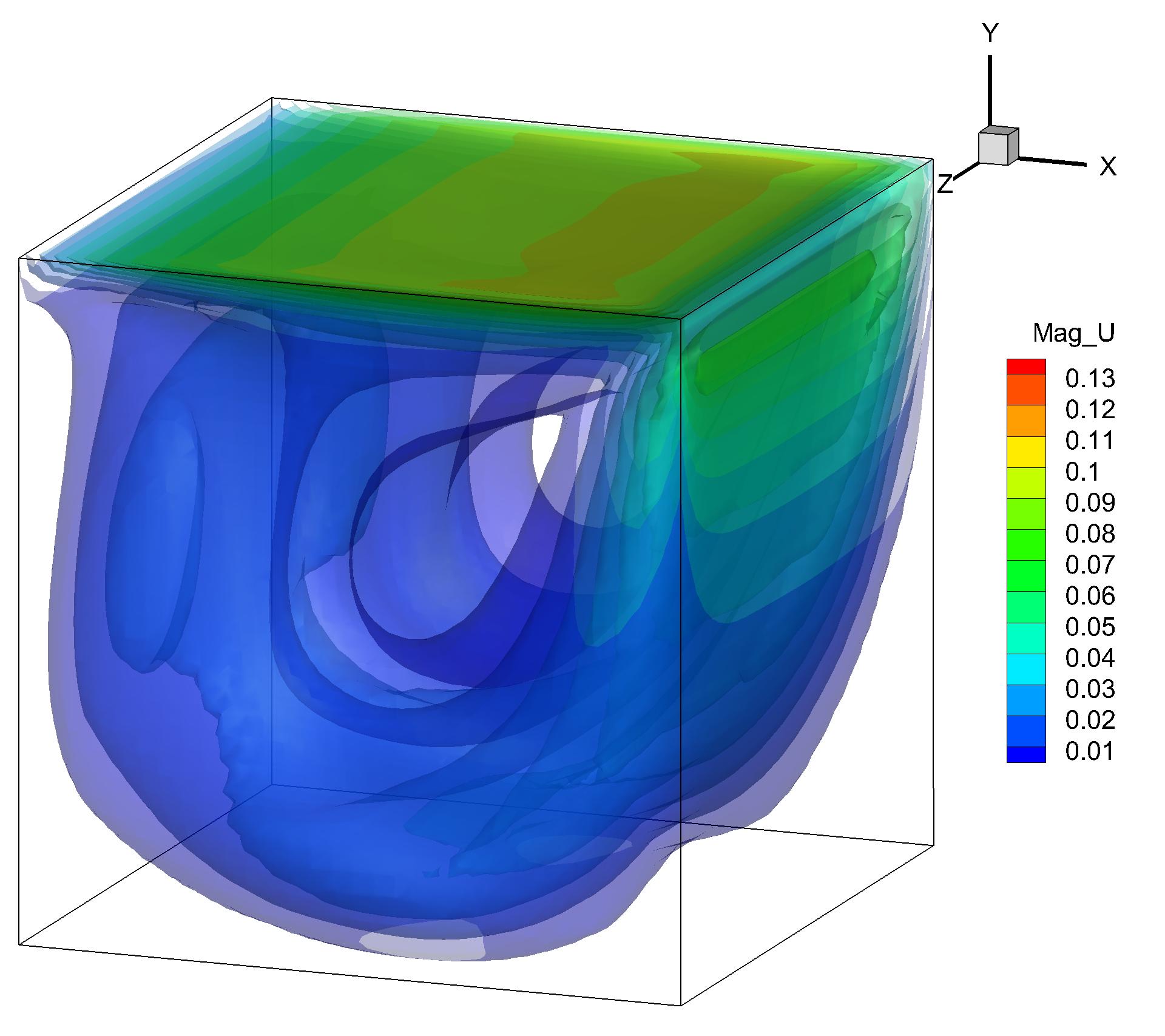}
	\includegraphics[width=0.4\textwidth]
	{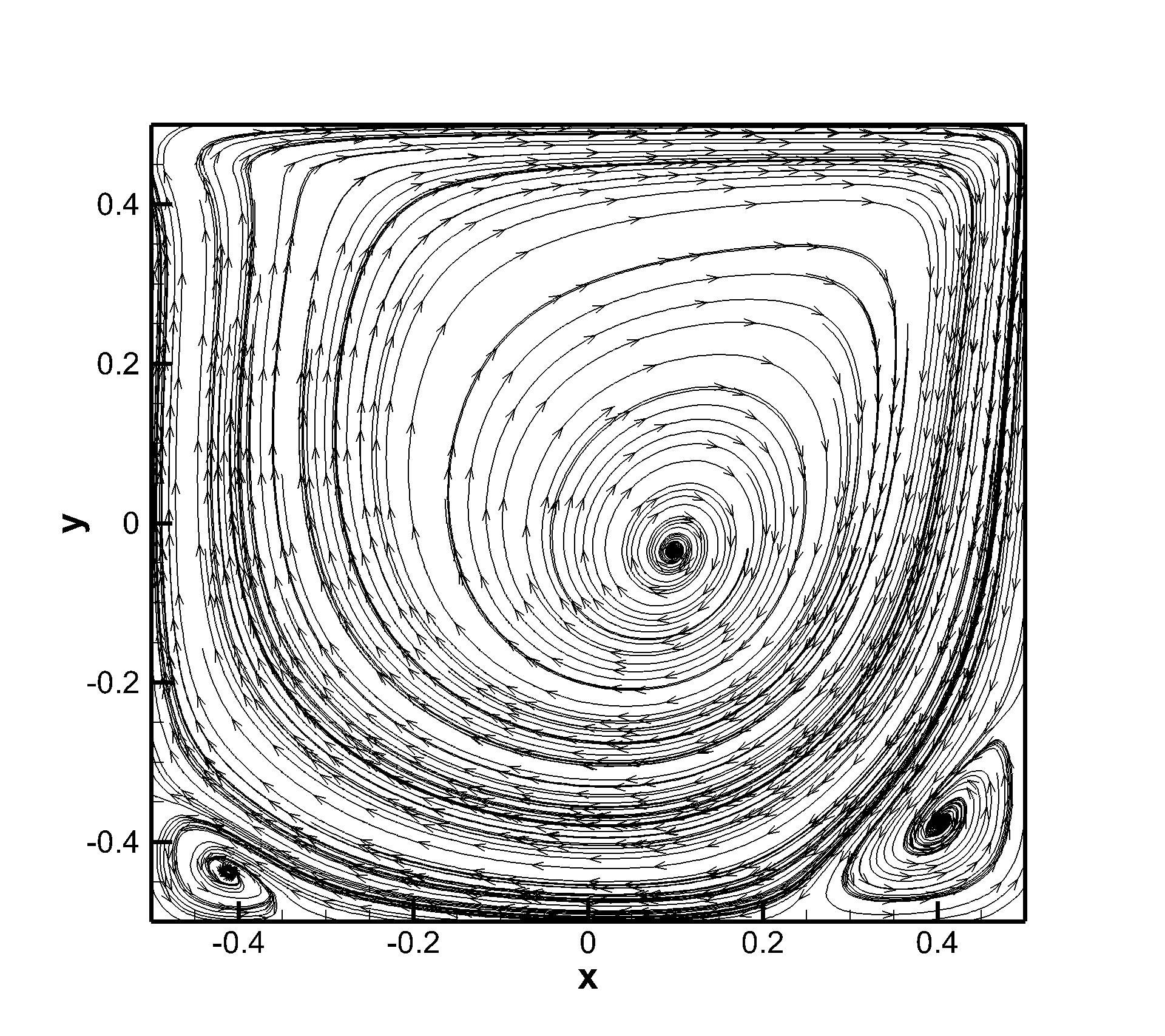}
	\includegraphics[width=0.4\textwidth]
	{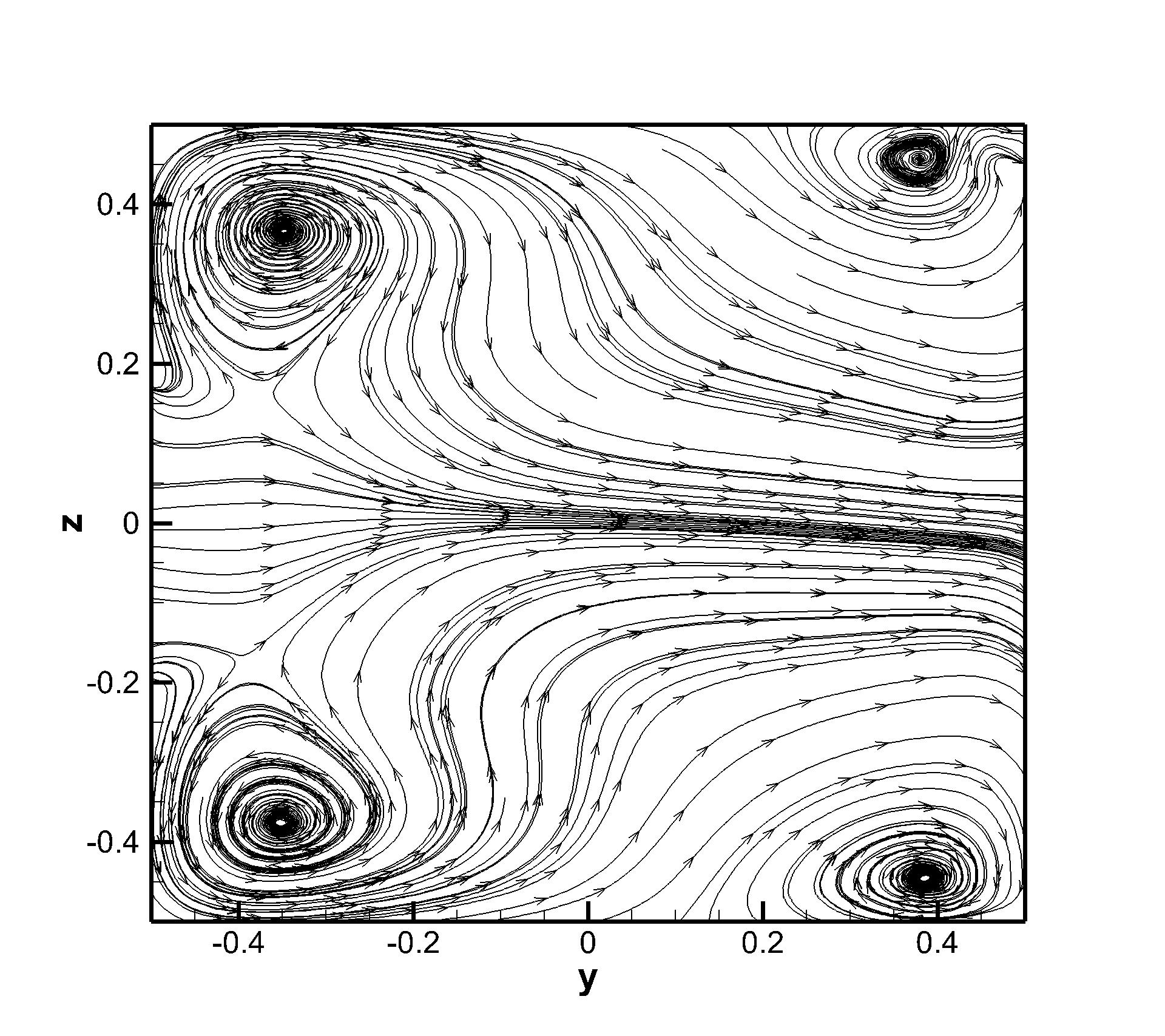}	
	\includegraphics[width=0.4\textwidth]
	{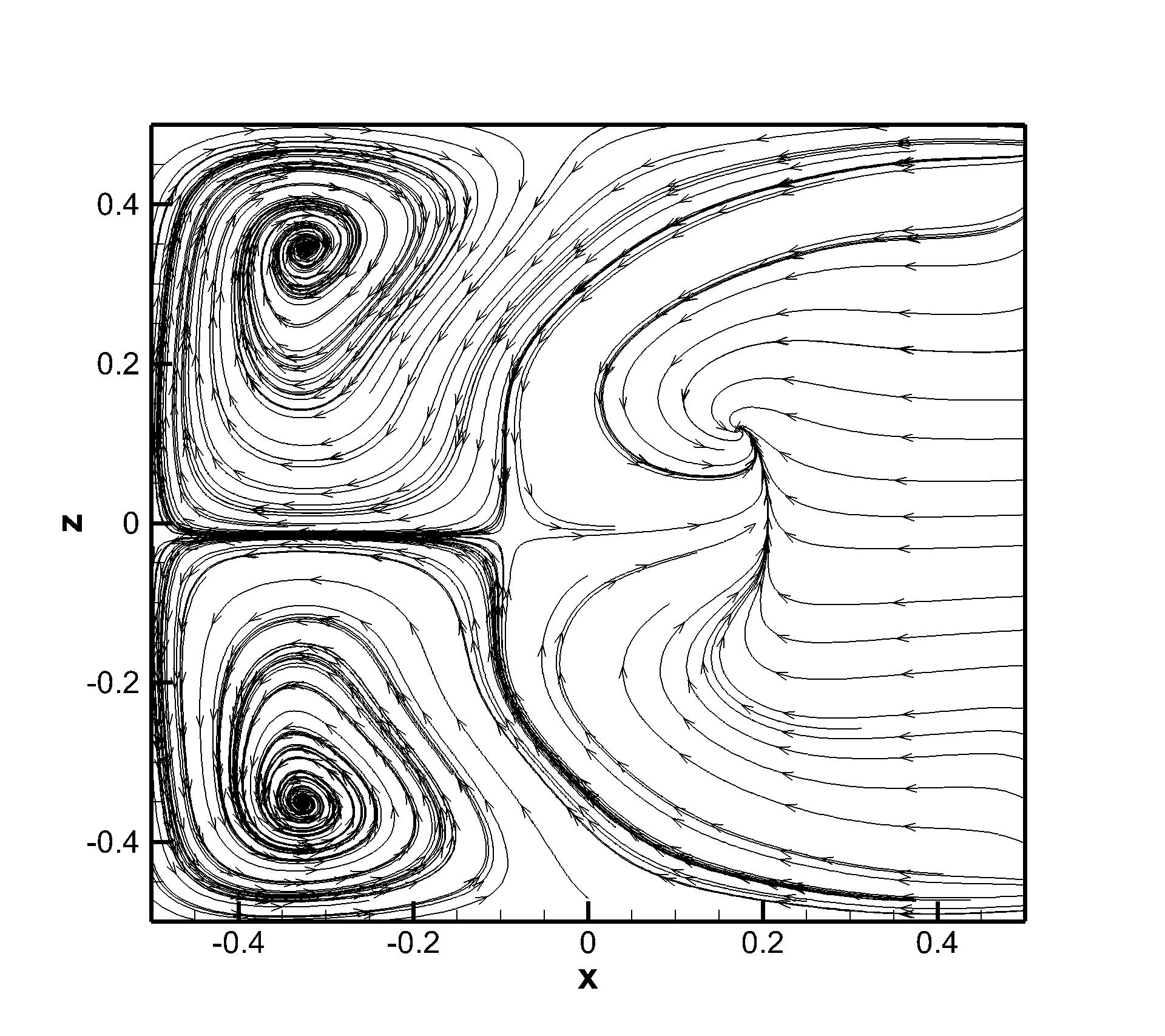}
	\vspace{-4mm} \caption{\label{cavity3d-1000-2}
		Lid-driven cavity flow: Re=1,000. Top left: The velocity magnitude contours.
		Others: The streamlines on $x=0$, $y=0$, and $z=0$ planes. }
\end{figure}

For the Reynolds number $Re=1,000$, both results from the second-order and third-order reconstruction are presented under the uniform mesh. A low-order boundary treatment is used for the second-order scheme, which ensures a stable solution.
The $U$-velocities
along the line $x=0, z=0$, and $V$-velocities along the line $y=0, z=0$, are shown in Fig.~\ref{cavity3d-1000-1}.
The velocity profiles from the third-order scheme match very well with the benchmark data \cite{shu2003numerical}.
The velocity magnitude contours and streamlines by the third-order scheme are shown in Fig.~\ref{cavity3d-1000-2}.
The cavity case demonstrates the high-order accuracy of the compact GKS.

\noindent{\sl{(b) Re=3,200}}

\begin{figure}[htp]	
	\centering
	\includegraphics[width=0.4\textwidth]
	{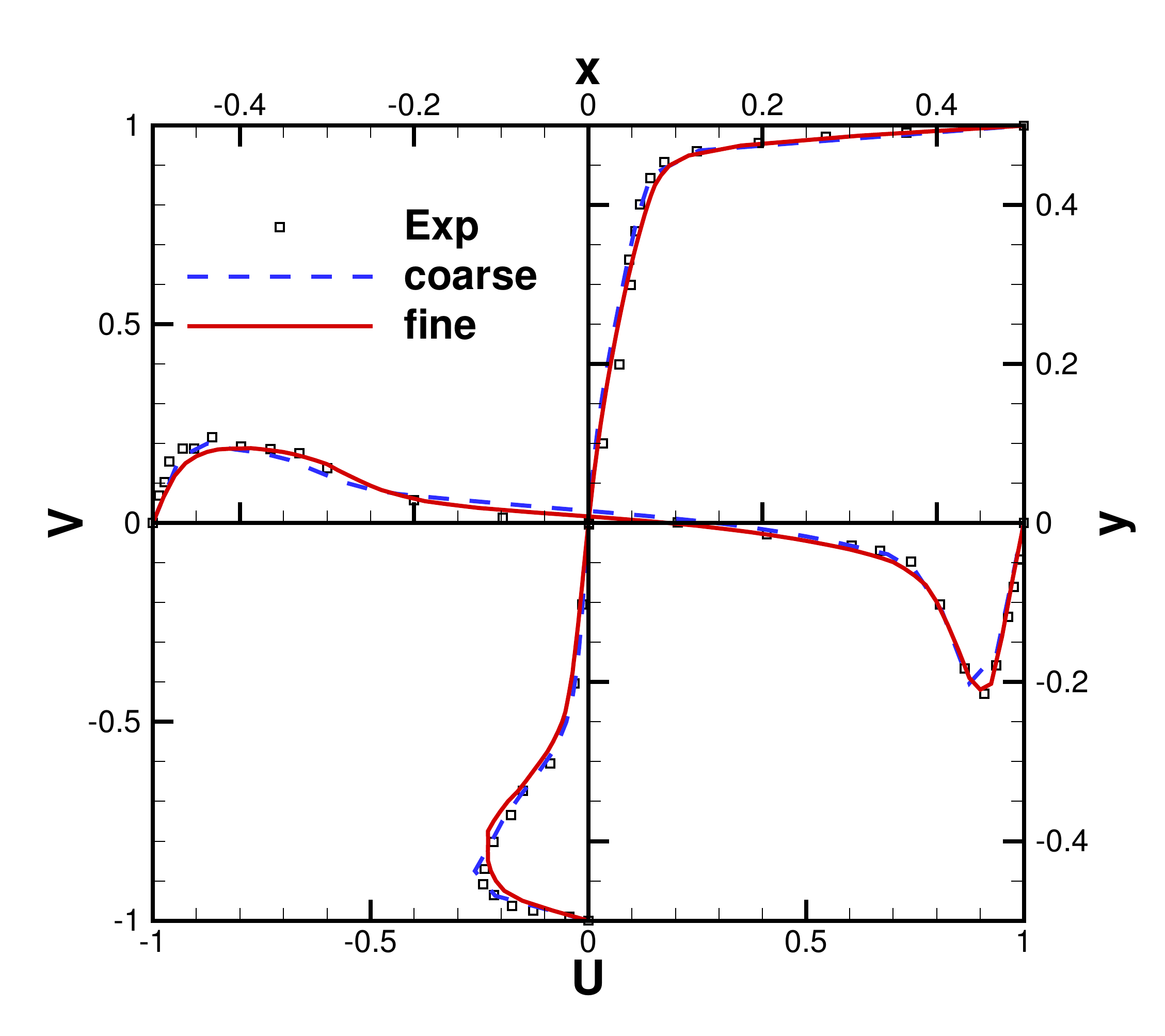}	
\includegraphics[width=0.4\textwidth]
{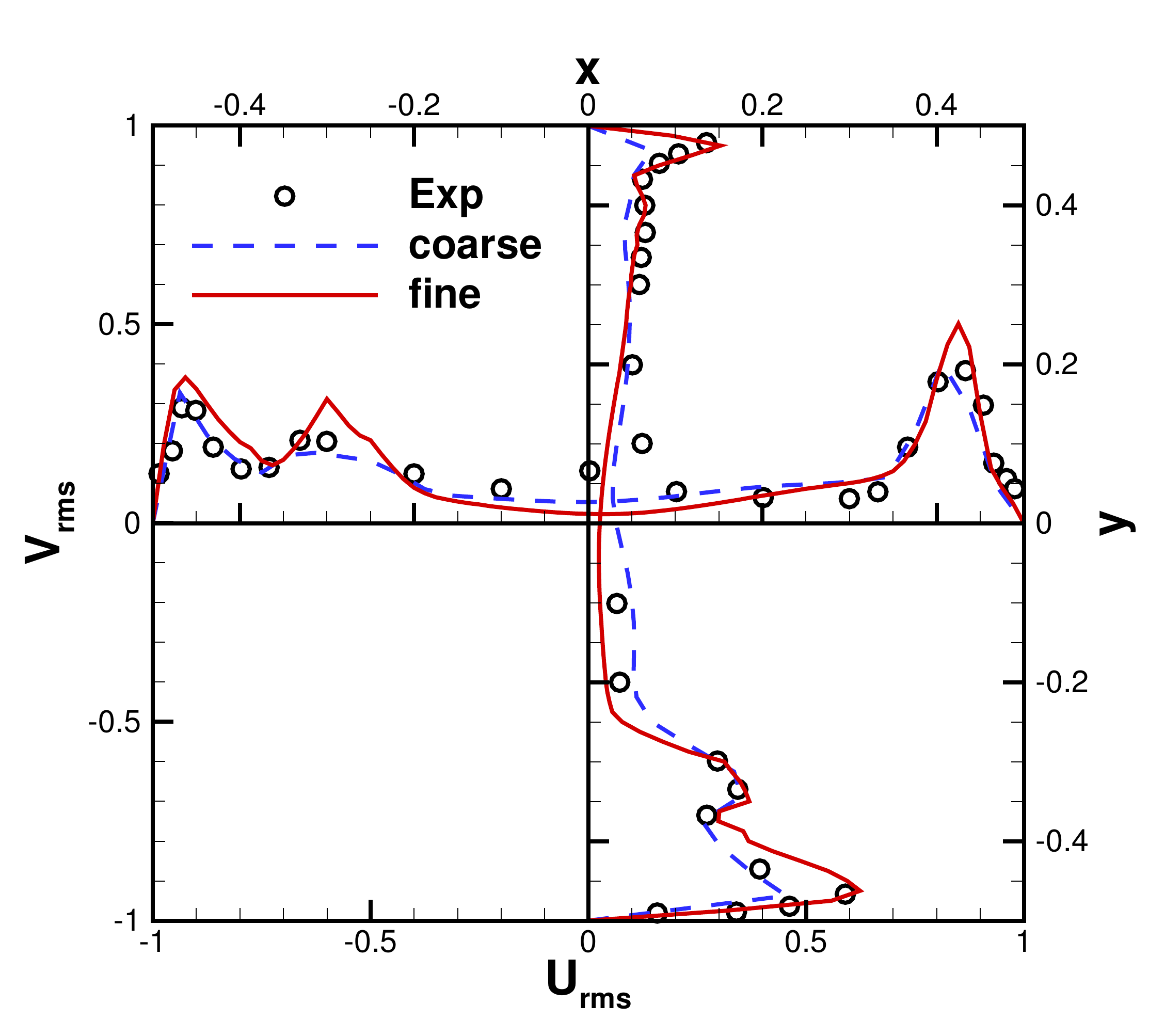}
	 \caption{\label{cavity3d-3200-1}
		Lid-driven cavity flow: Re=3,200. The mean  and RMS velocity profiles obtained by the third-order compact GKS are compared with the experimental data in \cite{prasad1988some,prasad1989reynolds}.}
\end{figure}

The flow becomes transient when $Re>2000$. Experimental results can be found in \cite{prasad1988some,prasad1989reynolds} at Re=3200.
The mean velocity and root-mean-square (RMS) velocity profiles are collected during a time interval, which corresponds to 7 to 10 minutes in the experiment \cite{prasad1988some} and 172 non-dimensional time in the simulation.
The results obtained from the third-order scheme for the $U$ velocity component
along the line $x=0, z=0$, and the $V$ velocity component along the line $y=0, z=0$ are presented in Fig.~\ref{cavity3d-3200-1}.
Both results under the uniform mesh and the refined non-uniform mesh agree well with the experimental data.
A better agreement in the RMS U-velocity  $U_{rms}$ can be observed in the refined mesh calculation.

\subsection{Subsonic viscous flow around a sphere at Re=118}

\begin{figure}[htp]	
	\centering	
	\includegraphics[width=0.4\textwidth]{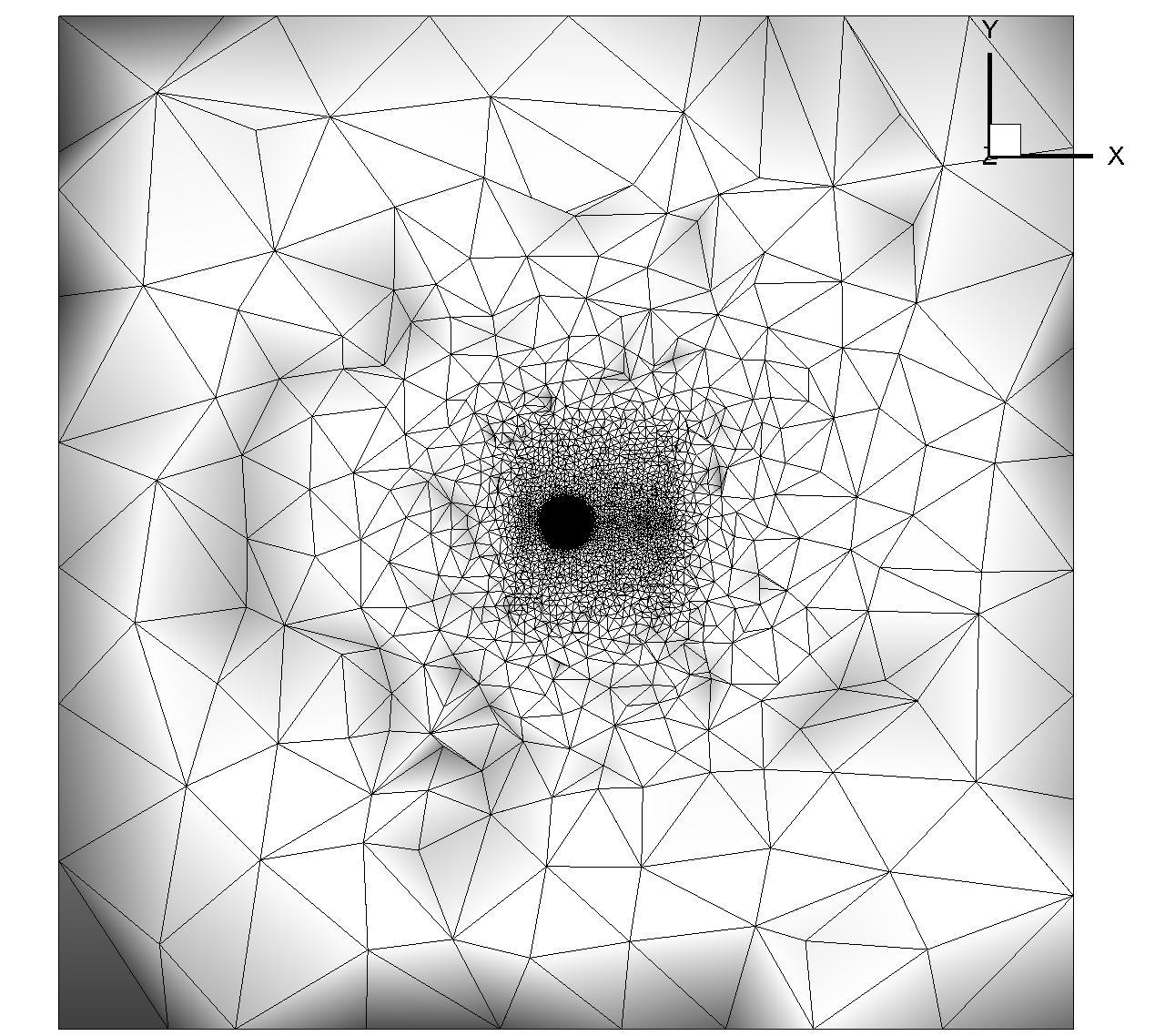}
	\includegraphics[width=0.4\textwidth]{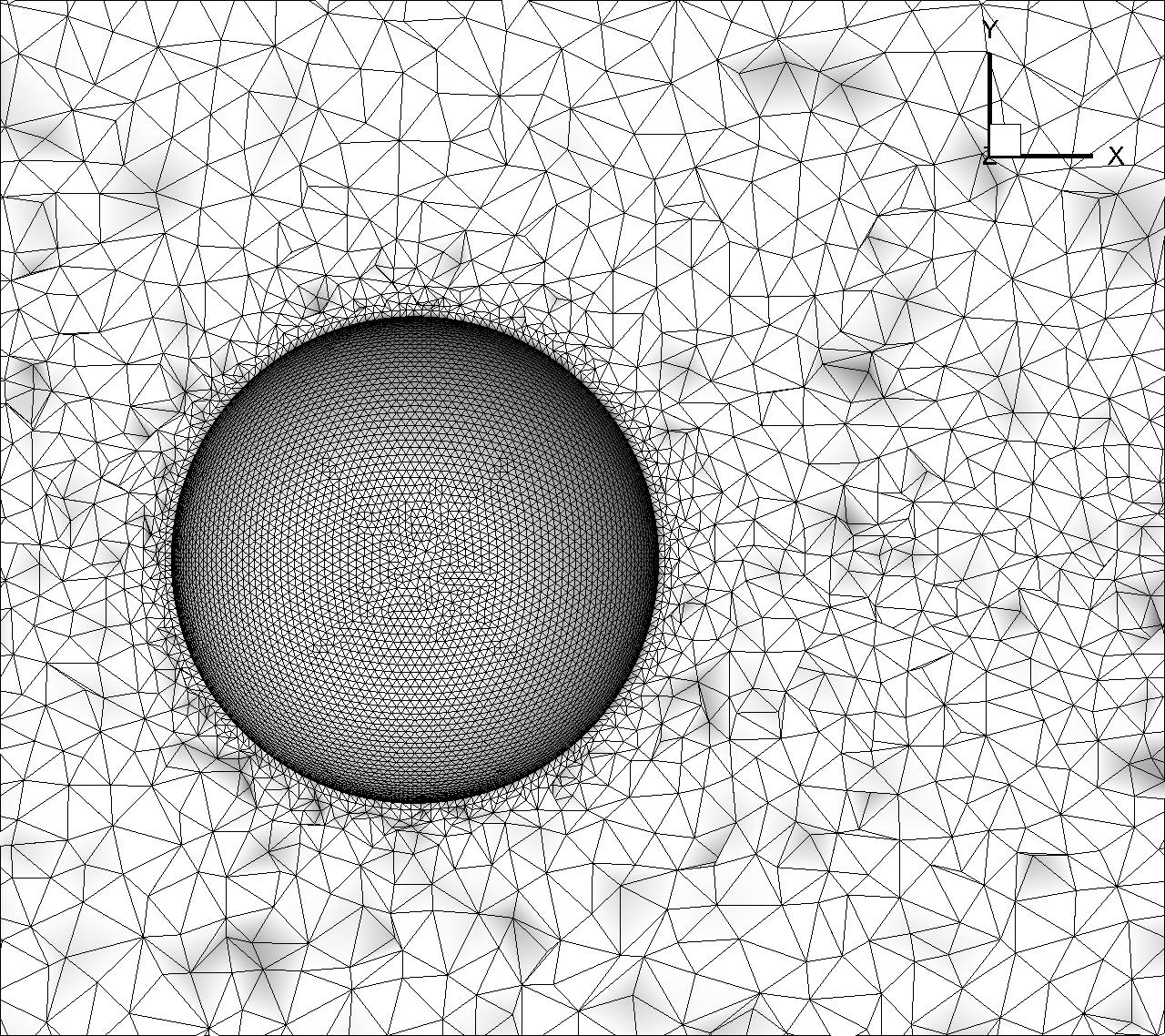}
	\caption{\label{viscous-subsonic-sphere-fine-mesh}
		Subsonic flow passing through a viscous sphere. Mesh number: 399,546.}
\end{figure}

\begin{figure}[htp]	
	\centering
	\includegraphics[height=0.4\textwidth]
	{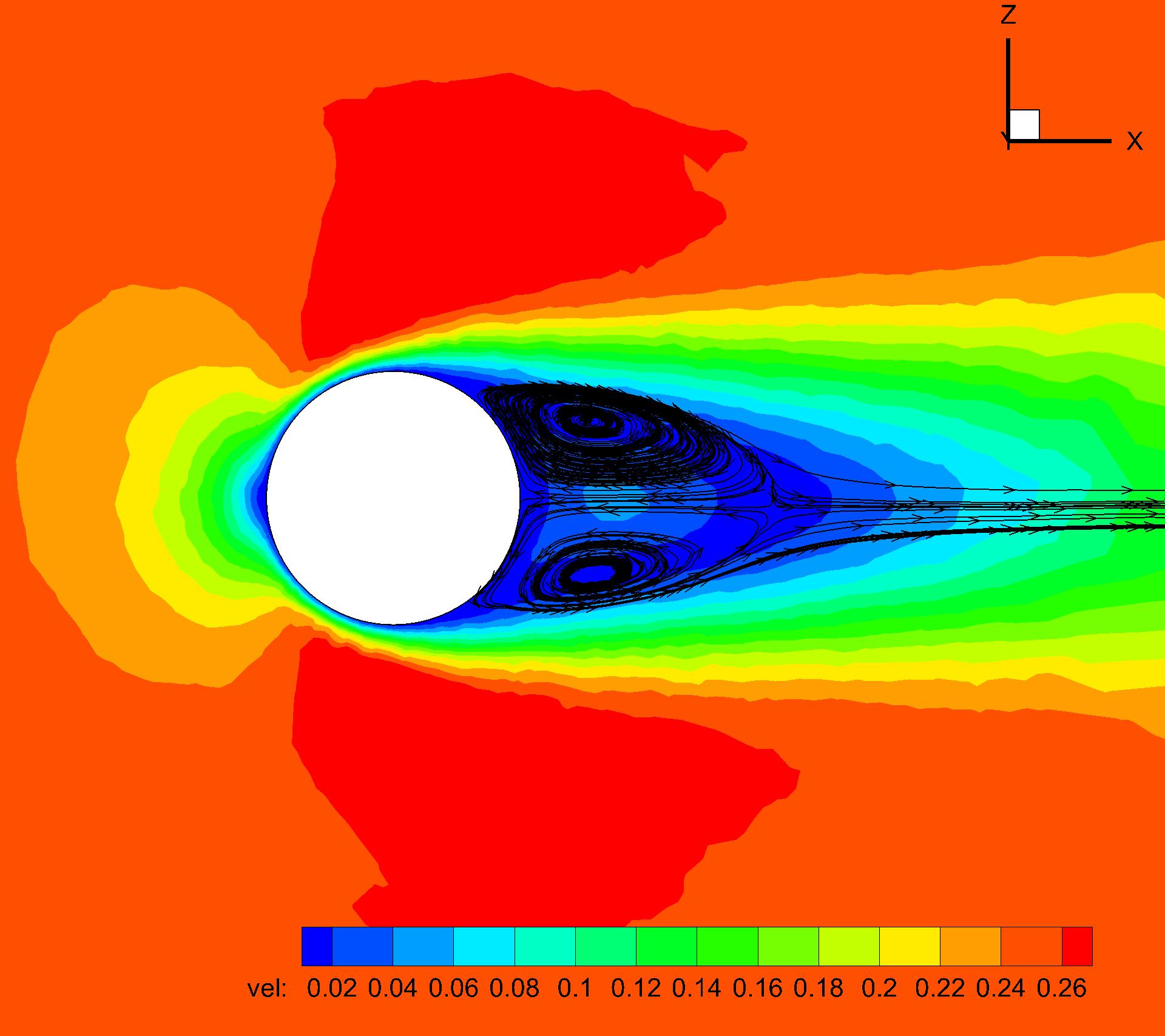}
	\includegraphics[height=0.4\textwidth]
	{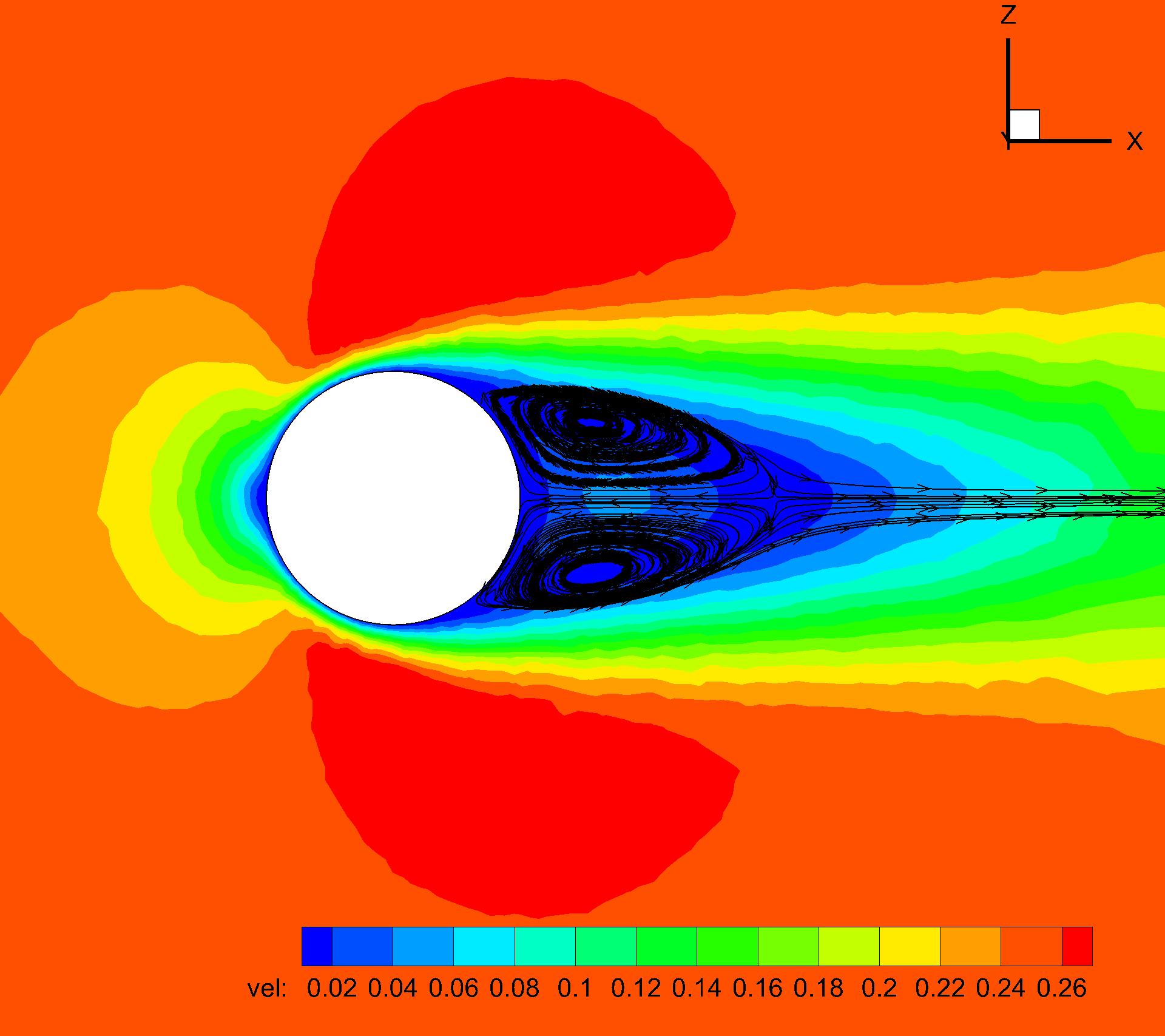}

	\caption{\label{viscous-subsonic-sphere-fine-vel-contour}
		Subsonic flow passing through a viscous sphere. Ma=0.2535. Re=118. Left: The second-order GKS. Right: The third-order GKS.}
\end{figure}

\begin{figure}[htp]	
	\centering
	\includegraphics[height=0.4\textwidth]
    {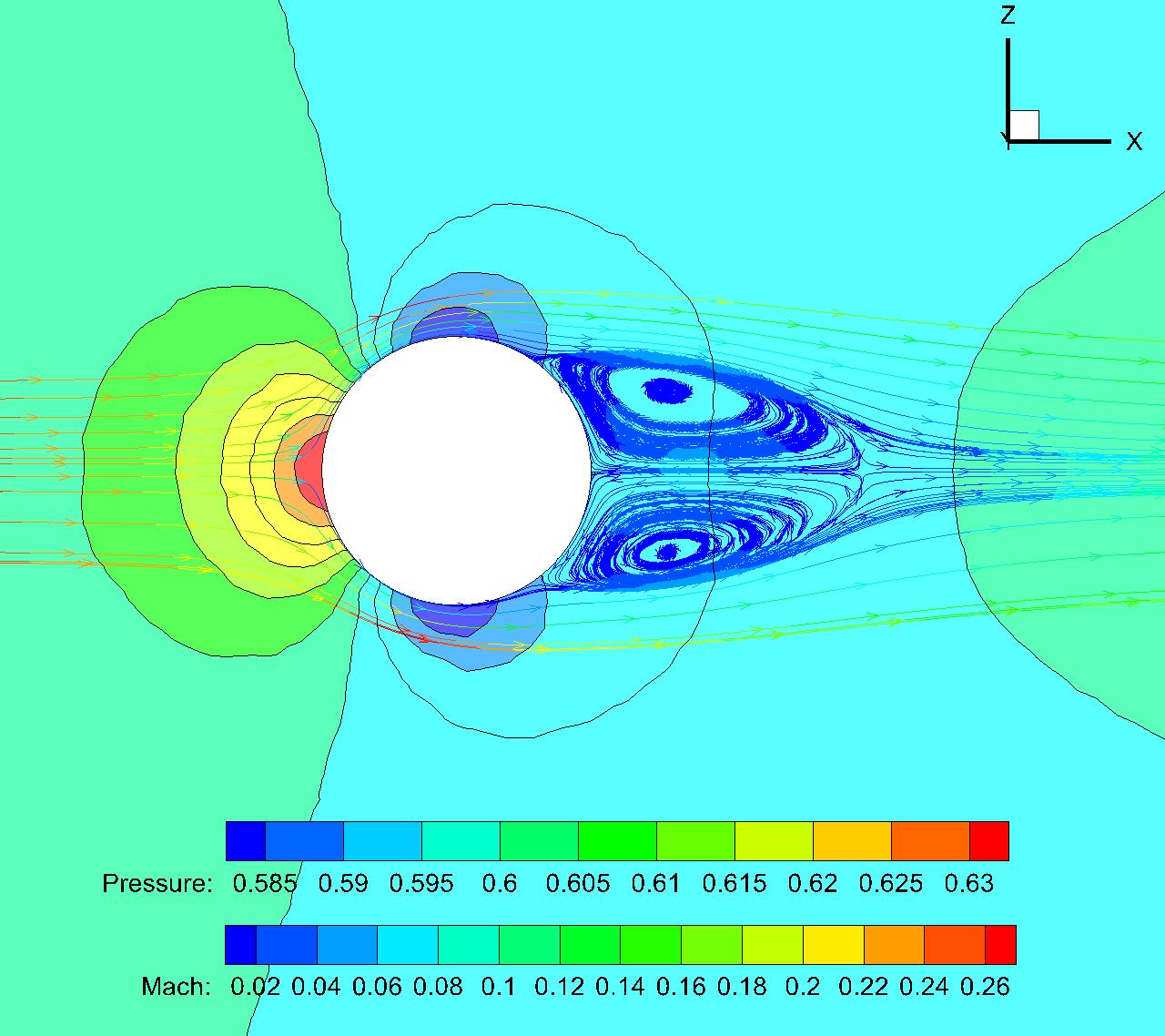}
	\includegraphics[height=0.4\textwidth]
	{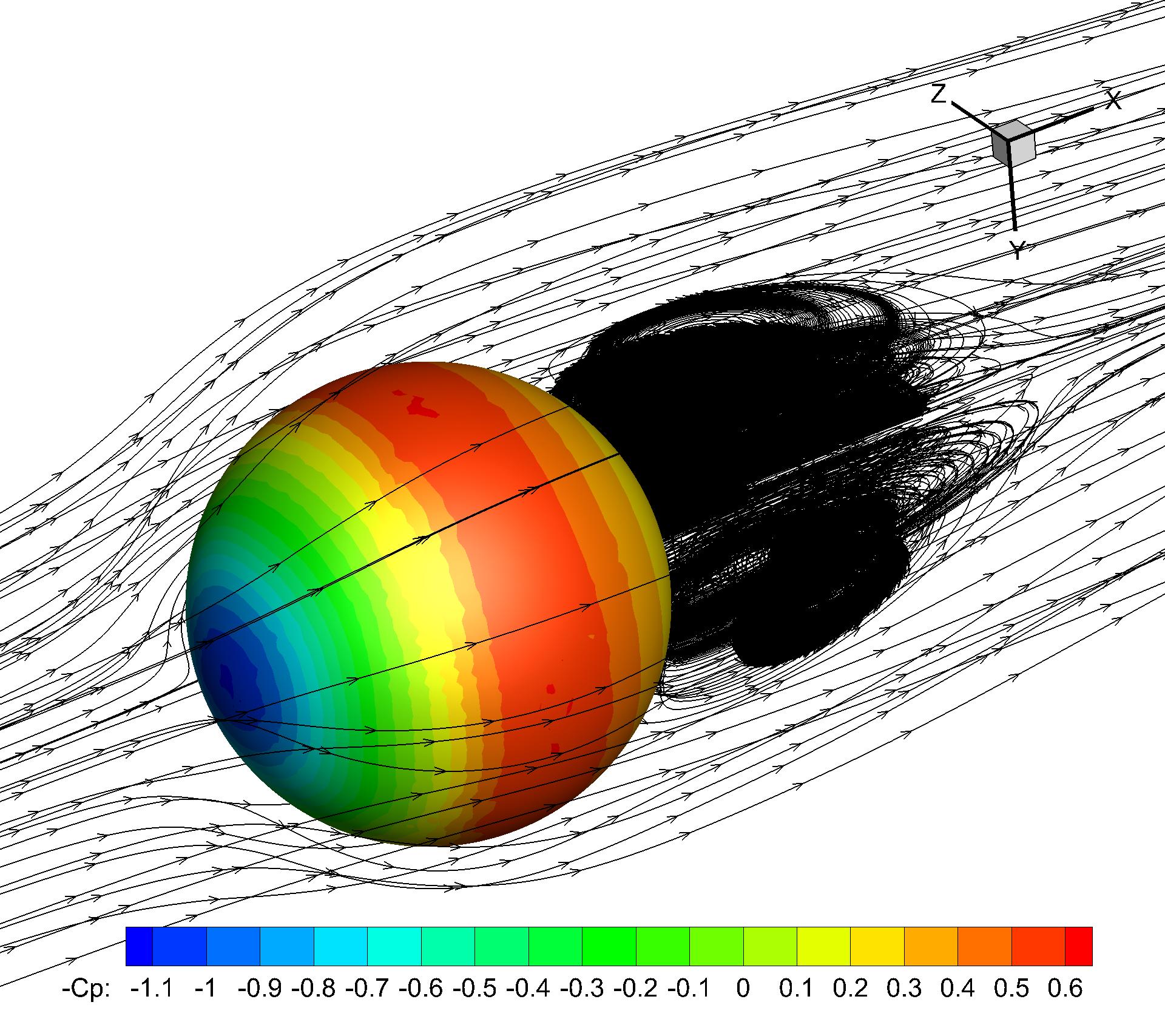}
	\caption{\label{viscous-subsonic-sphere-fine-contour-3rd}
		Subsonic flow passing through a viscous sphere by the third-order compact GKS. Ma=0.2535. Re=118. }
\end{figure}

A low-speed viscous flow passing through a sphere is tested.
The Reynolds number based on the diameter of the sphere $D=1$  is 118.
In such case, a drag coefficient $C_D=1$ was reported from the experiment in \cite{taneda1956experimental}.
The far-field flow condition outside boundary of the domain is set with the free stream condition
\begin{equation*}
\begin{split}
(\rho,U,V,W,p)_{\infty} =(1,0.2535,0,0,\frac{1}{\gamma}),
\end{split}
\end{equation*}
with $\gamma=1.4$, $Ma_{\infty}=0.2535$.
The surface of the sphere is set as a non-slip adiabatic wall.
The first mesh off the wall has a size $ h \approx 1.5 \times 10^{-2} D$, as shown in Fig.~\ref{viscous-subsonic-sphere-fine-mesh}.
Both second-order and third-order schemes with non-linear reconstructions are tested. A clean and symmetric velocity contour is observed from the third-order compact GKS, as shown in Fig.~\ref{viscous-subsonic-sphere-fine-vel-contour}.
The pressure contour and the 3-D streamline are also presented in Fig.~\ref{viscous-subsonic-sphere-fine-contour-3rd}, where the high resolution from
the non-linear compact reconstruction has been demonstrated, even with mesh irregularity. 
The quantitative results are given in Table \ref{viscous-subsonic-sphere}, including the drag coefficient $C_D$,  the separation angle $\theta$, and the closest wake length $L$, as defined in \cite{ji2020three}.

\begin{table}[htbp]
	\small
	\begin{center}
		\def\temptablewidth{1.0\textwidth}
		{\rule{\temptablewidth}{1pt}}
		\begin{tabular*}{\temptablewidth}{@{\extracolsep{\fill}}c|c|c|c|c|c}
			Scheme & Mesh number & Cd  & $\theta$  &L &Cl\\
			\hline
			Experiment \cite{taneda1956experimental}	&-- & 1.0  & 151 & 1.07 & -- \\ 	
			Current 2nd & 399,546 & 1.027  & 126.9 & 1.00 & 1.5e-2\\
			Current 3rd & 399,546 & 1.018  & 127.4 & 1.00 & 1.7e-3\\
			Implicit third-order DDG \cite{cheng2017parallel} & 160,868 & 1.016 & 123.7 & 0.96 & --\\
			Implicit fourth-order VFV \cite{wang2017thesis}  & 458,915 & 1.014 & --& -- & 2.0e-5\\
			Implicit third-order AMR-VFV \cite{pan2018high} & 621,440   & 1.016 &--& -- & --\\
		\end{tabular*}
		{\rule{\temptablewidth}{0.1pt}}
	\end{center}
	\vspace{-4mm} \caption{\label{viscous-subsonic-sphere} Quantitative comparisons among different compact schemes  for the viscous flow over a sphere.}
\end{table}

\subsection{Supersonic viscous flow passing through a sphere at Re=300}

\begin{figure}[htp]	
	\centering	
	\includegraphics[width=0.4\textwidth]
	{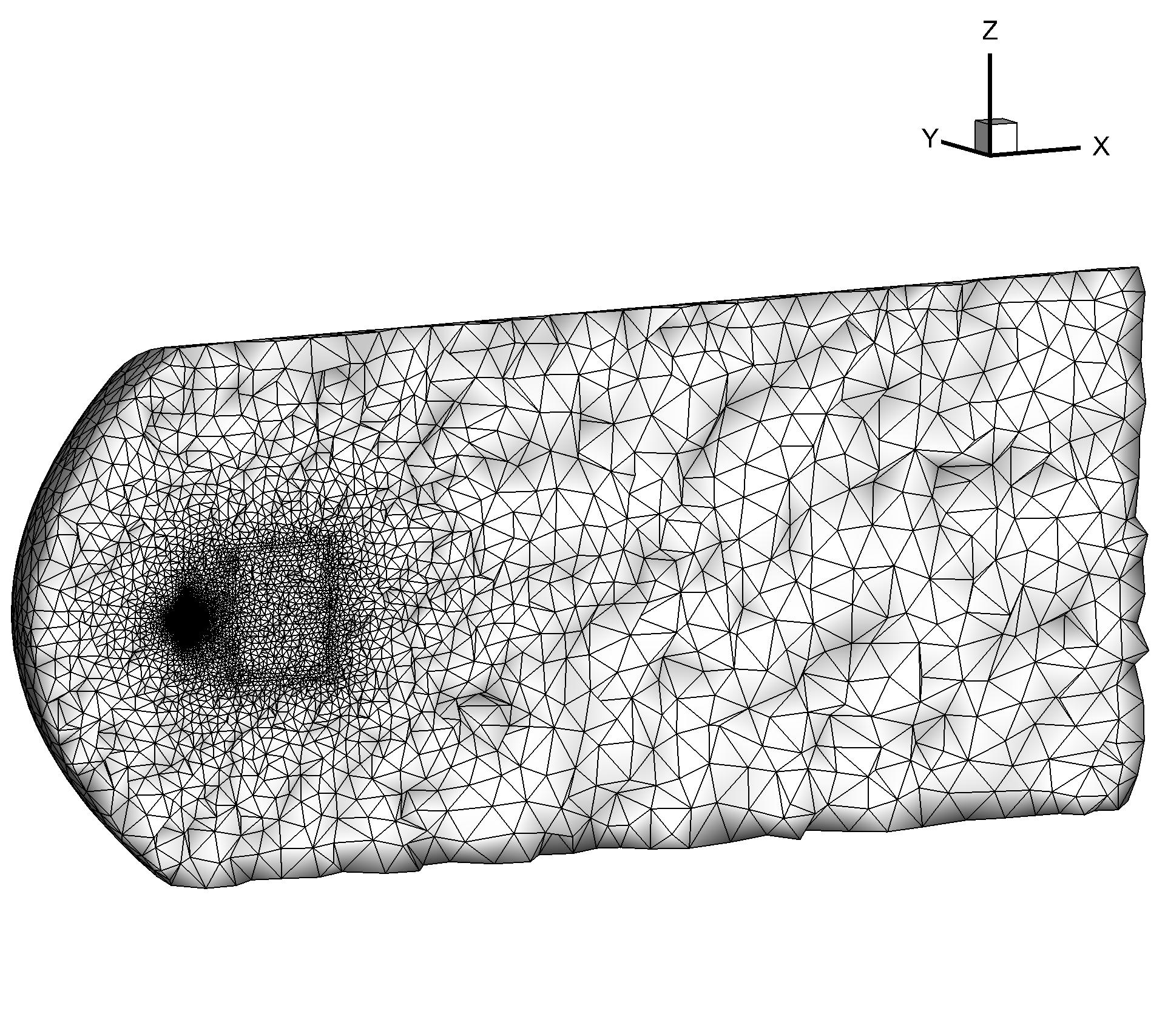}
	\includegraphics[width=0.4\textwidth]
	{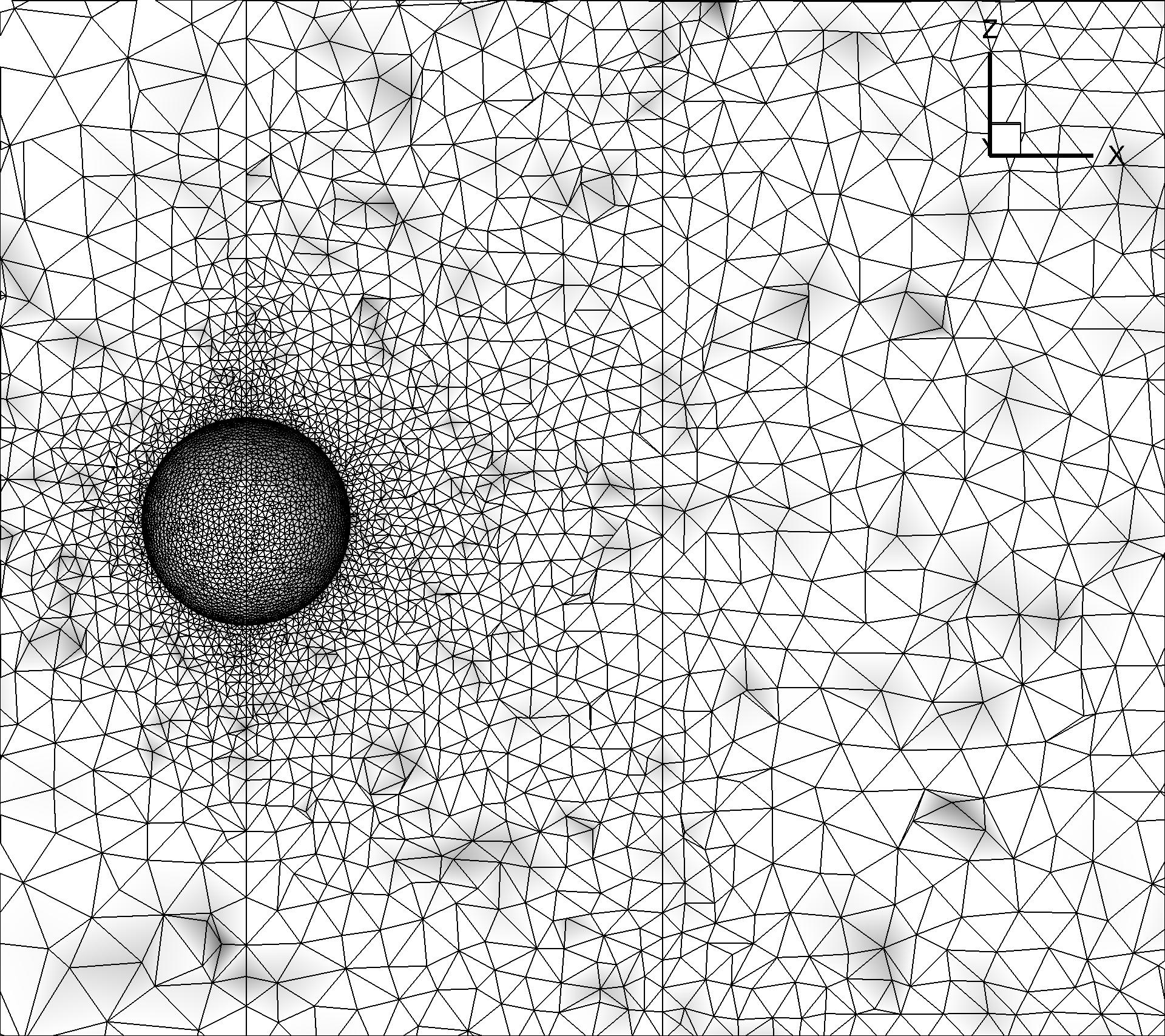}
	\caption{\label{viscous-supersonic-sphere-fine-mesh}
		Supersonic flow passing through a viscous sphere. Mesh number: 665,914.}
\end{figure}

\begin{figure}[htp]	
	\centering
	\includegraphics[height=0.4\textwidth]
	{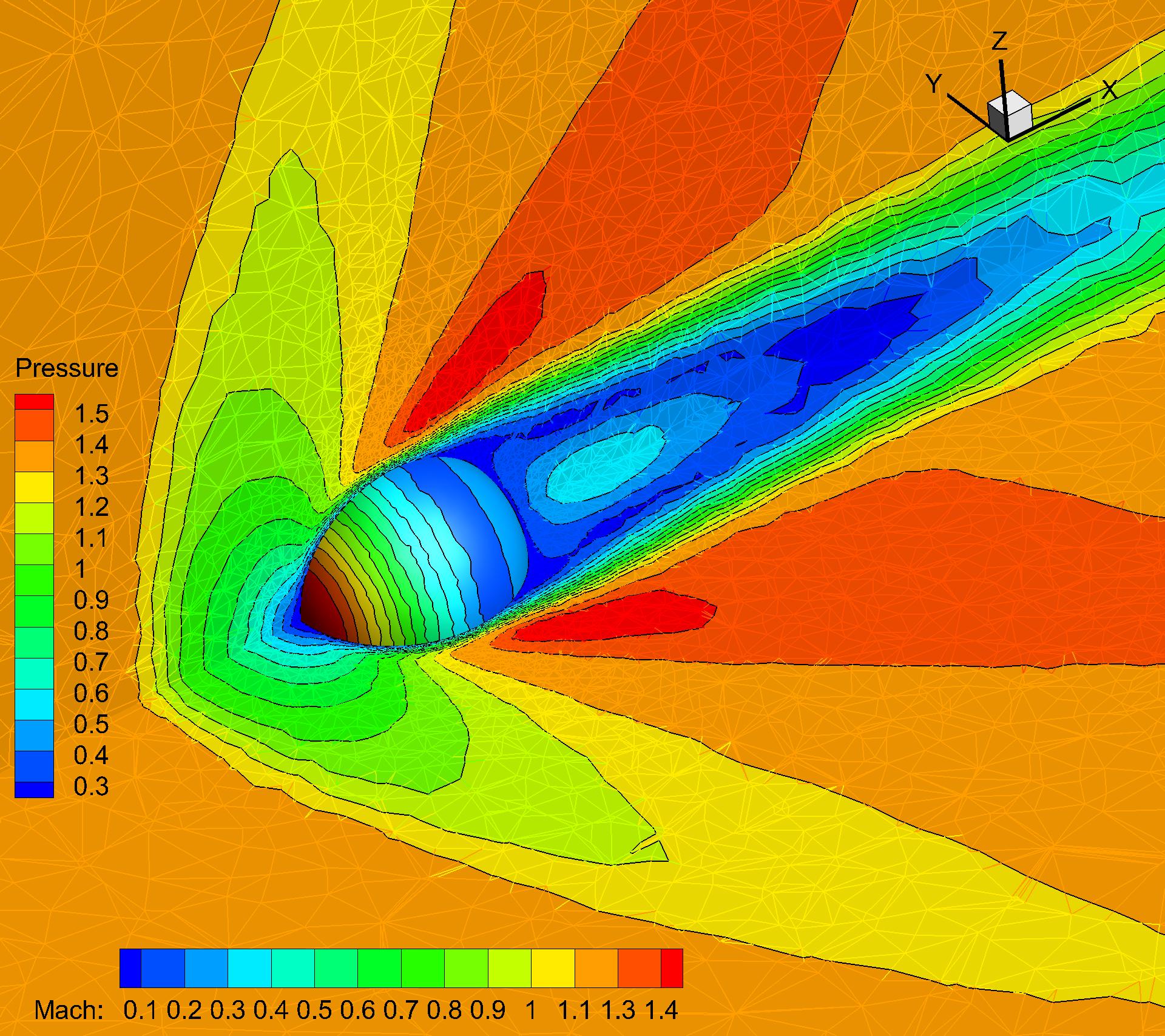}
	\includegraphics[height=0.4\textwidth]
	{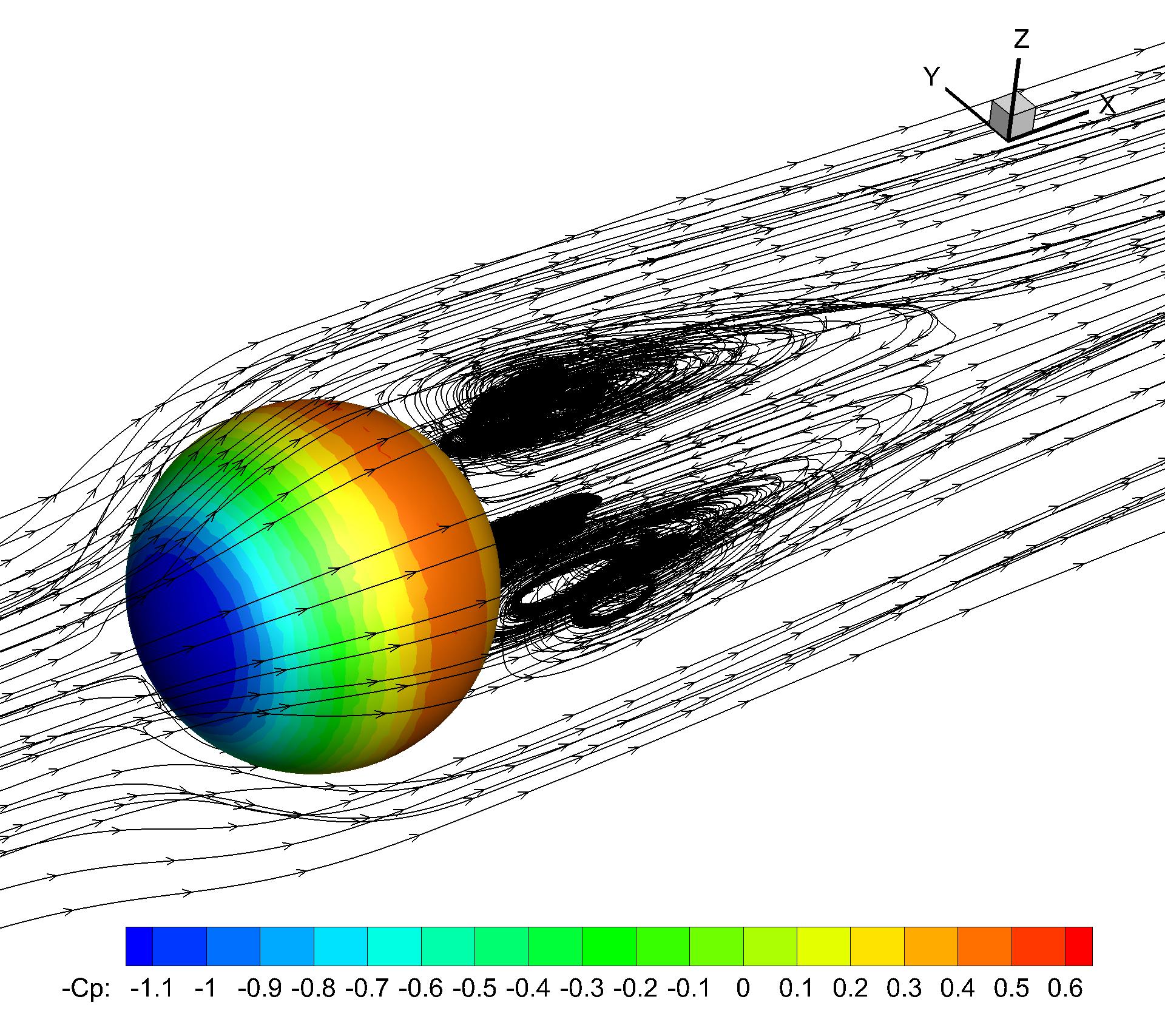}
	\caption{\label{viscous-supersonic-sphere-fine-contour-3rd}
		Supersonic flow passing through a viscous sphere by the third-order compact GKS. Ma=1.2. Re=300.}
\end{figure}

To validate the robustness of the current scheme for the high-speed viscous flow, a supersonic flow passing through a sphere with $Ma=1.2$ is tested.
The non-slip adiabatic boundary condition is imposed on the surface of the sphere. The Reynolds number is 300 based on the diameter $D=1$. The Prandtl number is $Pr=1$.
The tetrahedral mesh with an upstream length of 5 and a downstream length of 40 is shown in Fig.~\ref{viscous-supersonic-sphere-fine-mesh}.
The first mesh size at the wall has a thickness $ 2.3 \times 10^{-2} D$.
The numerical results obtained by the third-order compact GKS are shown in Fig.~\ref{viscous-supersonic-sphere-fine-contour-3rd}.
Quantitative results are listed in Table \ref{viscous-supersonic-sphere},
which have good agreement with those given by Nagata et al. \cite{Nagata2016sphere}. Note that the proposed second-order GKS cannot survive for this case.

\begin{table}[htbp]
	\small
	\begin{center}
		\def\temptablewidth{1.0\textwidth}
		{\rule{\temptablewidth}{1pt}}
		\begin{tabular*}{\temptablewidth}{@{\extracolsep{\fill}}c|c|c|c|c|c}
			Scheme & Grid Number & Cd  & $\theta$  &L & Shock stand-off\\
			\hline
			WENO6 \cite{Nagata2016sphere} 	&909,072 & 1.282  & 126.9 & 1.61 & 0.69 \\ 	
			Current 3rd & 665,914 & 1.274  & 126.3 & 1.64 & 0.66-0.82 (0.72) \\	
		\end{tabular*}
		{\rule{\temptablewidth}{0.1pt}}
	\end{center}
	\vspace{-4mm} \caption{\label{viscous-supersonic-sphere} Quantitative comparisons between the current scheme and the reference solution for the supersonic viscous flow over a sphere.}
\end{table}

\subsection{Transonic inviscid flow around ONERA M6 wing}

\begin{figure}[htp]	
	\centering	
	\includegraphics[width=0.4\textwidth]{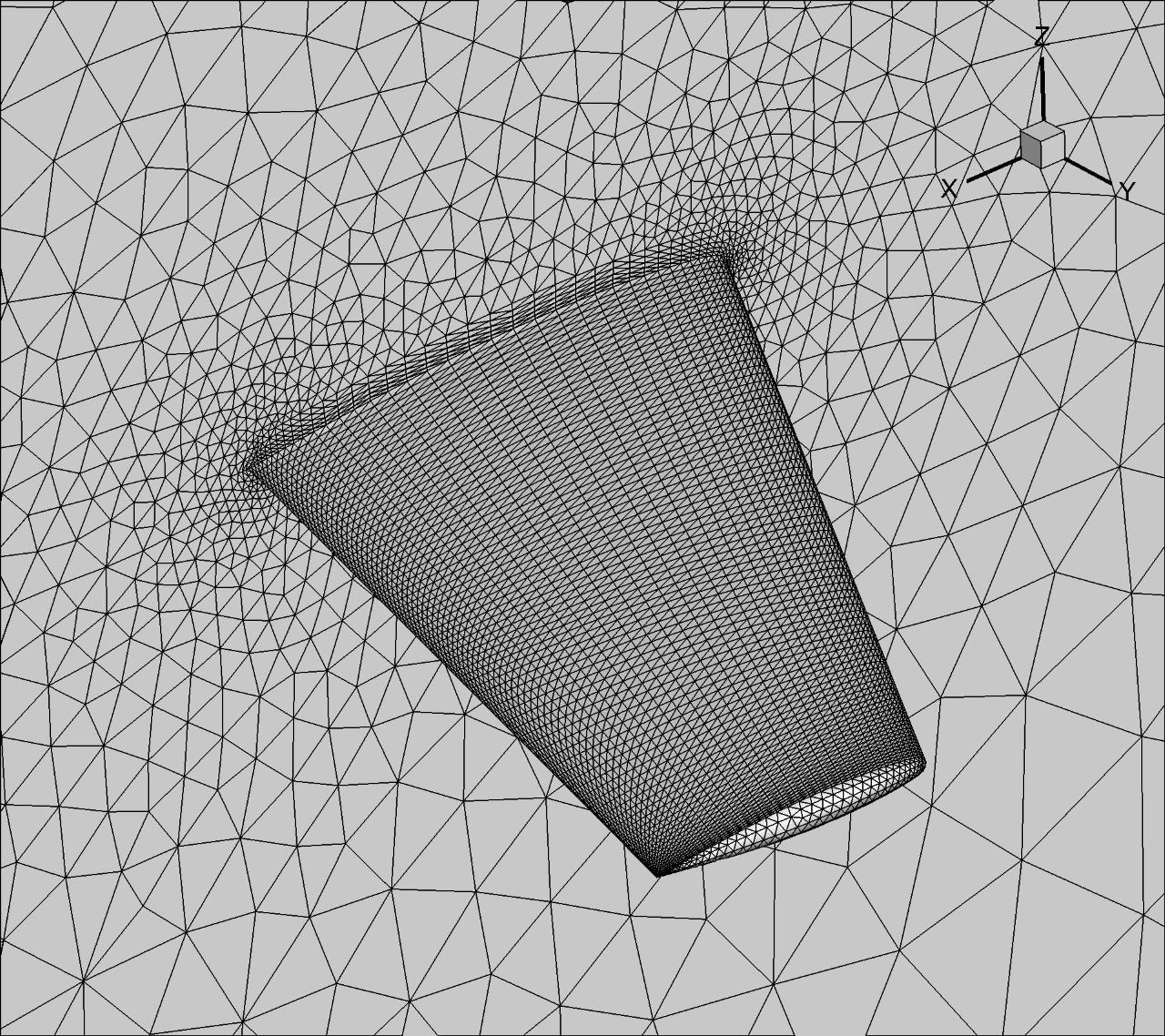}
	\includegraphics[width=0.4\textwidth]{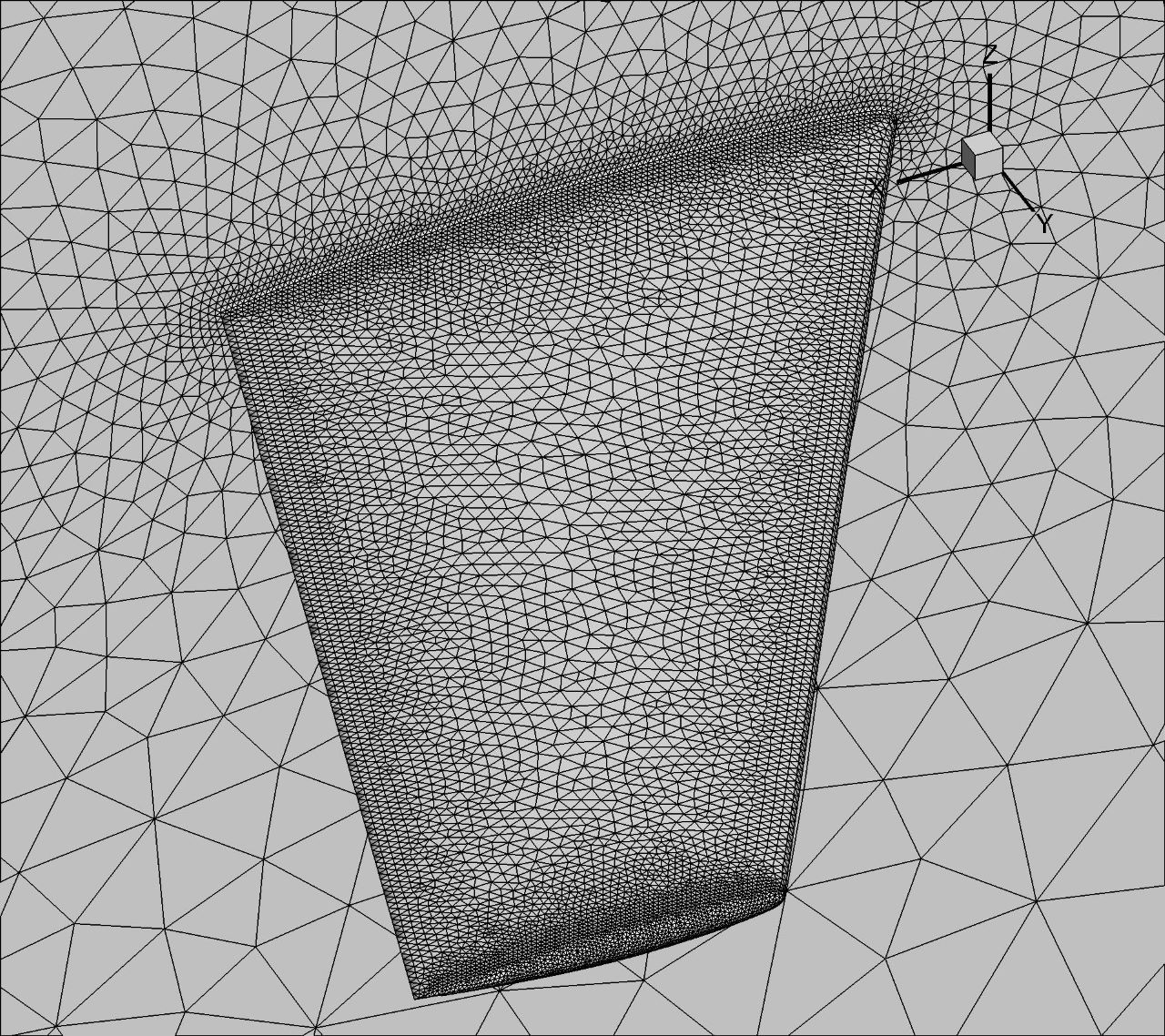}
	 \caption{\label{m6-wing-inviscid-mesh}
		Mesh for the inviscid ONERA M6 wing. Left: Mesh I with 294,216 cells. Right: Mesh II with 347,094 cells. }
\end{figure}

\begin{figure}[htp]	
	\centering	
	\includegraphics[width=0.24\textwidth]{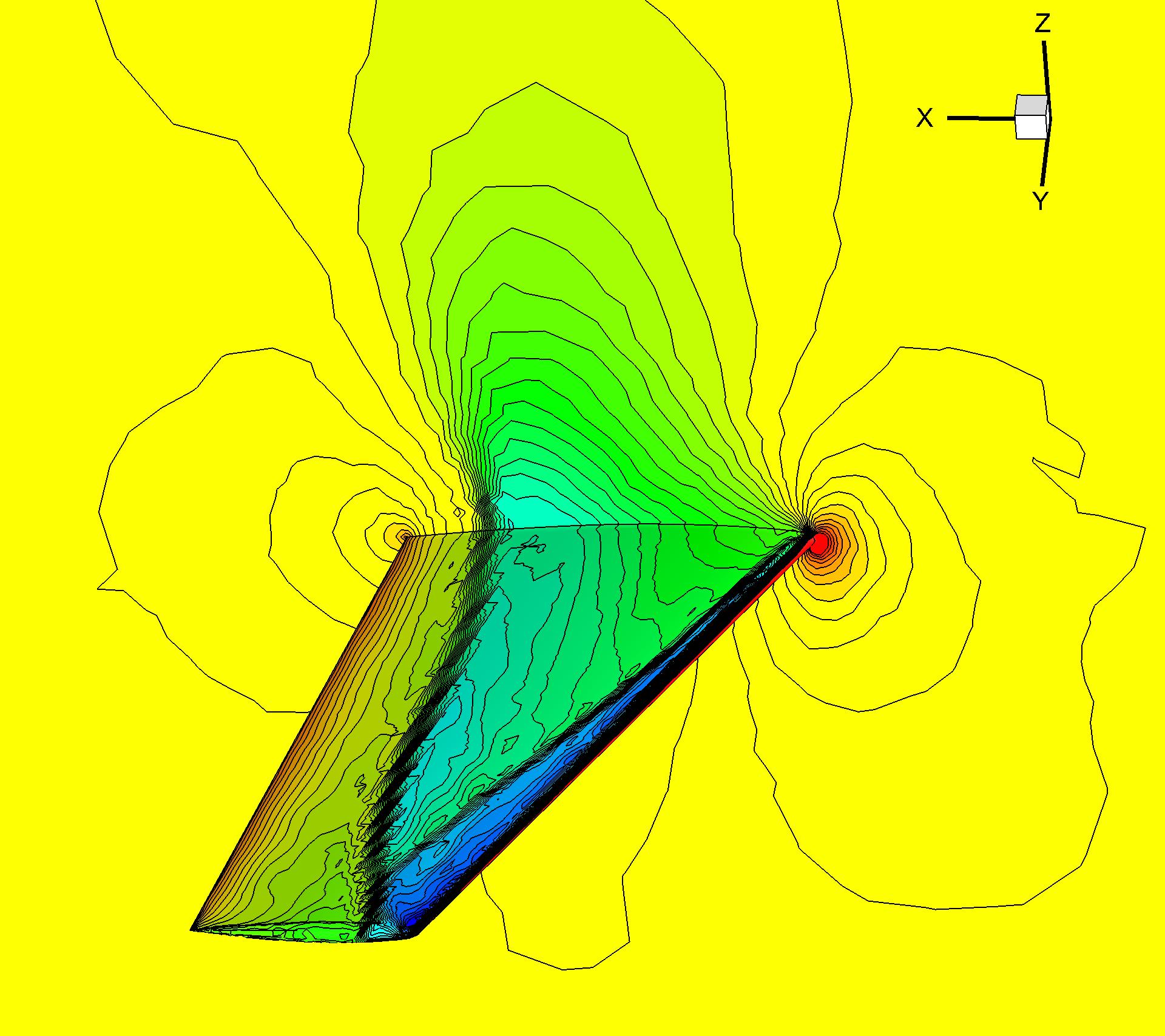}
	\includegraphics[width=0.24\textwidth]{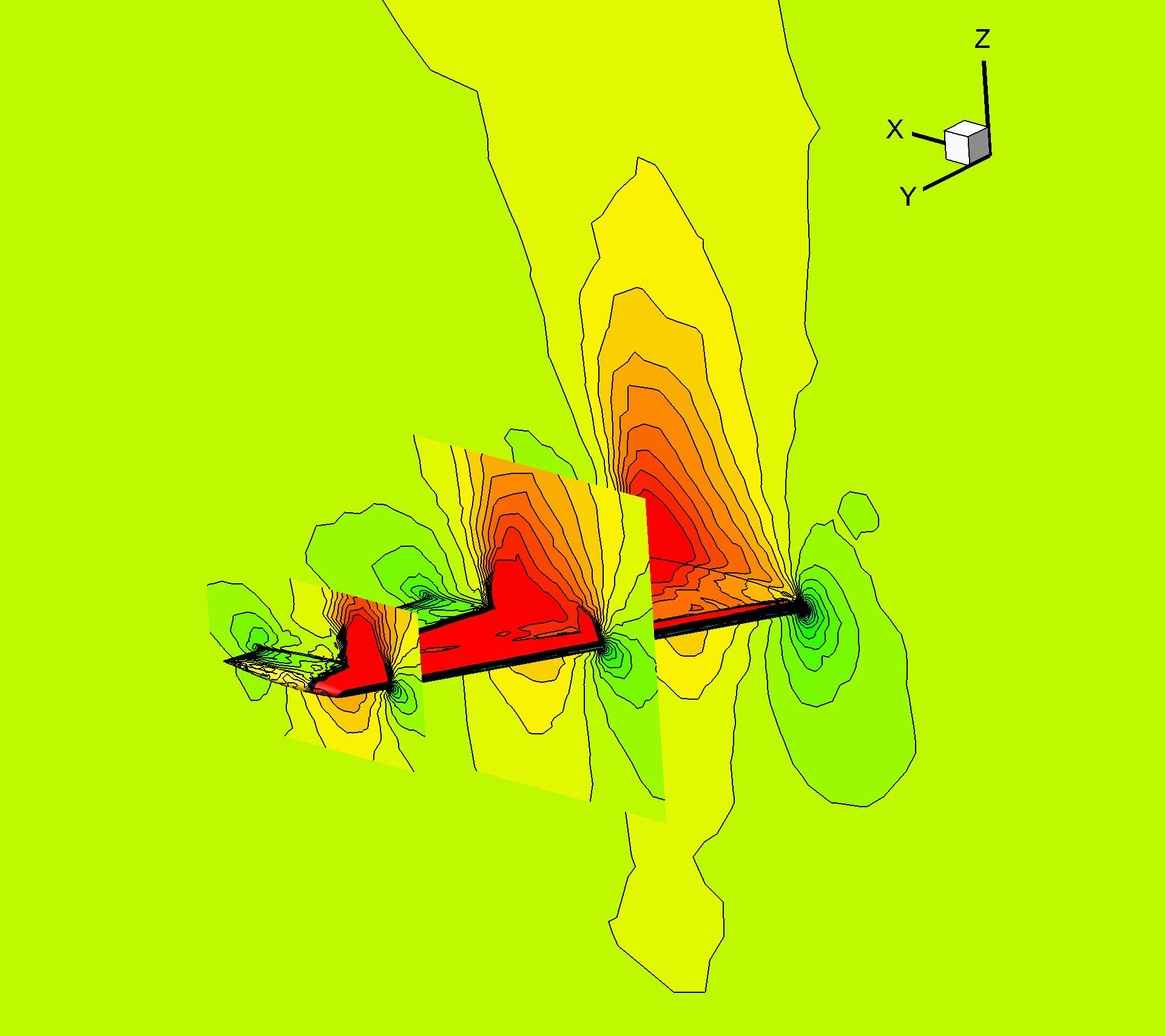}
	\includegraphics[width=0.24\textwidth]{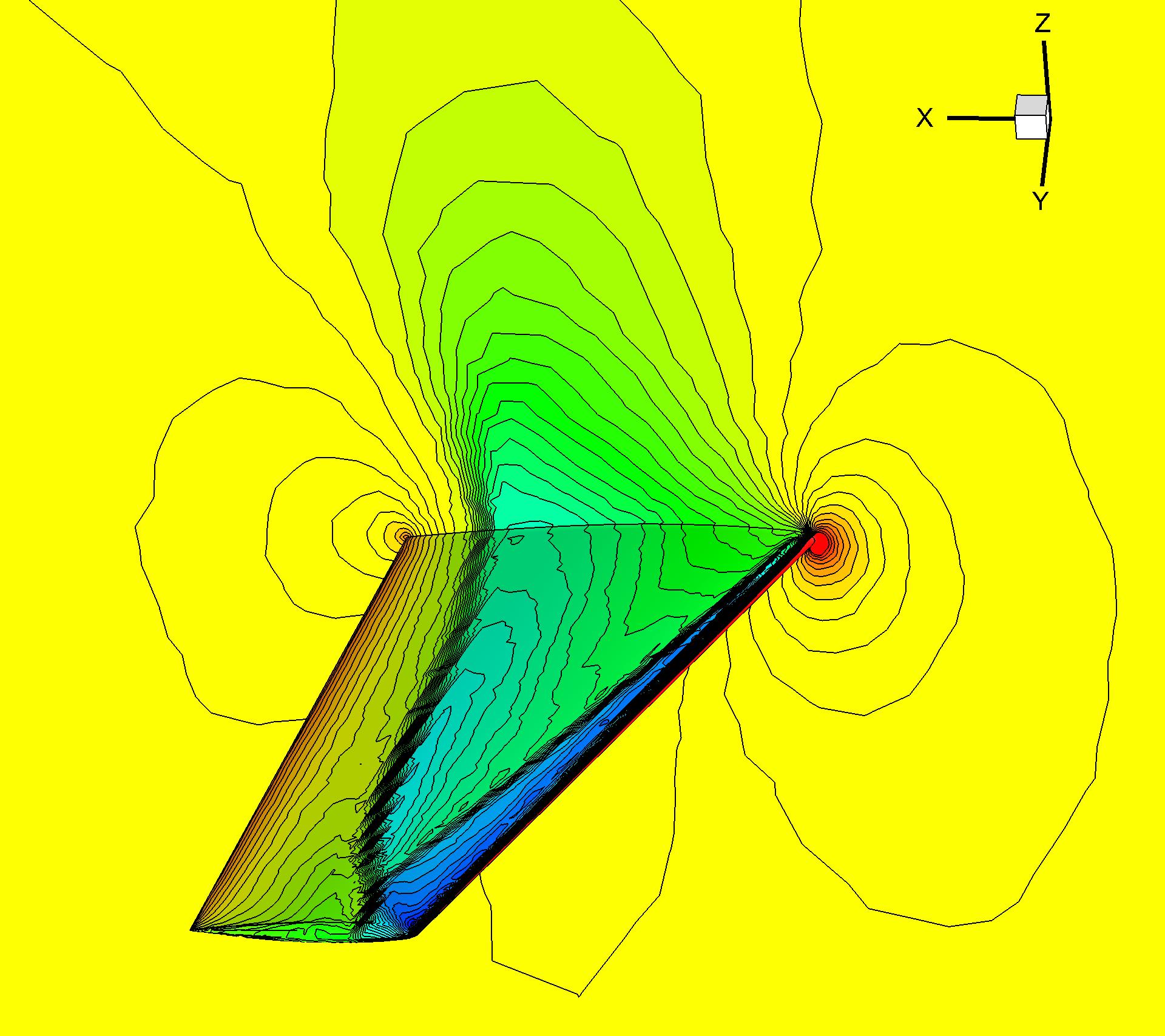}
	\includegraphics[width=0.24\textwidth]{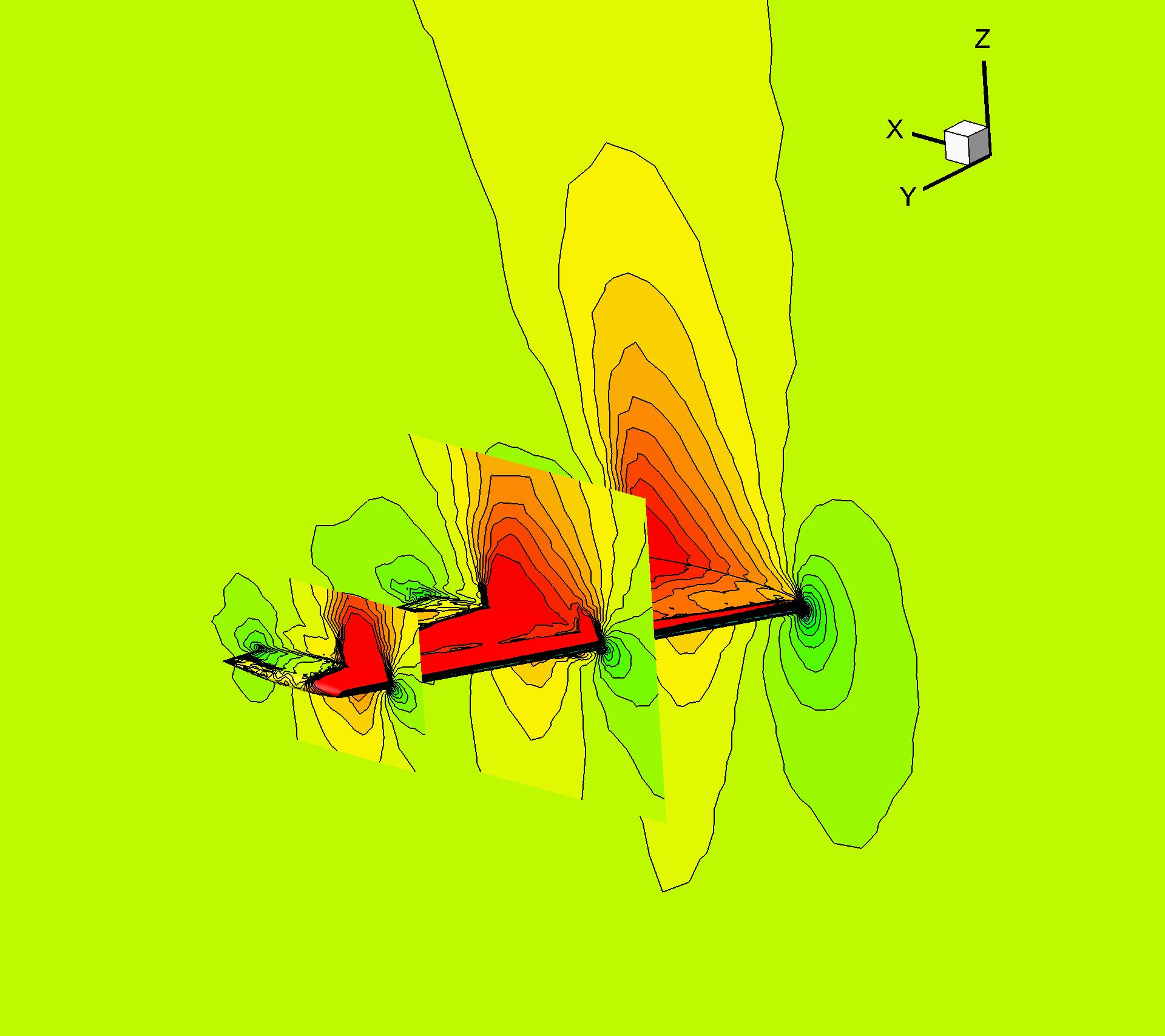}
	 \caption{\label{m6-wing-inviscid-refine}
		Transonic flow over an inviscid ONERA  M6 wing under Mesh I. Ma=0.8935. AOA=3.06$^{\circ}$. Left: The second-order GKS. Right: The third-order GKS.}
\end{figure}

\begin{figure}[htp]	
	\centering	
	\includegraphics[width=0.32\textwidth]{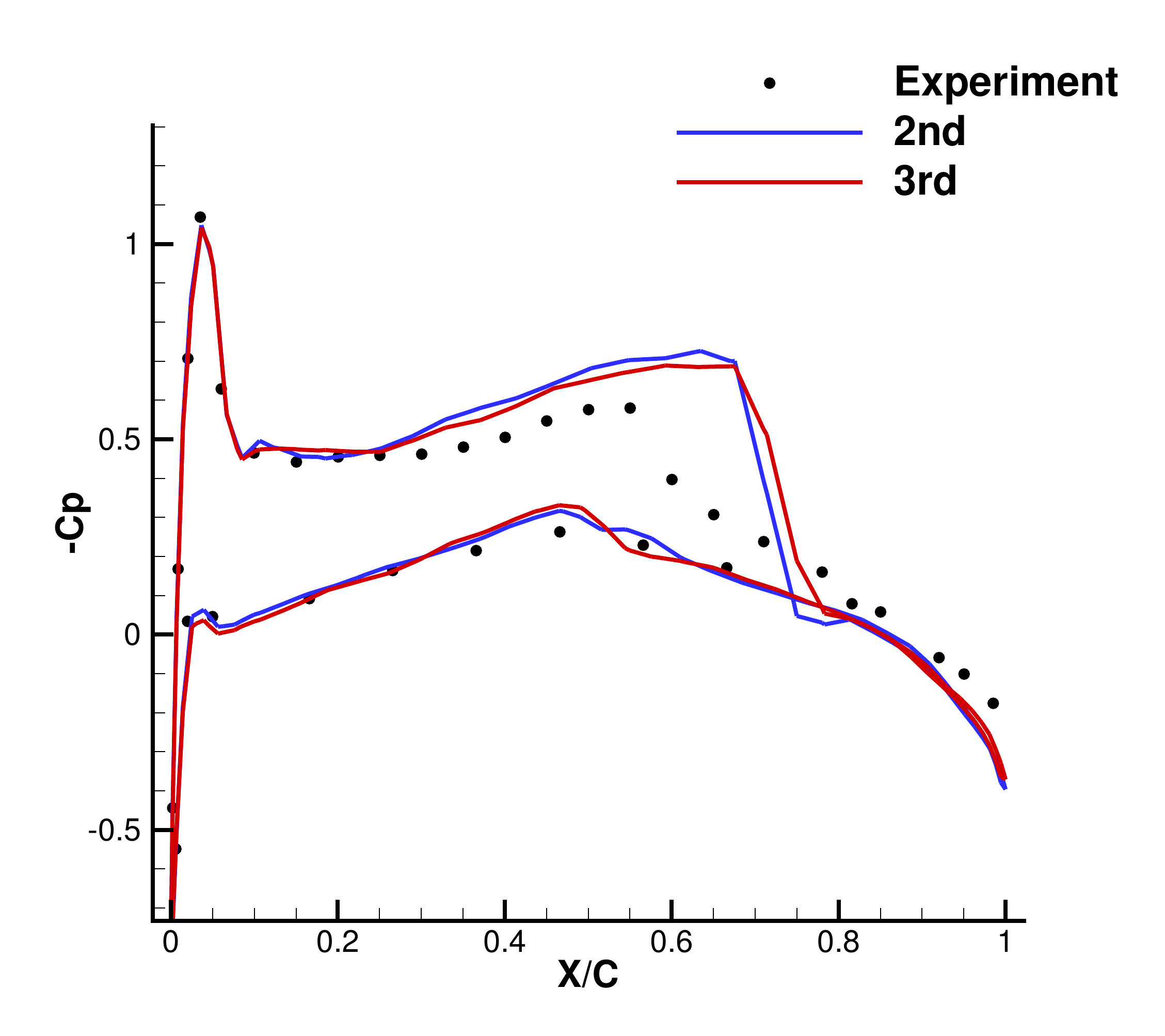}
	\includegraphics[width=0.32\textwidth]{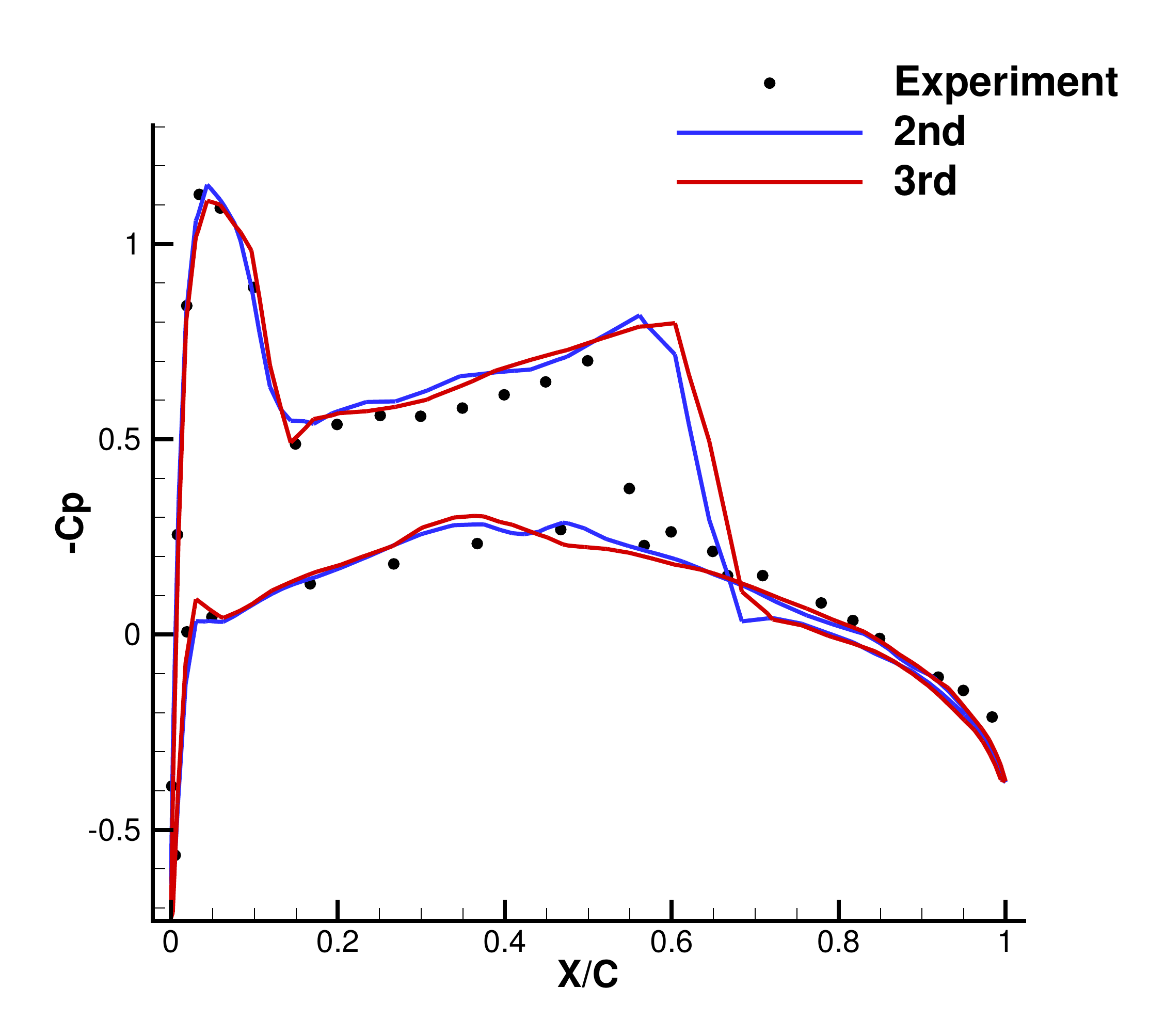}
	\includegraphics[width=0.32\textwidth]{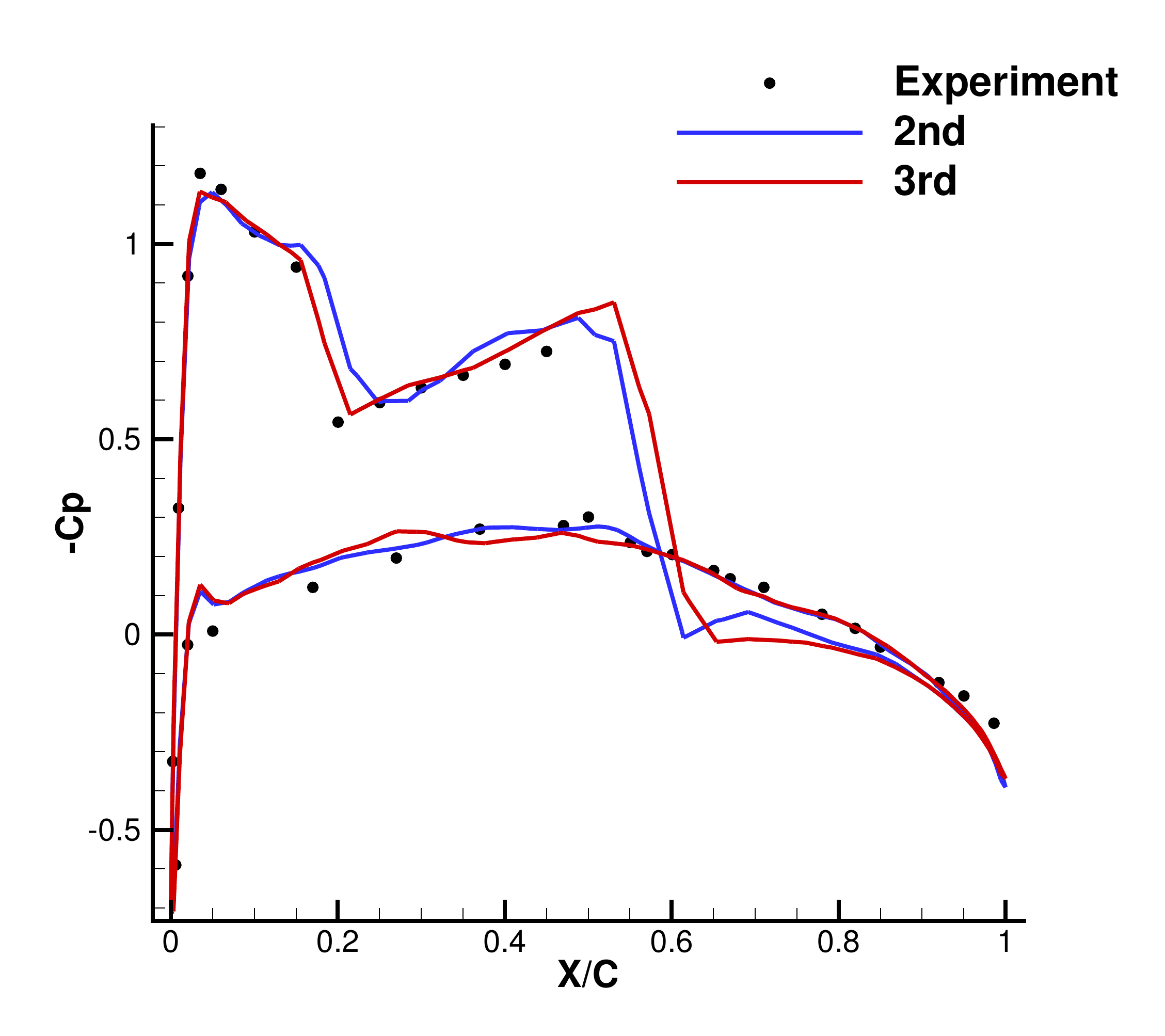}
	\includegraphics[width=0.32\textwidth]{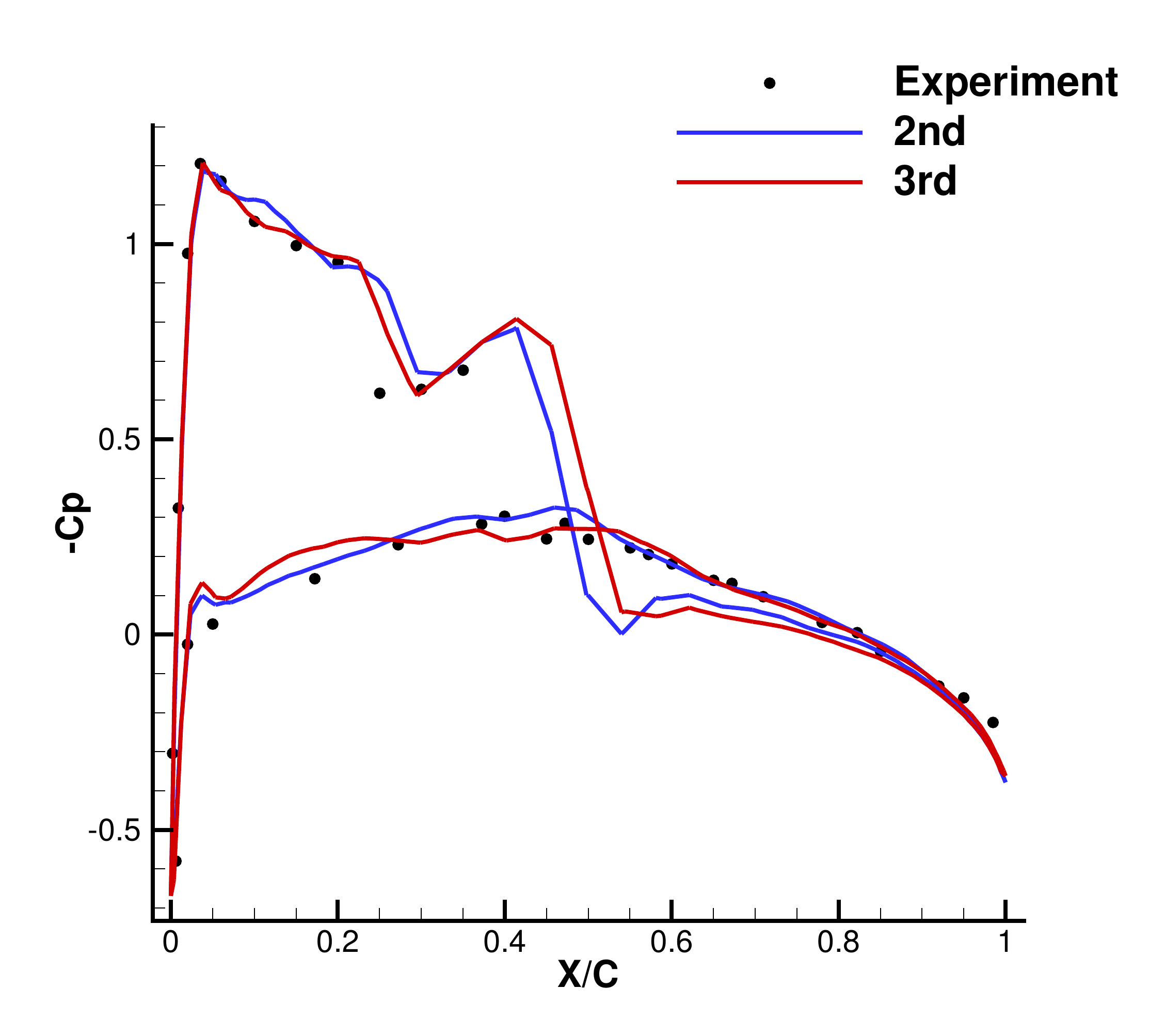}
	\includegraphics[width=0.32\textwidth]{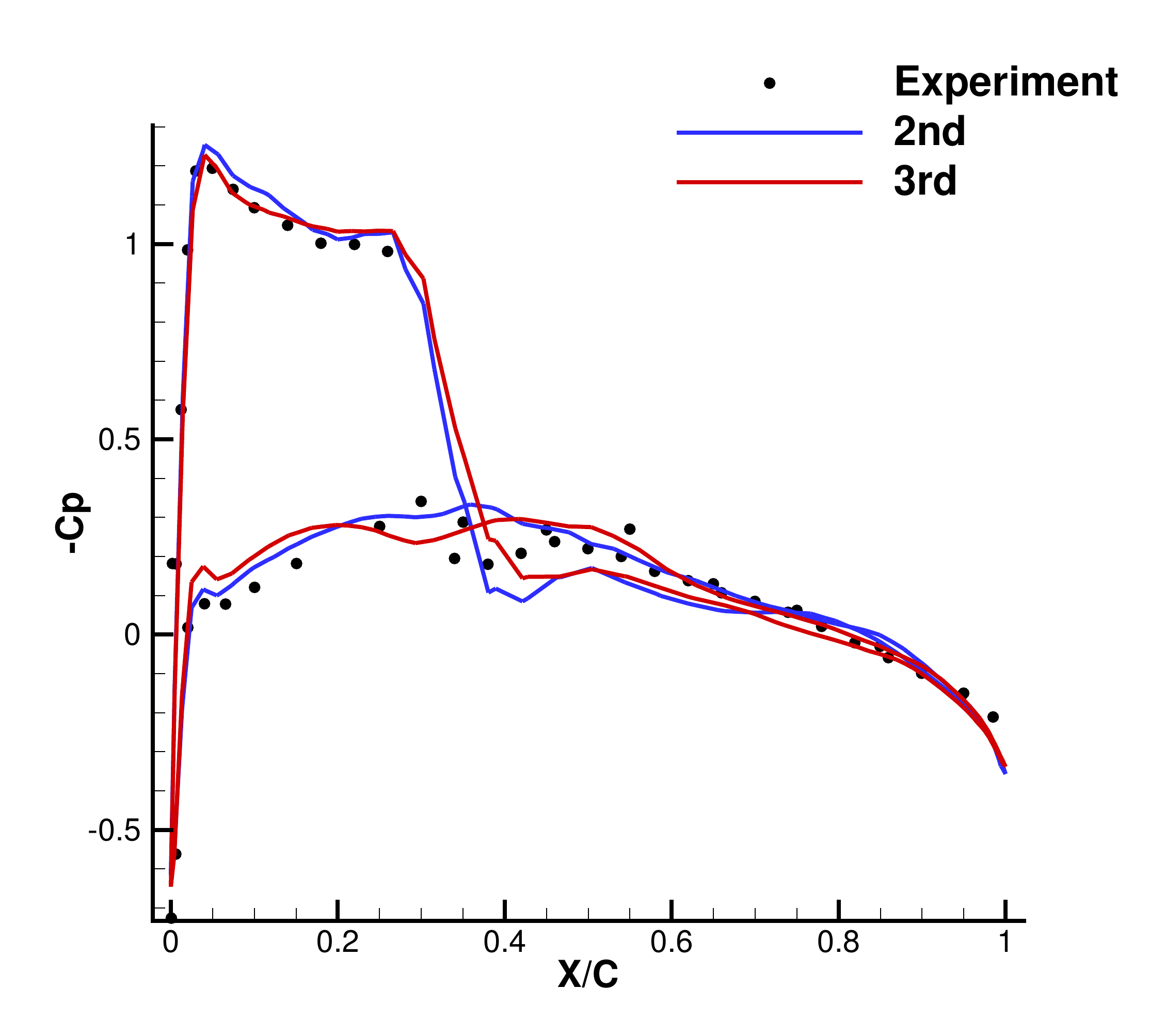}
	\includegraphics[width=0.32\textwidth]{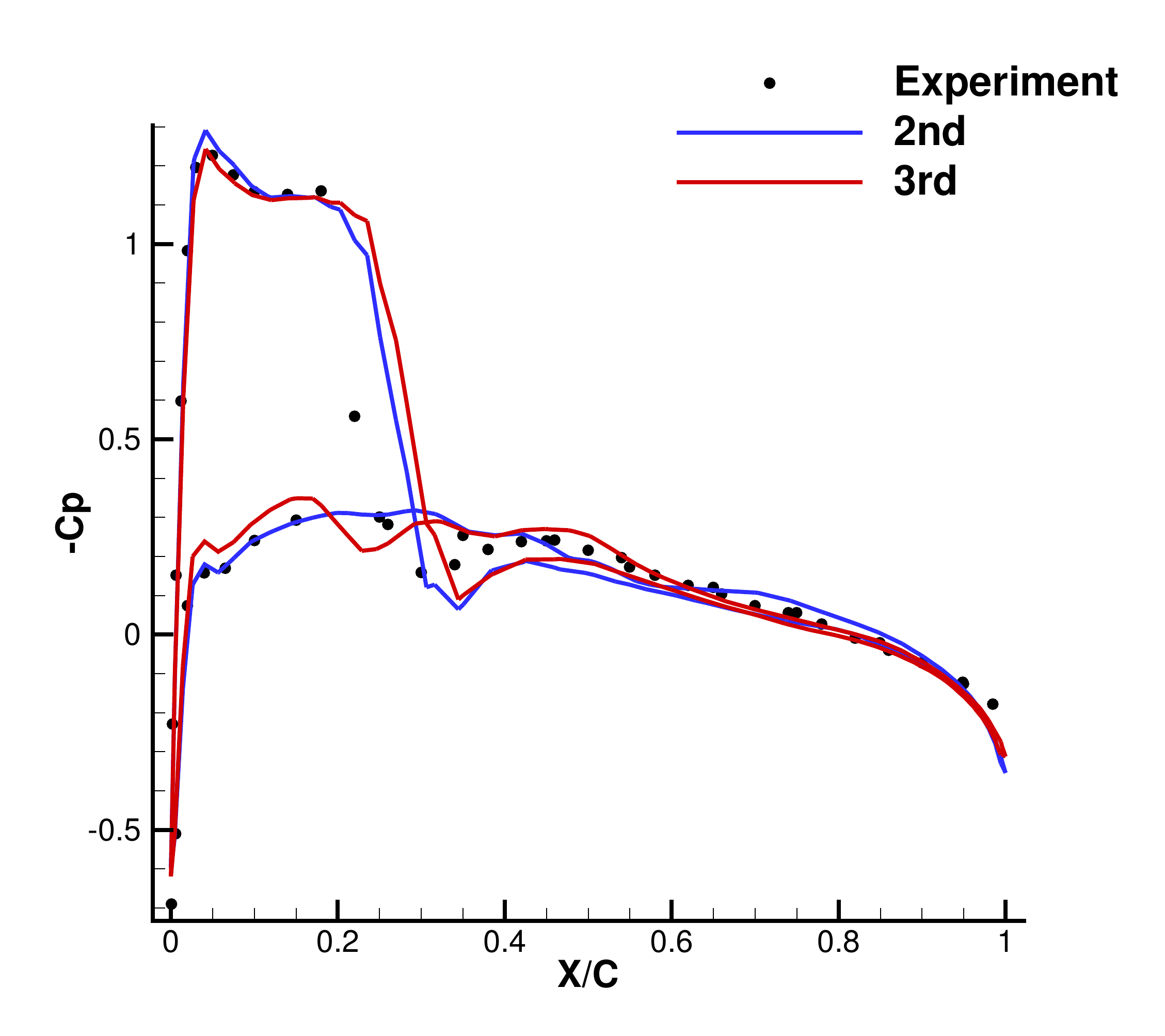}
	\vspace{-4mm} \caption{\label{m6-wing-inviscid-refine-cp}
		Pressure distributions for wing section at different semi-span locations Y/B on the ONERA  M6 wing under Mesh I. Ma=0.8935. AOA=3.06$^{\circ}$. Top: Y/B=0.20, 0.44, 0.65 from left to right. Bottom: Y/B=0.80, 0.90, 0.95 from left to right.}
\end{figure}

\begin{figure}[htp]	
	\centering	
	\includegraphics[width=0.24\textwidth]{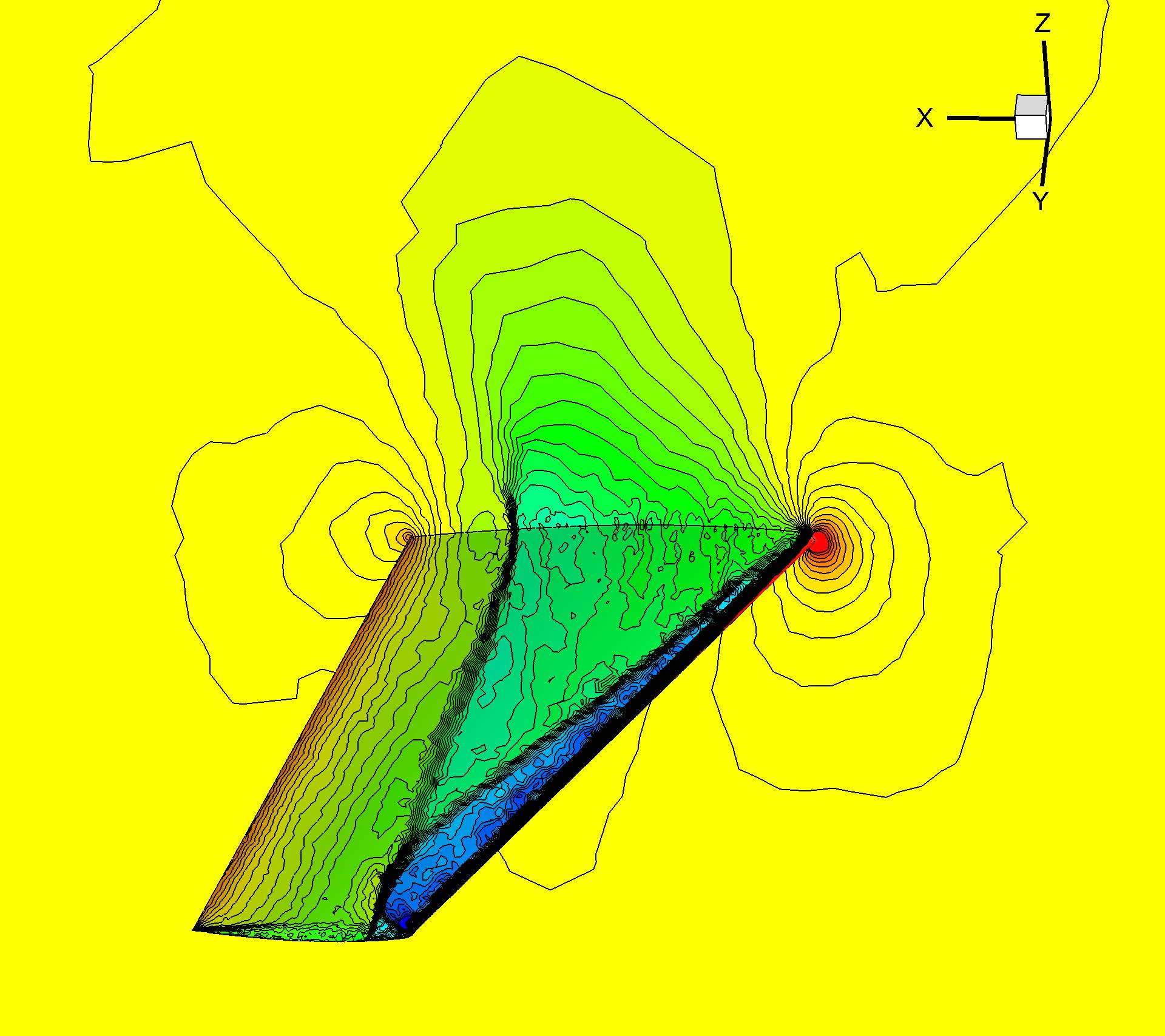}
	\includegraphics[width=0.24\textwidth]{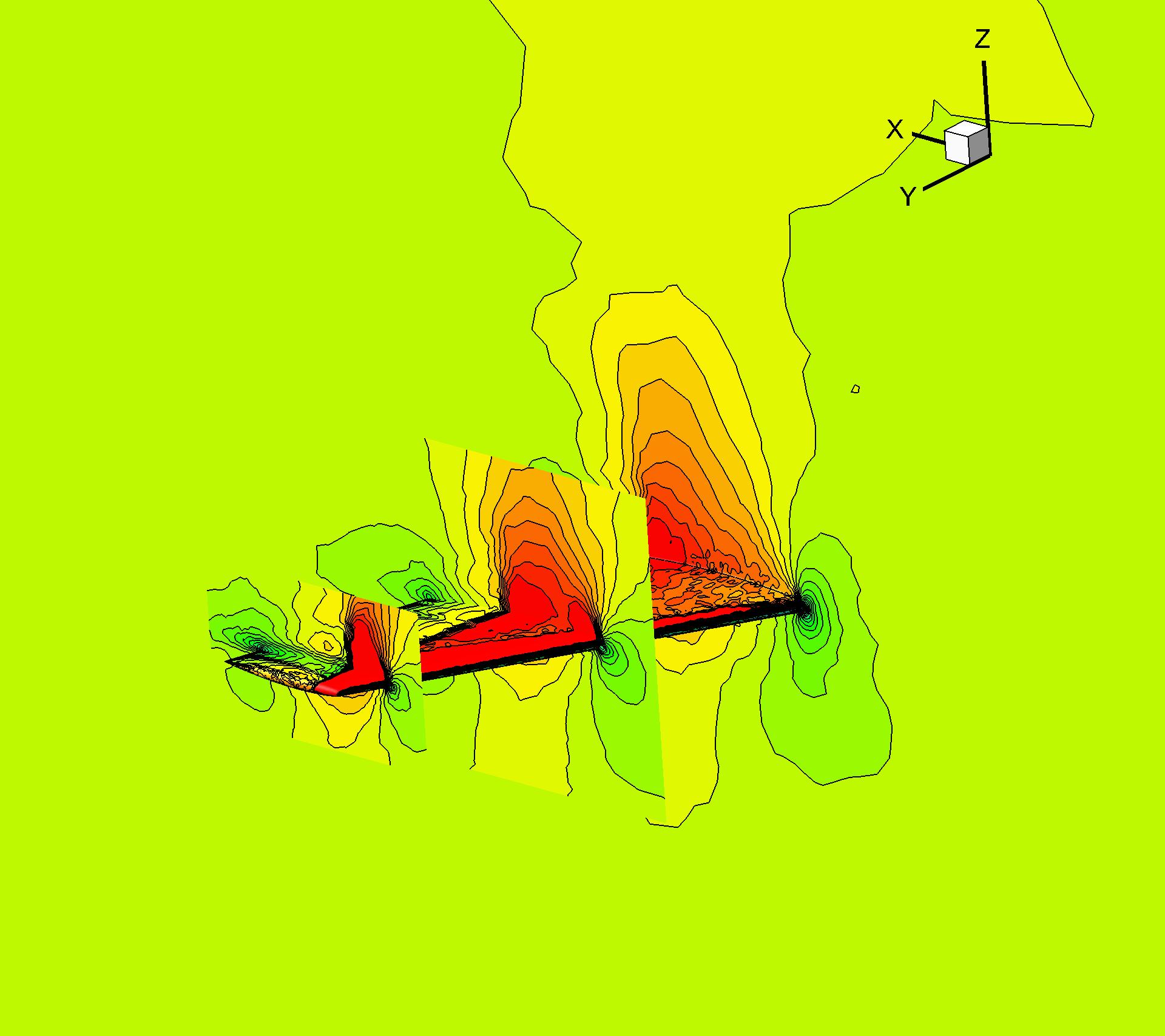}
	\includegraphics[width=0.24\textwidth]{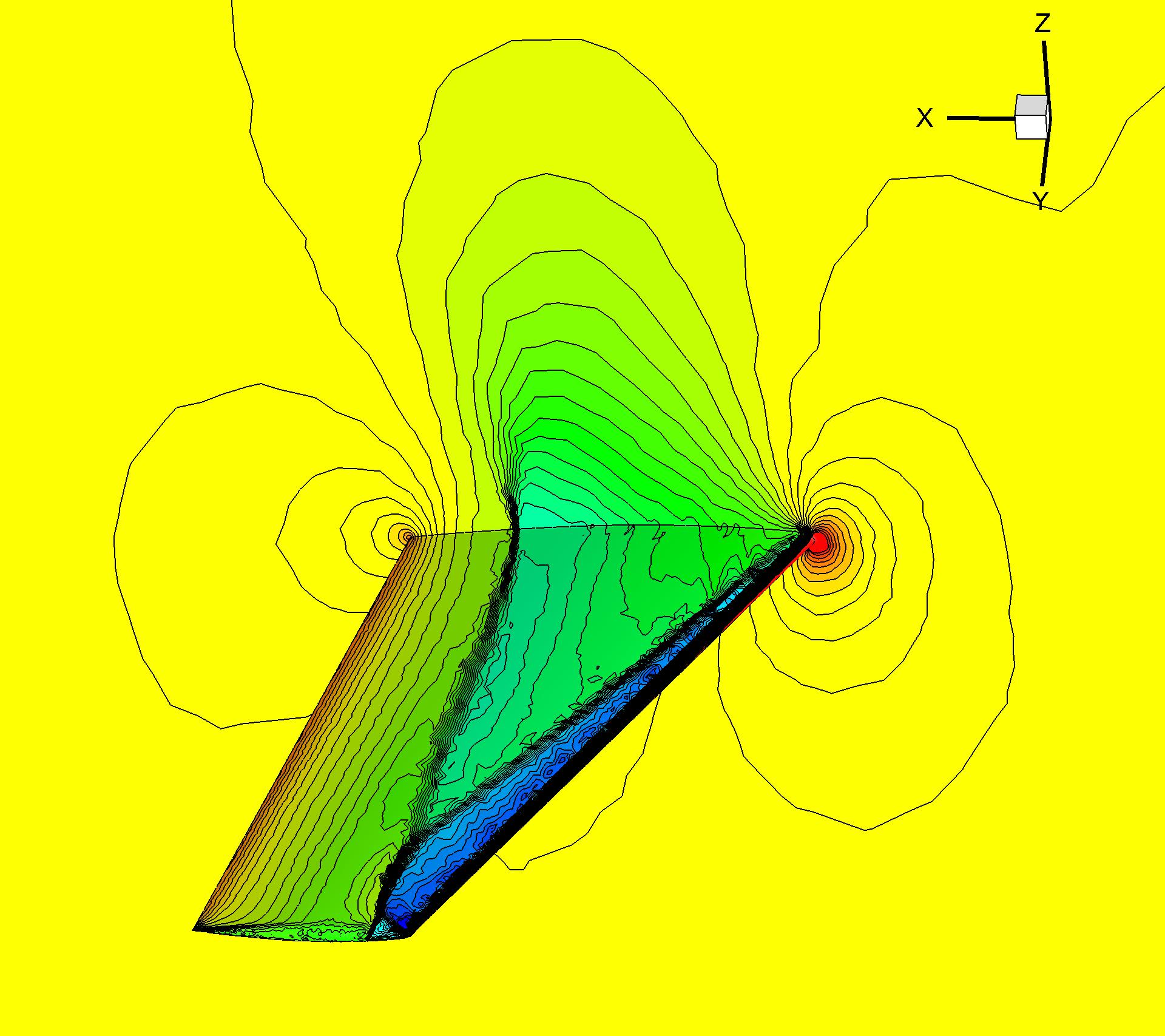}
	\includegraphics[width=0.24\textwidth]{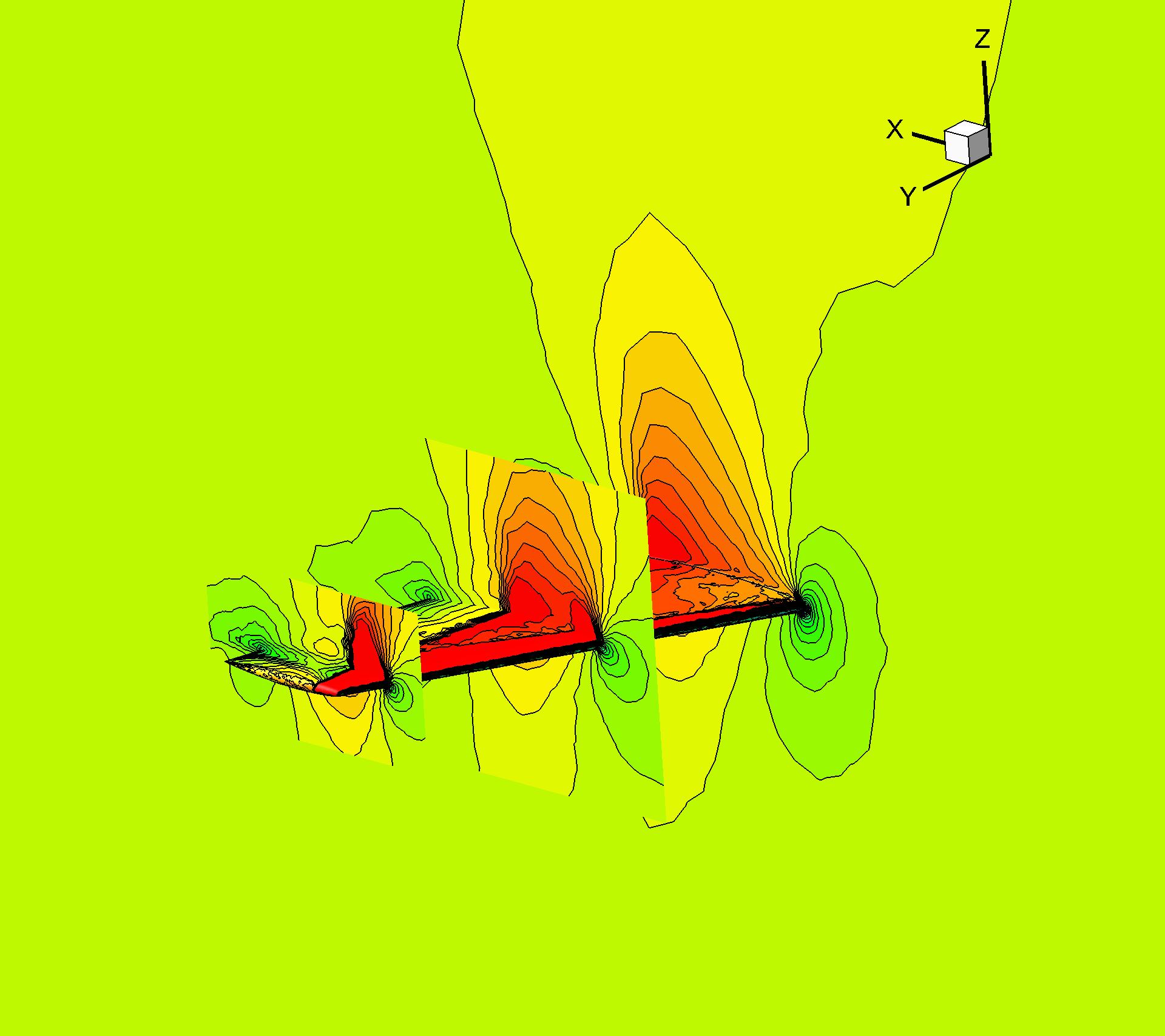}
	\caption{\label{m6-wing-inviscid-lyy}
		Transonic flow over an inviscid ONERA M6 wing under Mesh II. Ma=0.8935. AOA=3.06$^{\circ}$. Left: The second-order GKS. Right: The third-order GKS. }
\end{figure}

\begin{figure}[htp]	
	\centering	
	\includegraphics[width=0.32\textwidth]{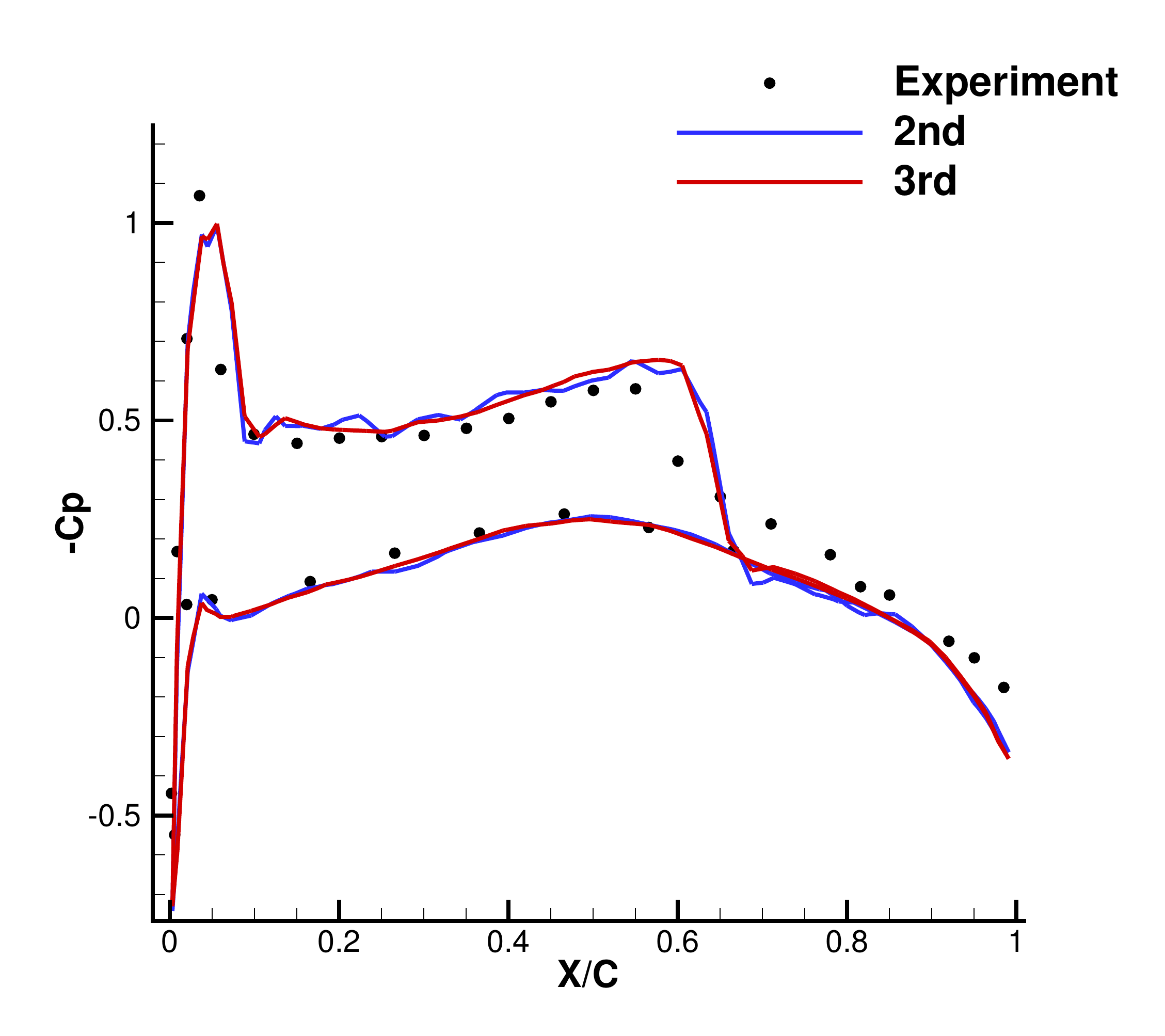}
	\includegraphics[width=0.32\textwidth]{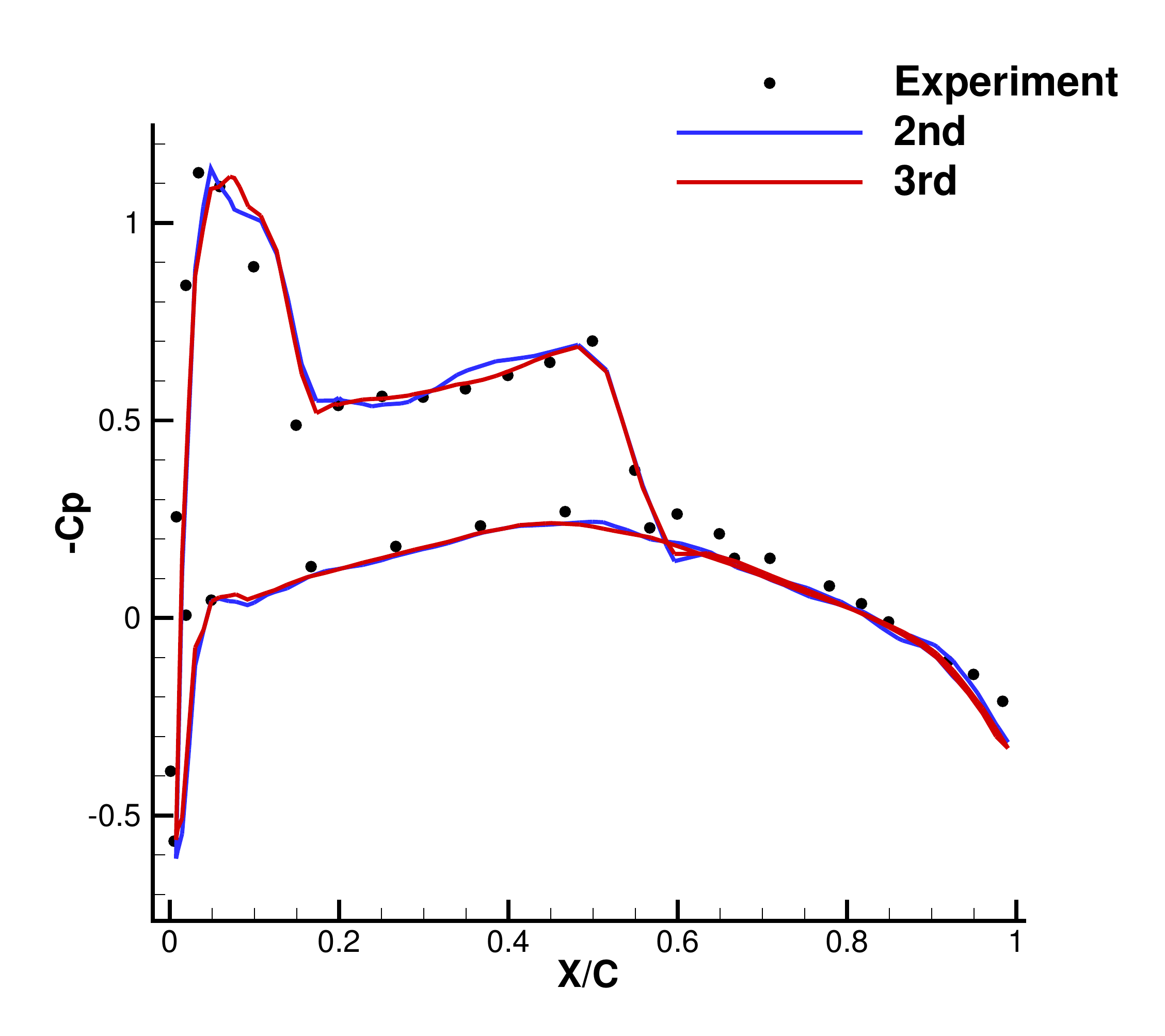}
	\includegraphics[width=0.32\textwidth]{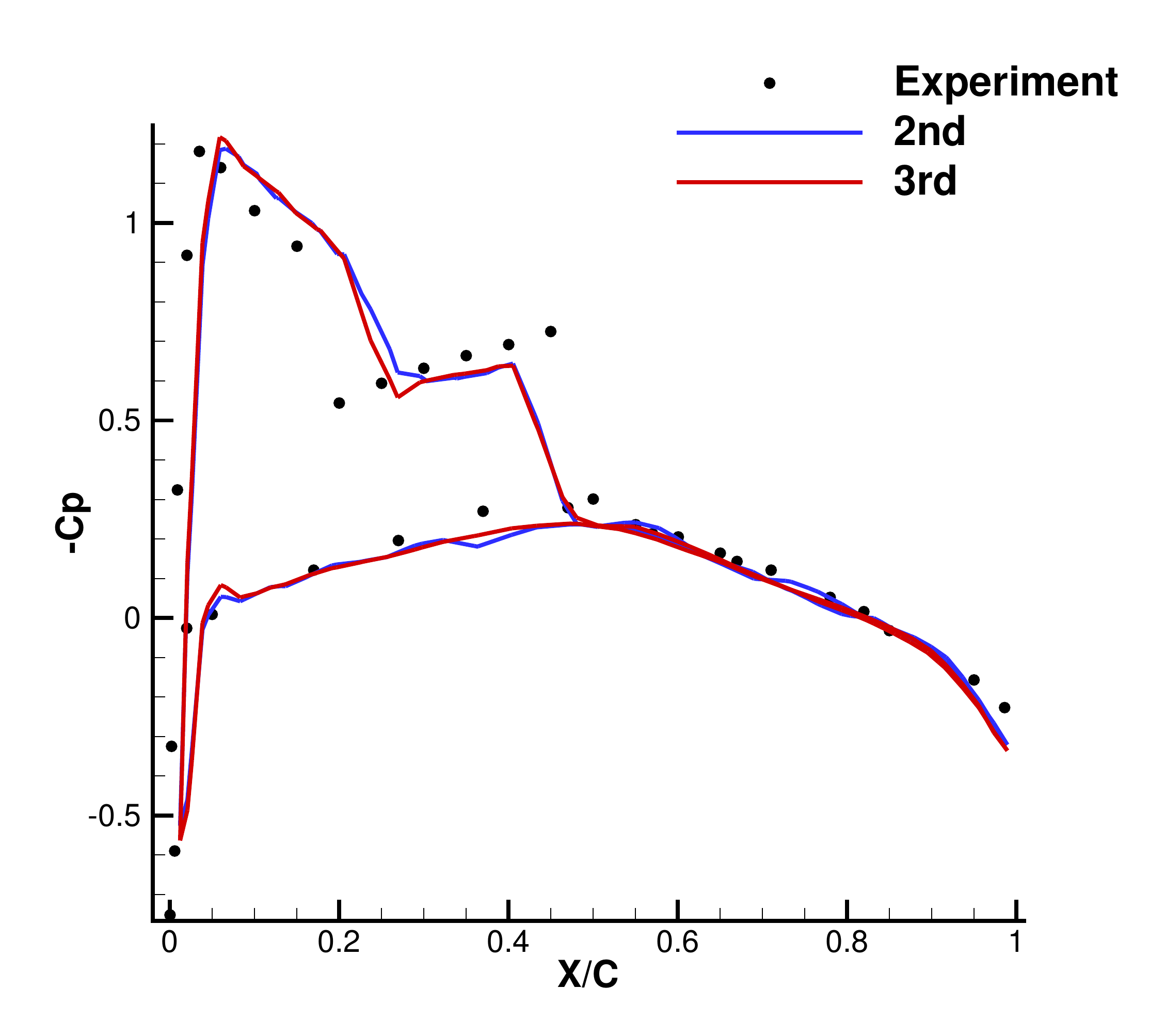}
	\includegraphics[width=0.32\textwidth]{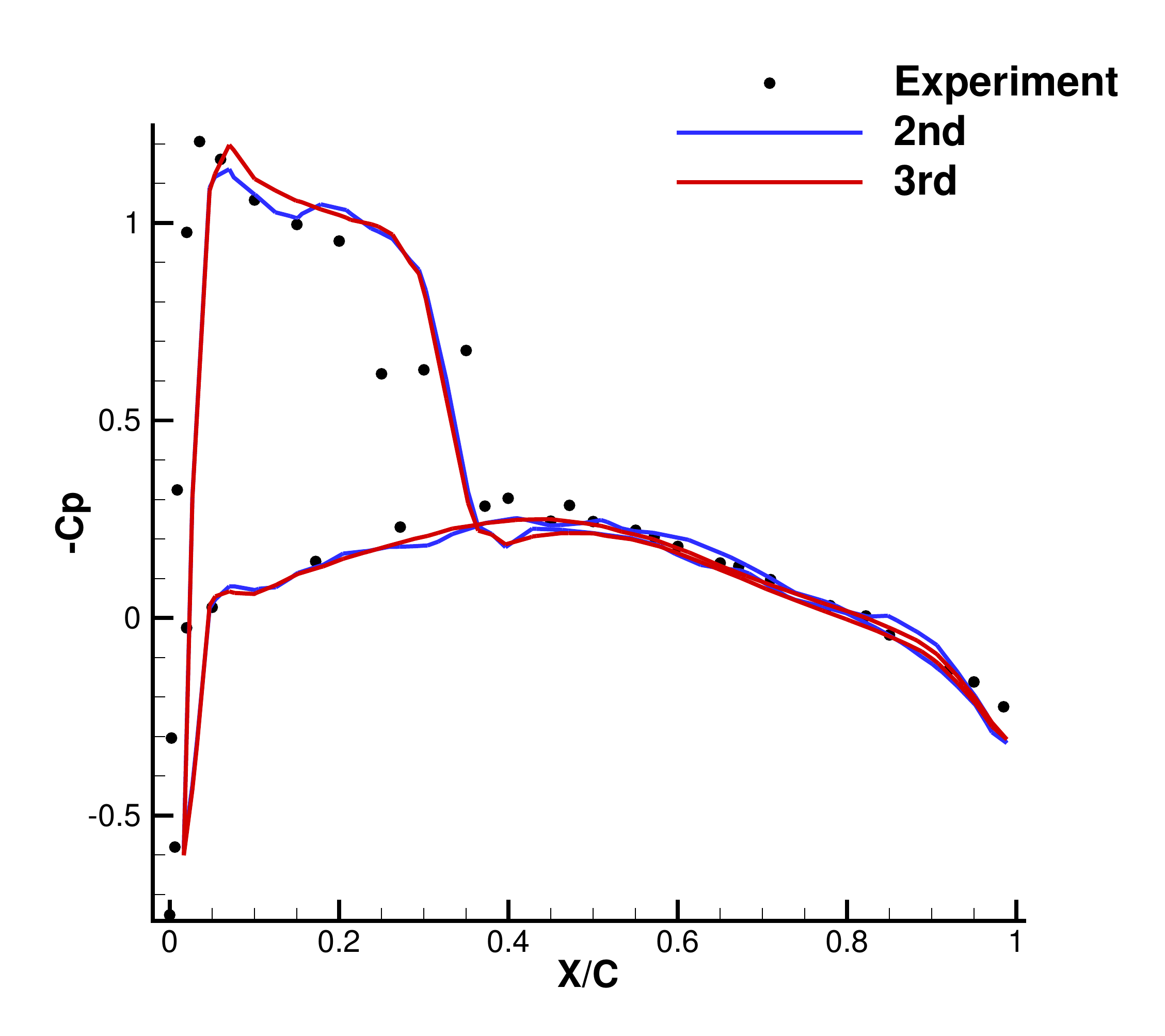}
	\includegraphics[width=0.32\textwidth]{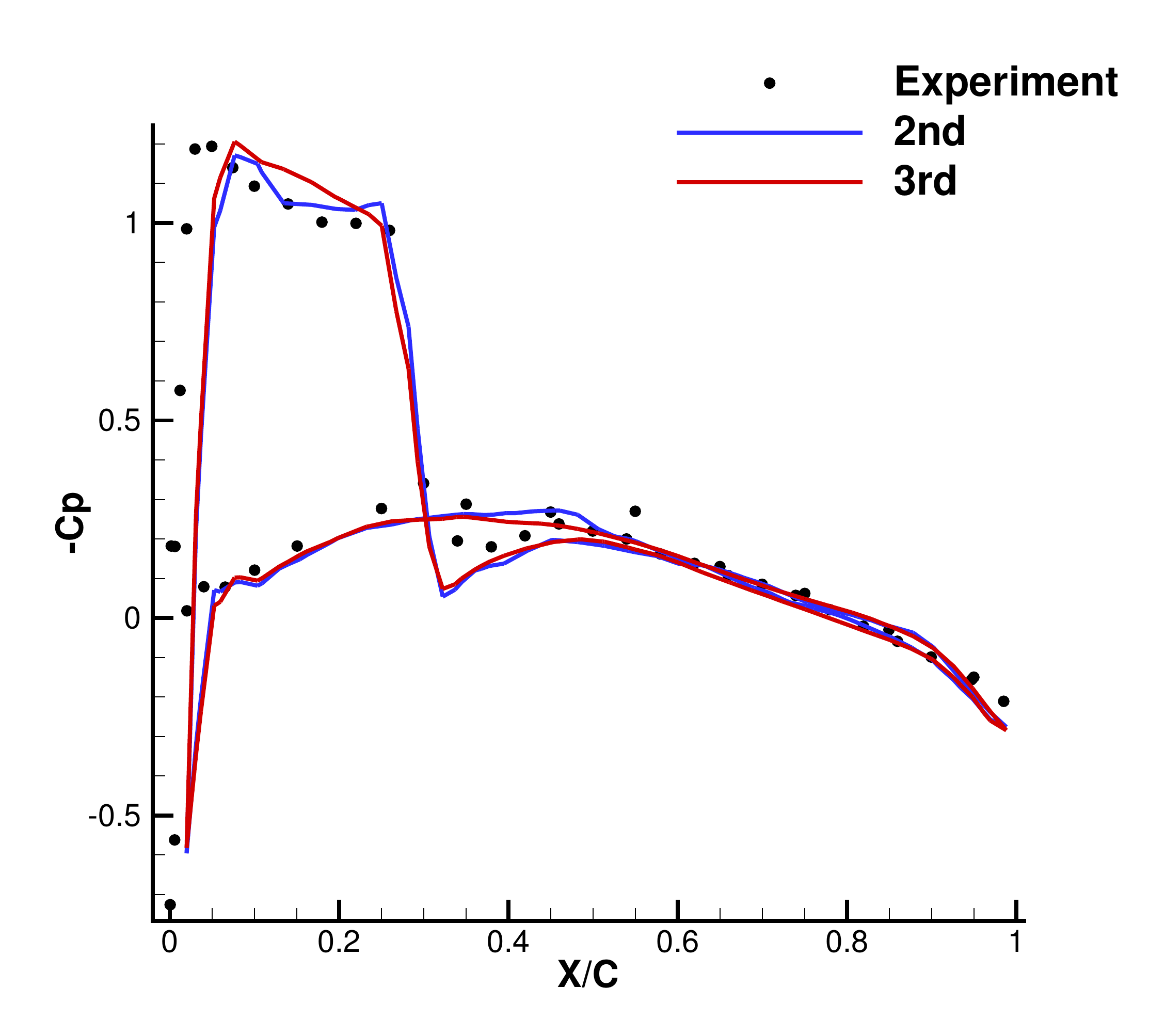}
	\includegraphics[width=0.32\textwidth]{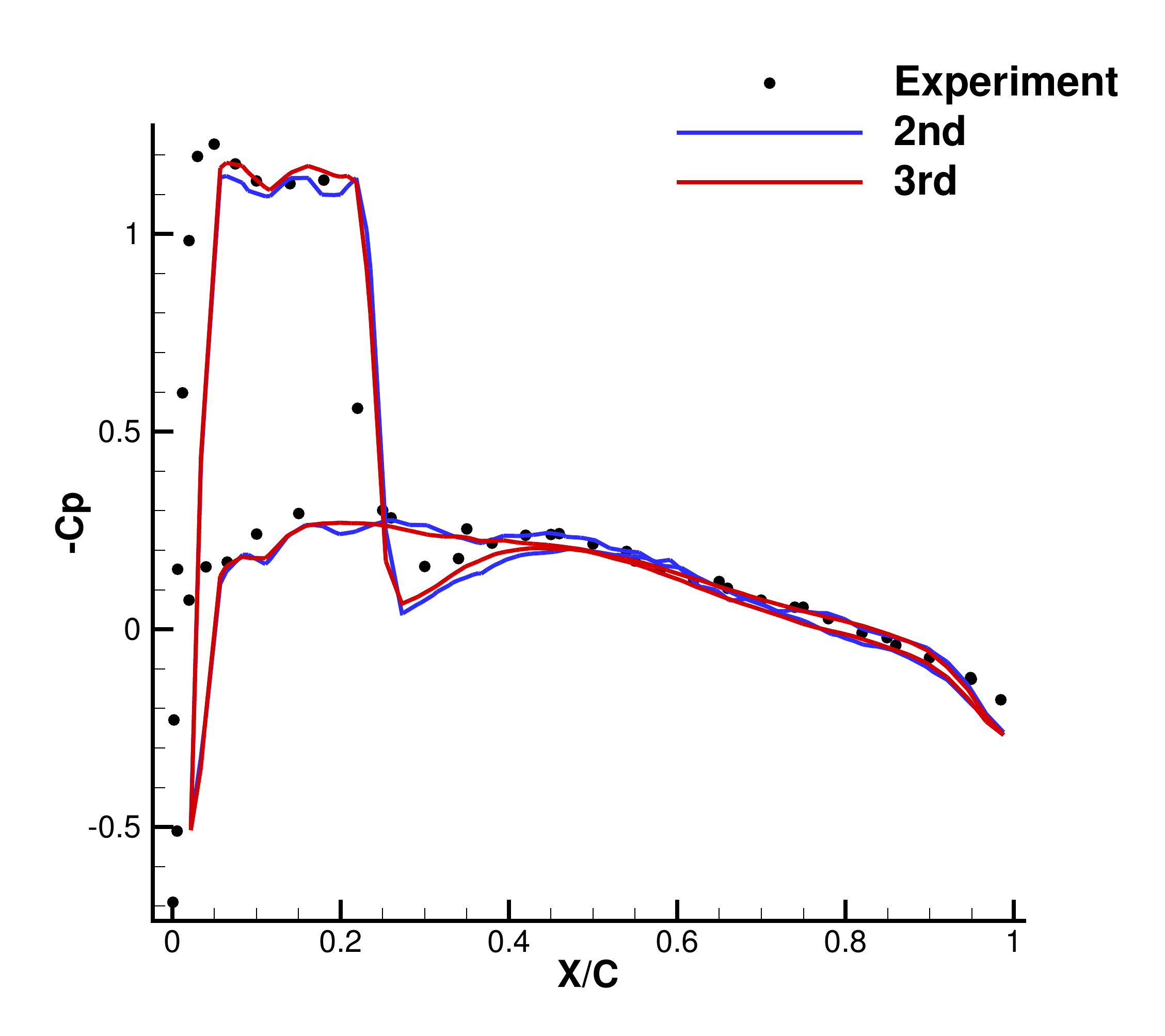}
	\vspace{-4mm} \caption{\label{m6-wing-inviscid-lyy-cp}
		Pressure distributions for wing section at different semi-span locations Y/B on the ONERA M6 wing under Mesh II. Ma=0.8935. AOA=3.06$^{\circ}$. Top: Y/B=0.20, 0.44, 0.65 from left to right. Bottom: Y/B=0.80, 0.90, 0.95 from left to right. }
\end{figure}

As a classic validation case for compressible external flow, the  transonic flow over the ONERA M6 wing is tested.
Experimental data are reported in \cite{schmitt1979pressure}, where the flow is fully turbulent. Same as the inviscid calculation in \cite{liu2020threedimension}, an incoming Mach number Ma=0.8395 and an angle of attack  AOA=3.06$^{\circ}$ are used, which correspond to
a rough prediction of the flow field under a very high Reynolds number.
In the computation, the wing has a slip wall boundary condition, and the Riemann invariants  are applied 10 times of the root chord length away from the wing. Two sets of meshes are used to  test the mesh sensitivity, as shown in Fig.~\ref{m6-wing-inviscid-mesh}.
For each mesh, the results from the second and third-order GKS are presented.
The surface pressure distributions and Mach number slices at different wing sections under Mesh I for both schemes are shown in Fig.~\ref{m6-wing-inviscid-refine}.
The ``Lambda'' shock is resolved from both schemes.
Third-order scheme presents accurate solutions with high resolution in pressure and Mach contours in smooth region.
Similar conclusions can be drawn from the results obtained from Mesh II, as shown in Fig.~\ref{m6-wing-inviscid-lyy}.
Quantitatively comparisons on the pressure distributions at six different locations on the wing are given in Fig.~\ref{m6-wing-inviscid-refine-cp} and Fig.~\ref{m6-wing-inviscid-lyy-cp}.
A better agreement in the secondary shock position is obtained with Mesh II at the semi-span locations Y/B = 0.20, 0.44, and 0.65.

\subsection{Supersonic flow over a YF-17 fighter}

\begin{figure}[htp]	
	\centering	
	\includegraphics[height=0.4\textwidth]{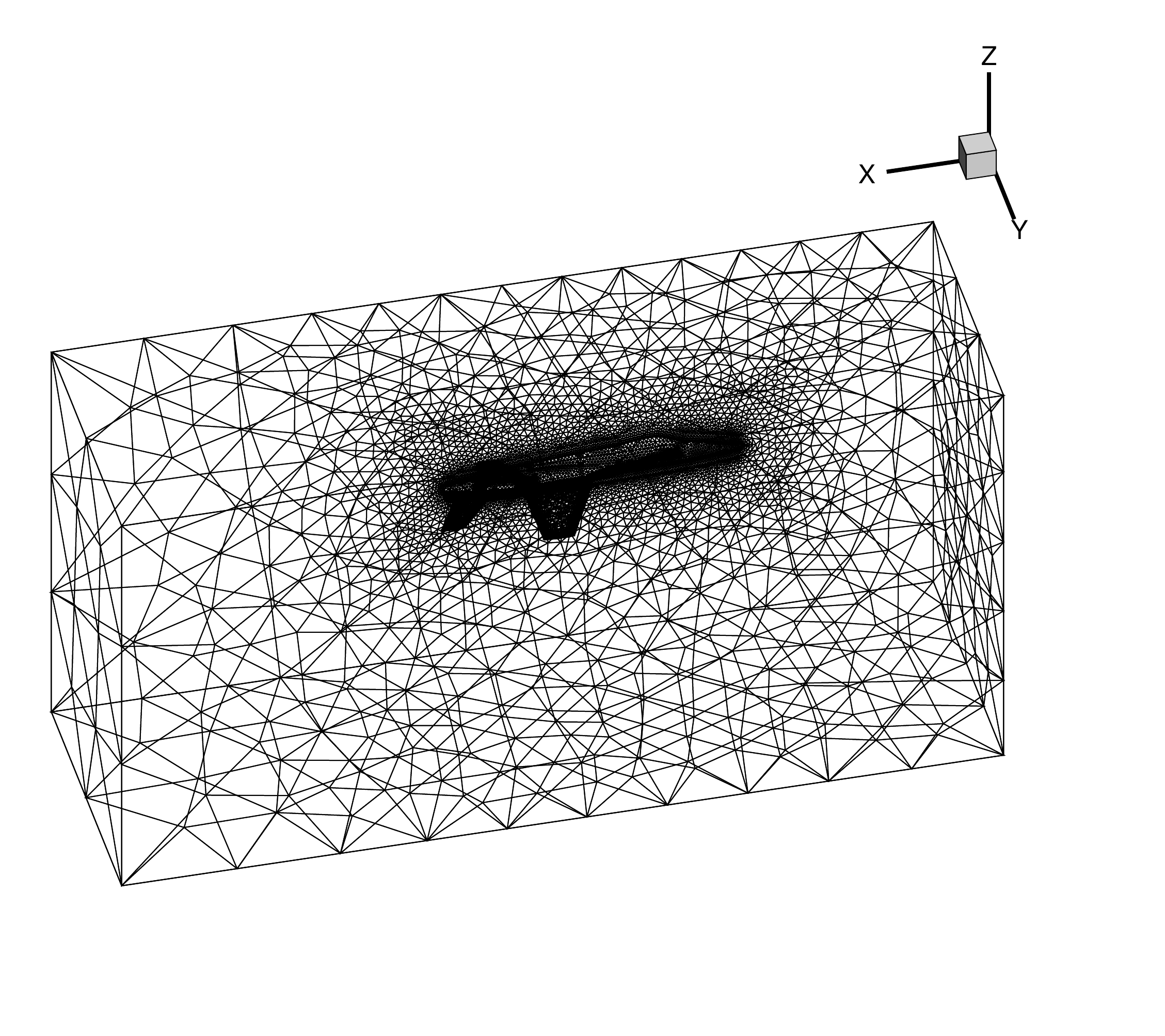}
	\includegraphics[height=0.4\textwidth]{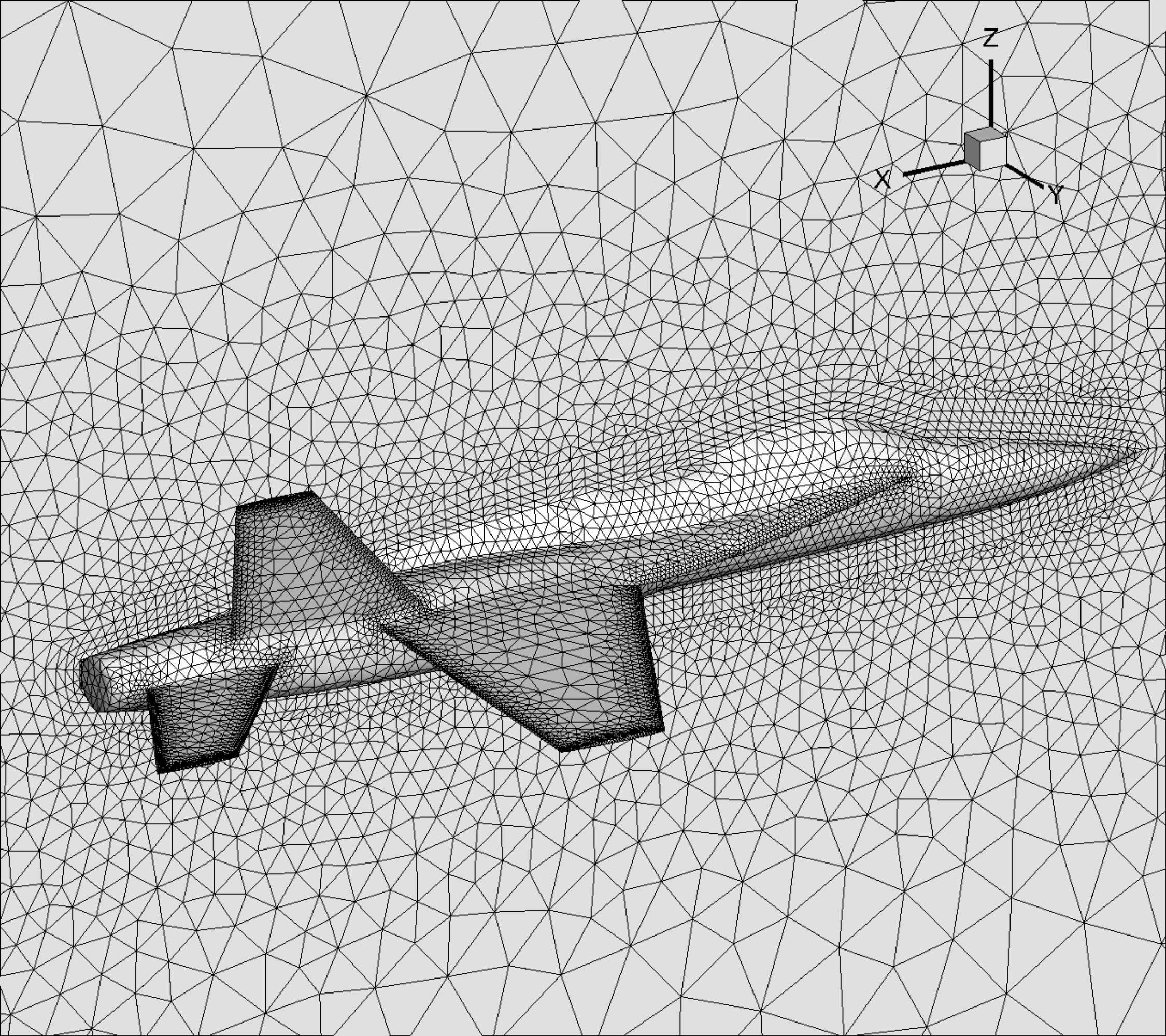}
     \caption{\label{yf-17-ma18-mesh}
		Supersonic flow passing through a YF-17 ("Cobra") model. Ma=1.8. AOA=1.25. Mesh number: 325,096. }
\end{figure}

\begin{figure}[htp]	
	\centering	
	\includegraphics[width=0.24\textwidth]{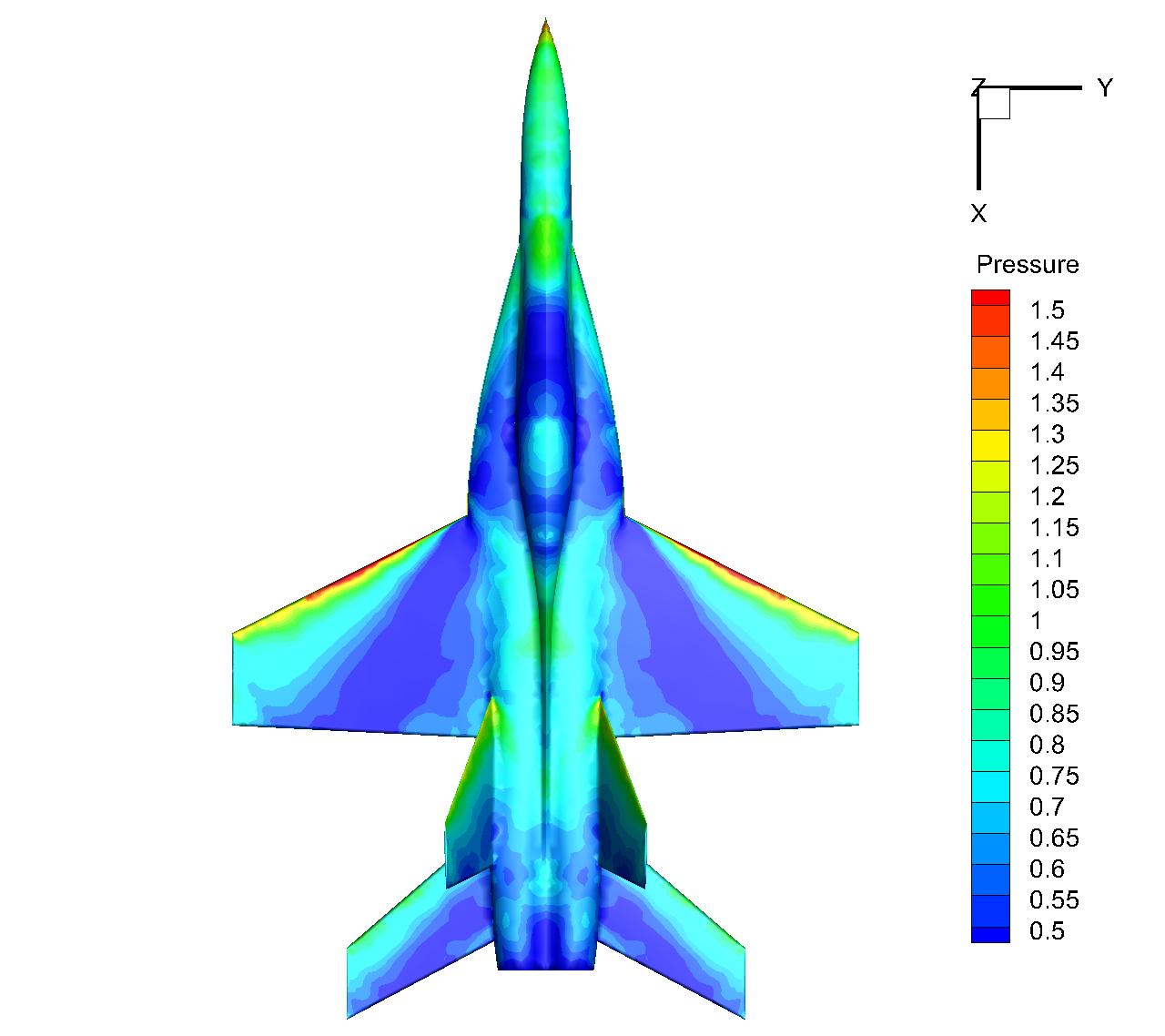}
	\includegraphics[width=0.24\textwidth]{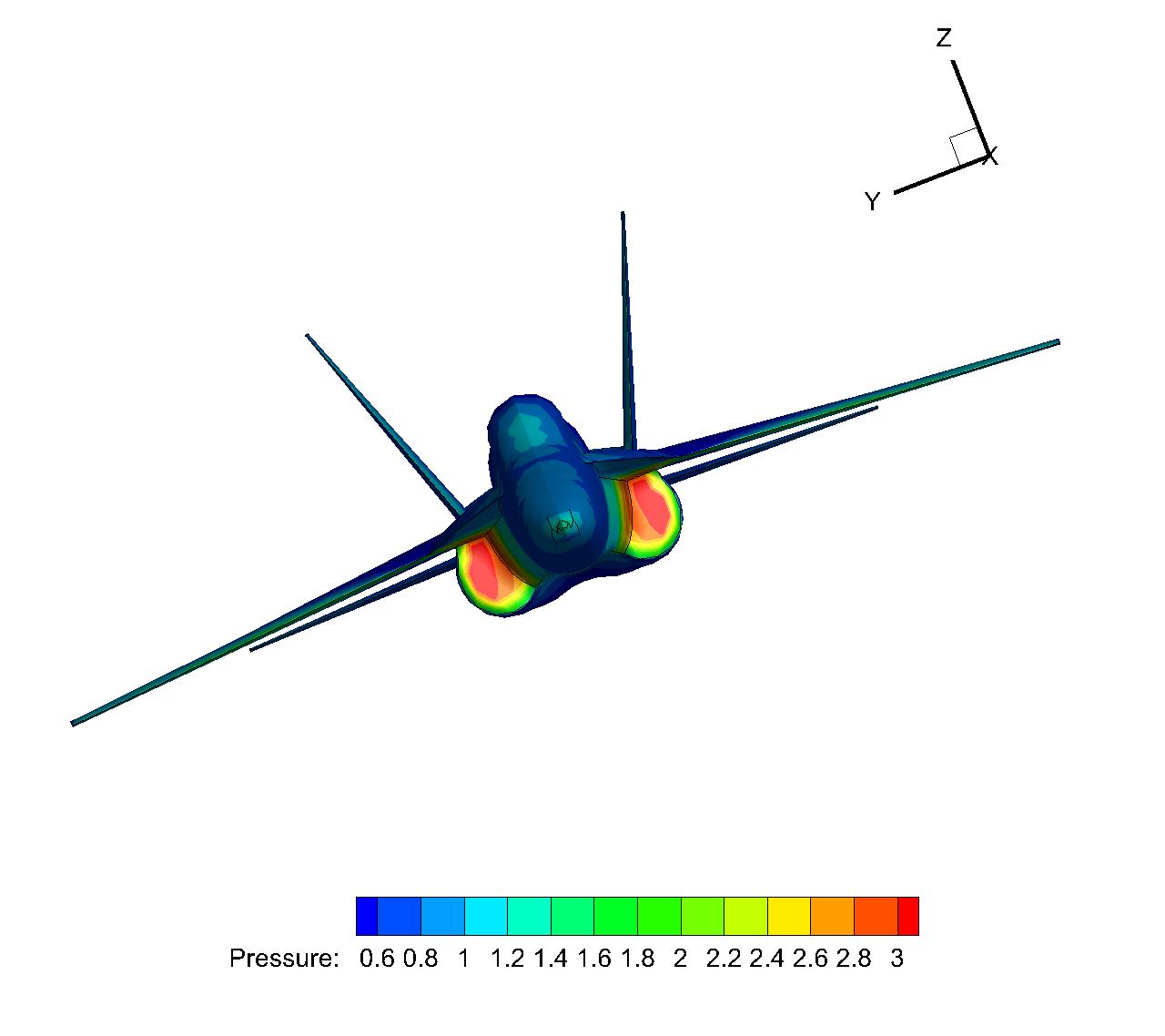}	
	\includegraphics[width=0.24\textwidth]{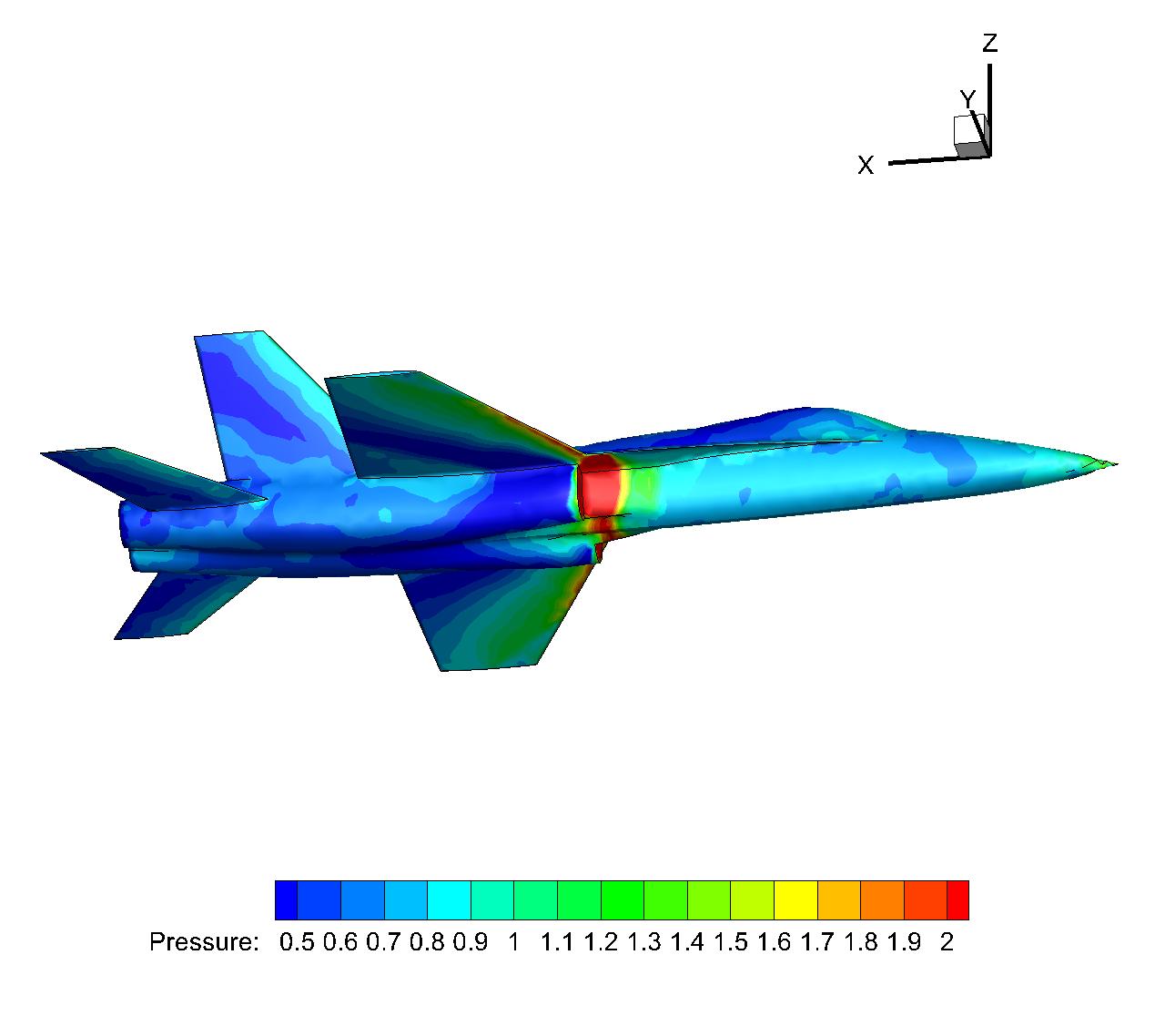}
	\includegraphics[width=0.24\textwidth]{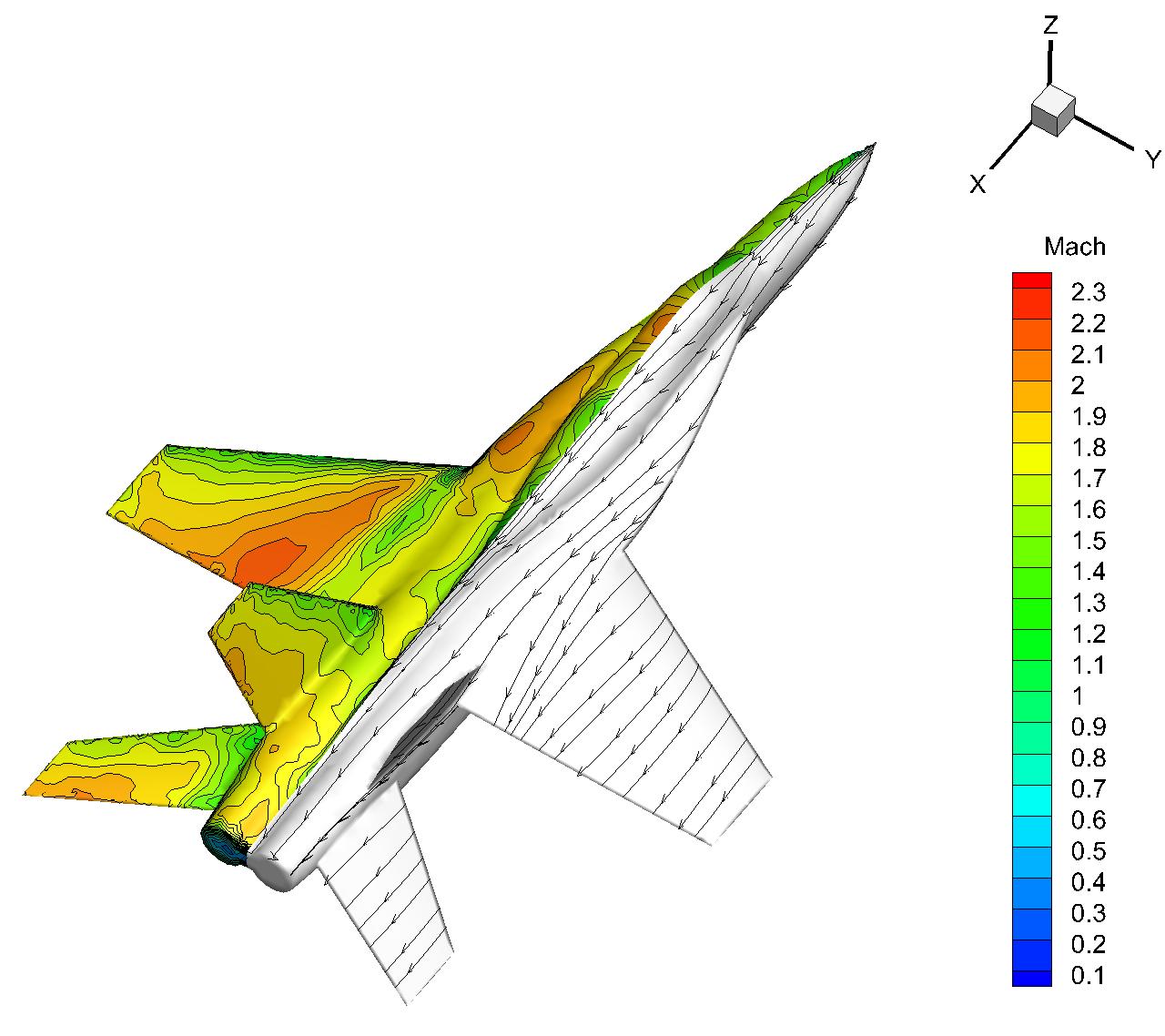}
	\vspace{-4mm} \caption{\label{yf-17-ma18-2nd}
		Supersonic flow passing through a YF-17 ("Cobra") model by the second-order GKS. Ma=1.8. AOA=$1.25^{\circ}$.  }
\end{figure}

\begin{figure}[htp]	
	\centering	
	\includegraphics[width=0.24\textwidth]{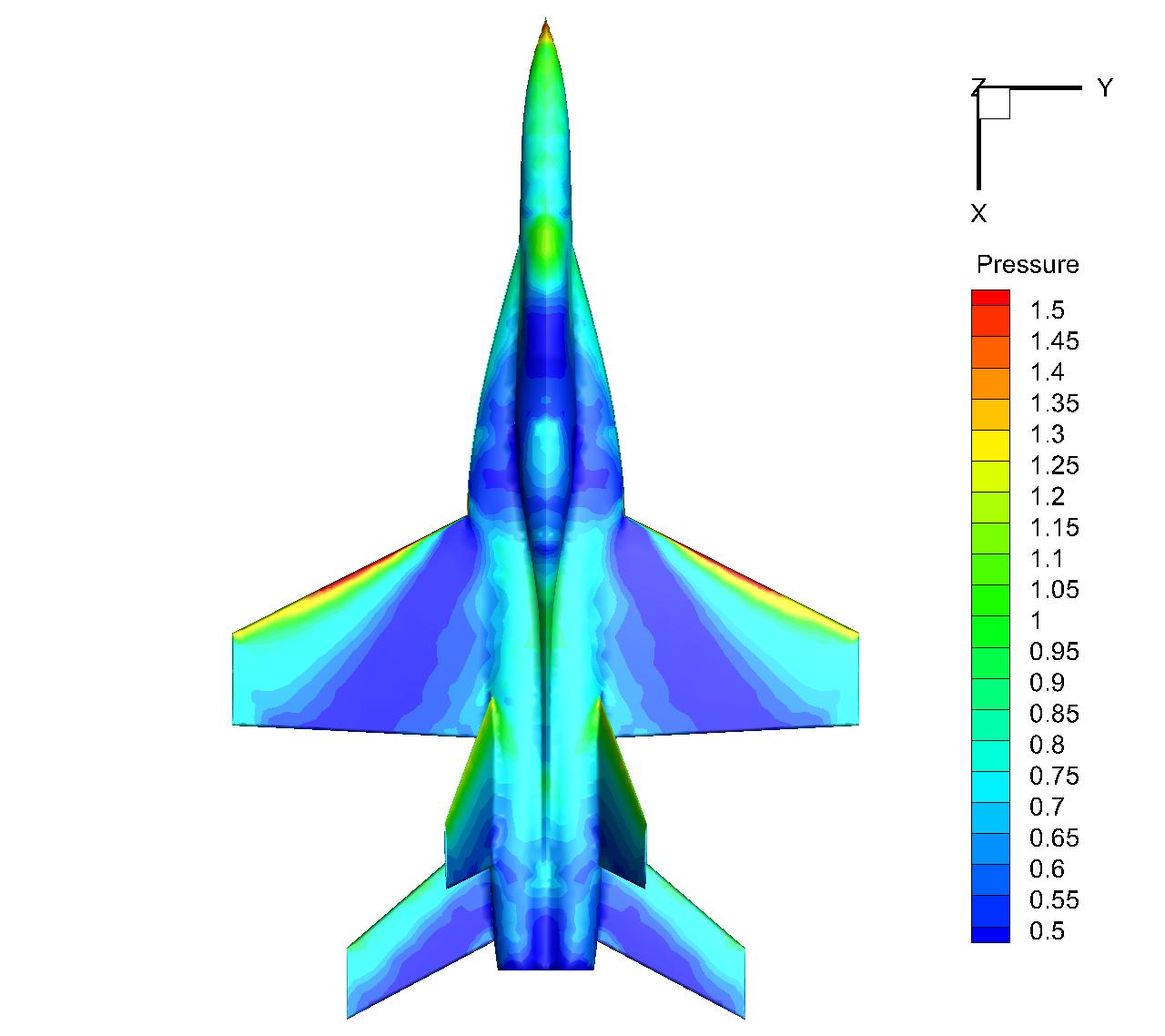}
	\includegraphics[width=0.24\textwidth]{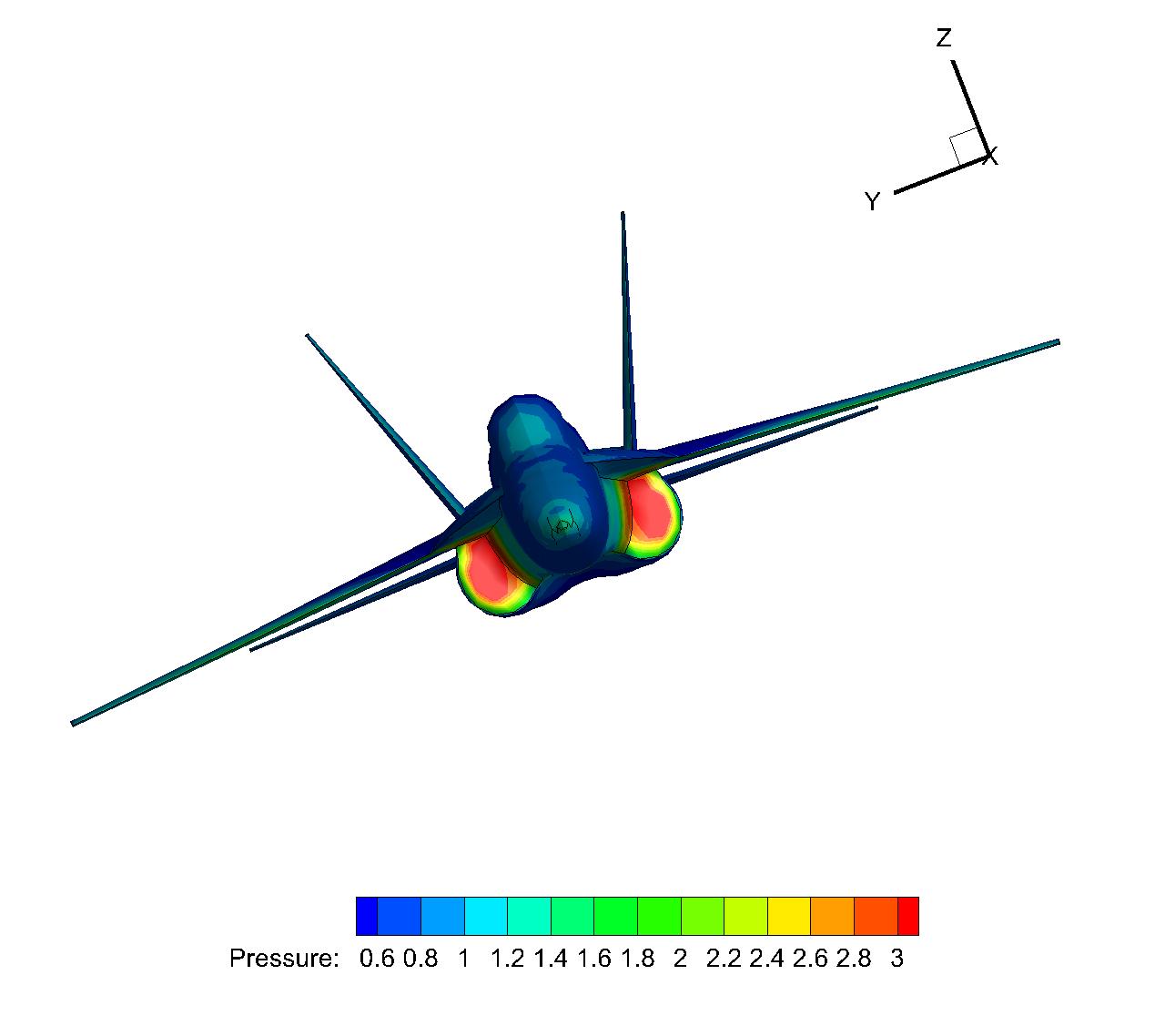}	
	\includegraphics[width=0.24\textwidth]{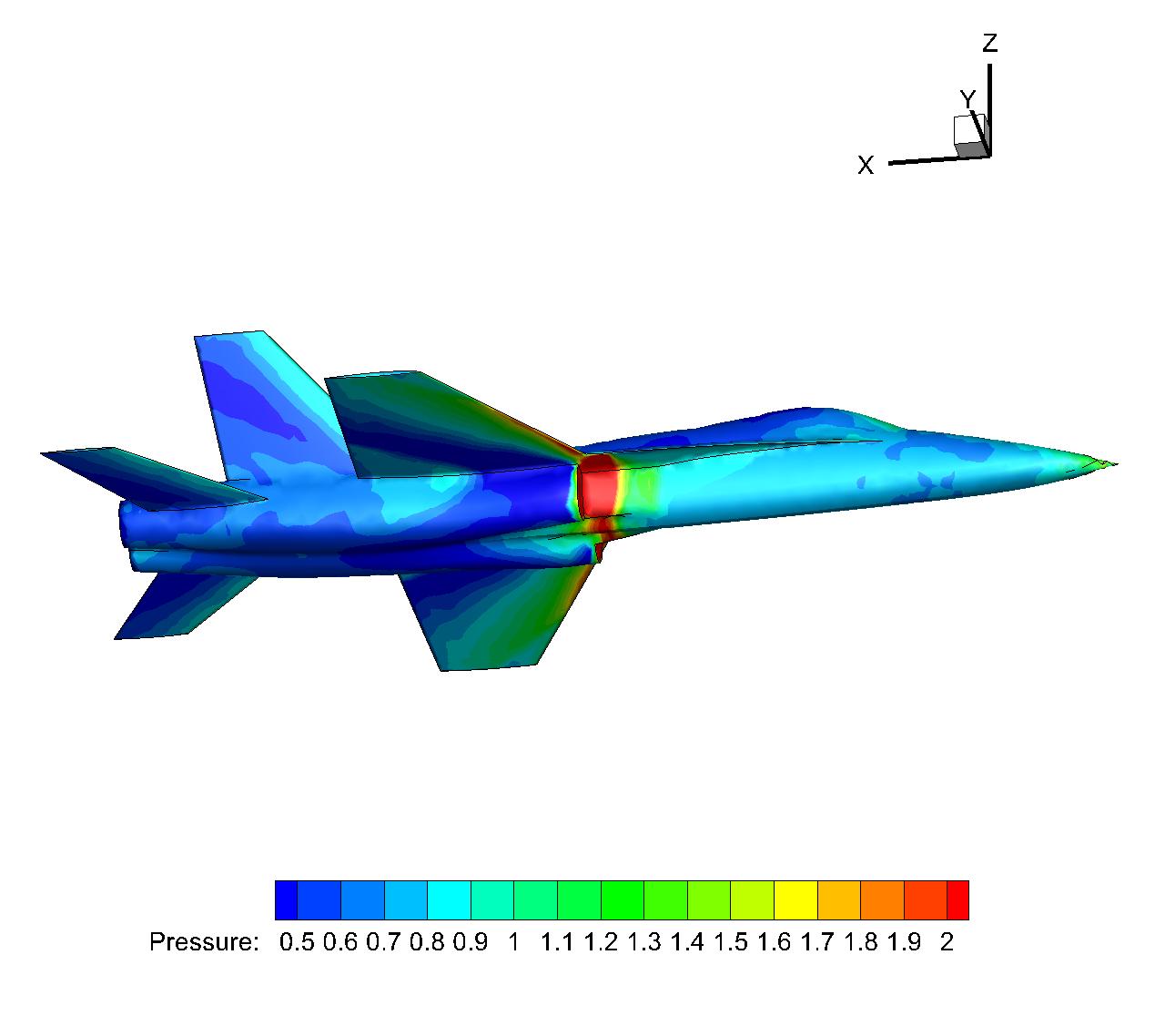}
	\includegraphics[width=0.24\textwidth]{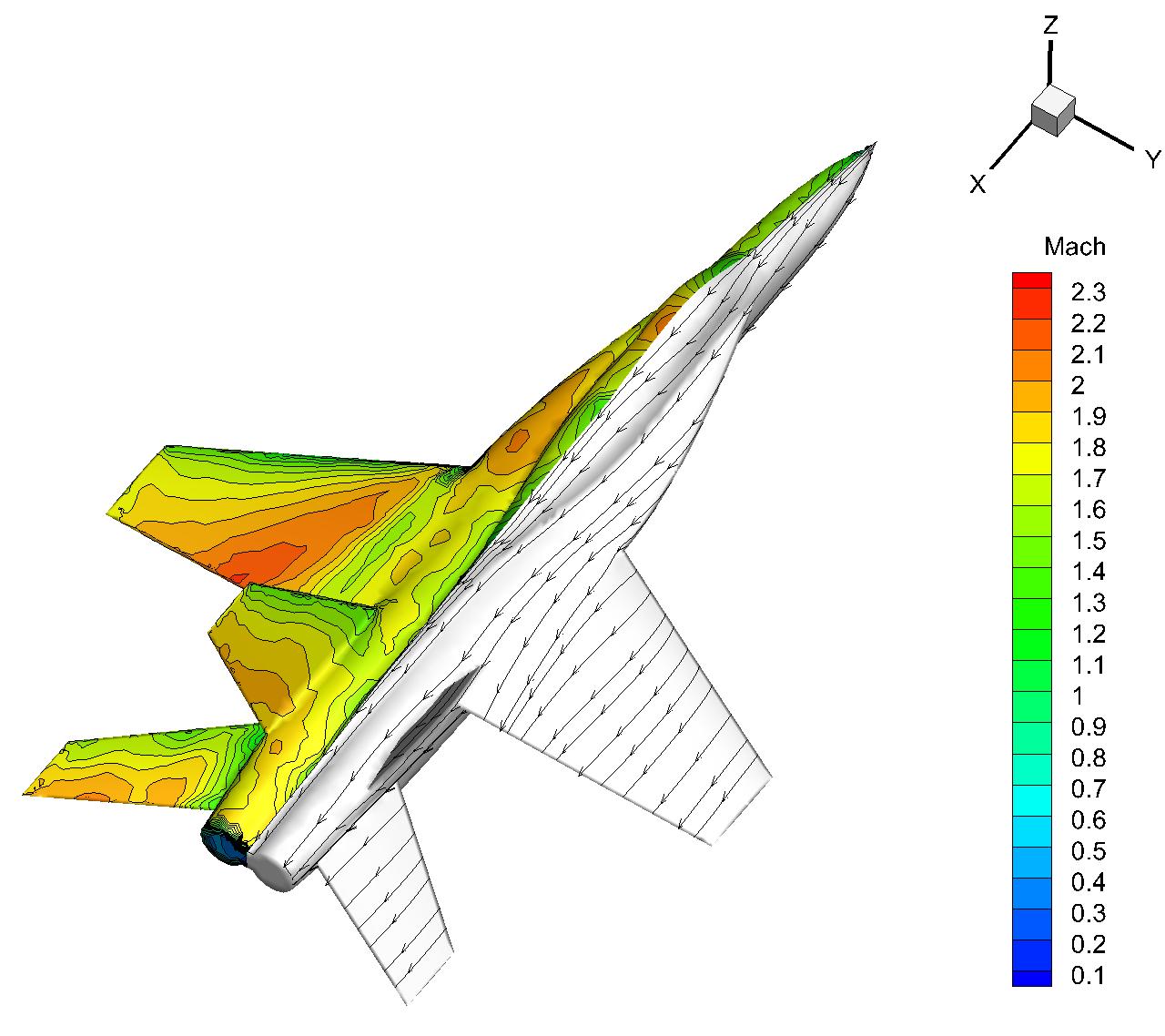}
	\vspace{-4mm} \caption{\label{yf-17-ma18-3rd}
		Supersonic flow passing through YF-17 ("Cobra") model by the third-order GKS. Ma=1.8. AOA=$1.25^{\circ}$.}
\end{figure}

The inviscid supersonic flow passing through a complete aircraft model is computed.
The computational mesh for a YF-17 ("Cobra") fighter model is shown in Fig.~\ref{yf-17-ma18-mesh} which is provided at ``https://cgns.github.io/CGNSFiles.html''.
A free stream at a Mach number Ma=$1.8$ and an angle of attack AOA=$1.25$ are adopted as the initial conditions.
The surface pressure,  Mach number distributions, and streamlines are presented in Fig.~\ref{yf-17-ma18-2nd} for the GKS with the second-order WENO reconstruction.
Complicated shocks appear in the locations including the nose, cockpit-canopy wing, horizontal stabilizer, and vertical stabilizer.
A slightly smoother solution is obtained by the compact GKS with the third-order HWENO reconstruction, as shown in Fig.~\ref{yf-17-ma18-3rd}.
The maximum Mach number on the surface is 2.4 for the second-order scheme  and 2.28 for the third-order one.
The current algorithm can handle complicated geometry, such as the mesh skewness near the wing tips and the lack of neighboring cell for the cell near boundary corners. The compact GKS demonstrates good mesh adaptability in the computation.

\subsection{Hypersonic flow around a blunt body}

\begin{figure}[htp]	
	\centering	
	\includegraphics[width=0.48\textwidth]
	{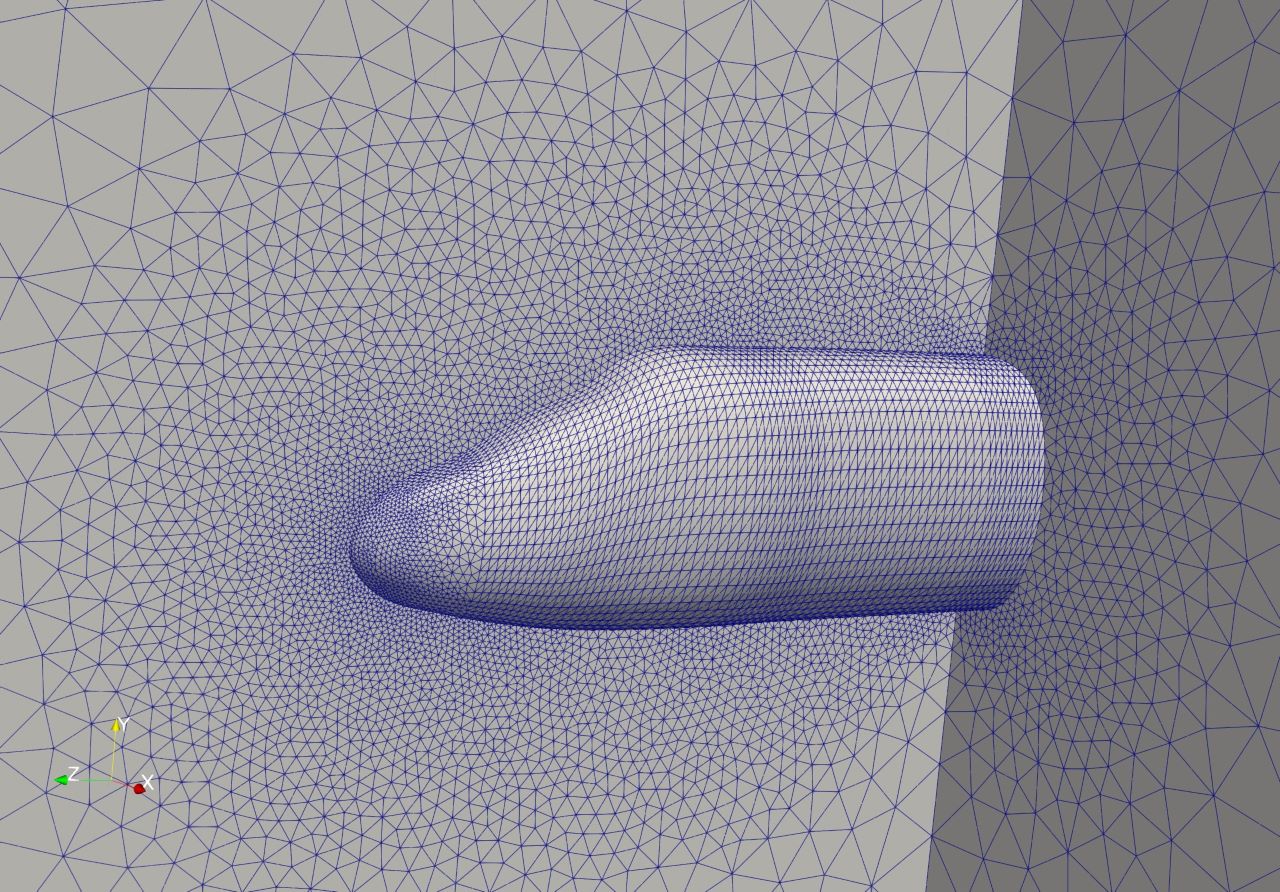}
	\includegraphics[width=0.48\textwidth]
	{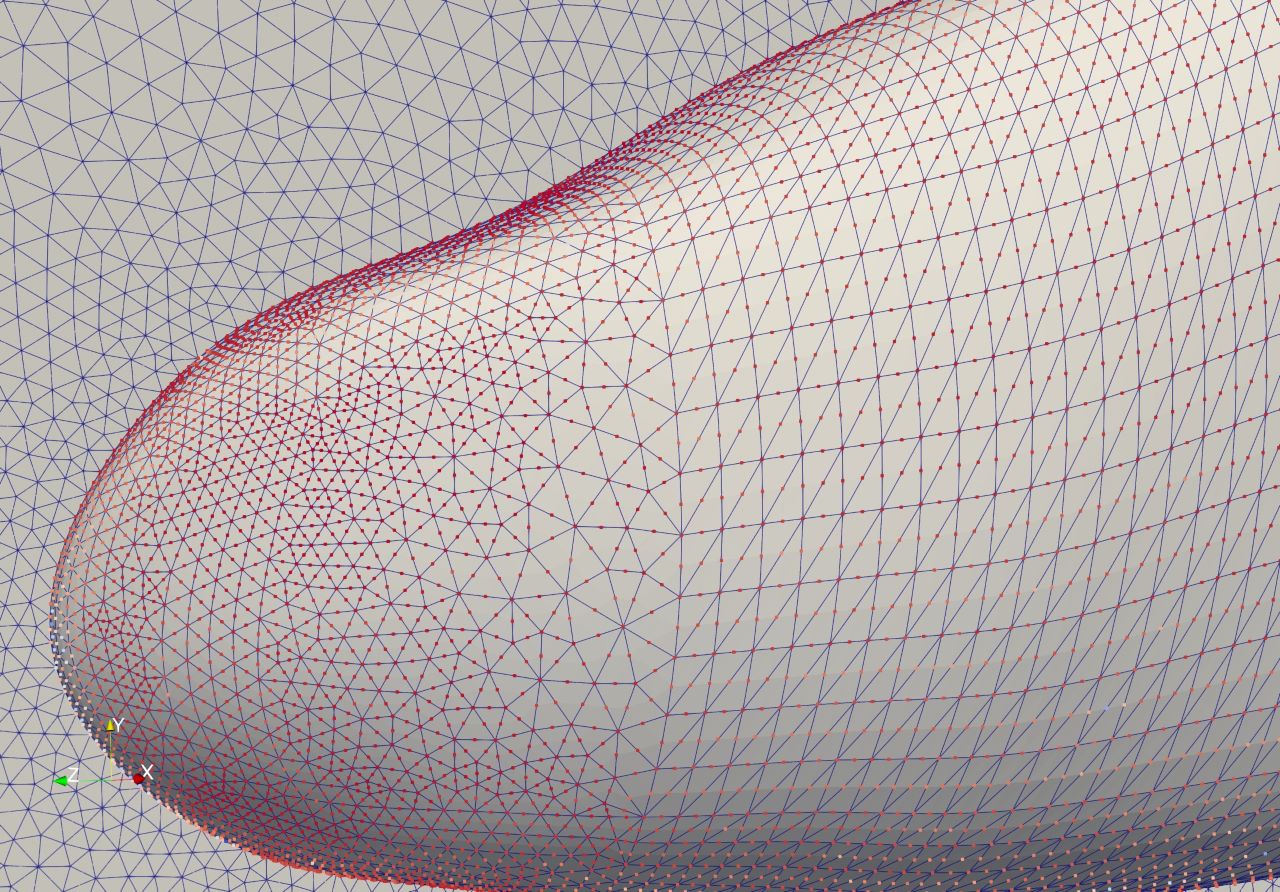}
	 \caption{\label{space-shuttle-mesh}
		Hypersonic inviscid flow over a blunt body. Mesh number: 117,221. }
\end{figure}

\begin{figure}[htp]	
	\centering	
	\includegraphics[height=0.4\textwidth]
	{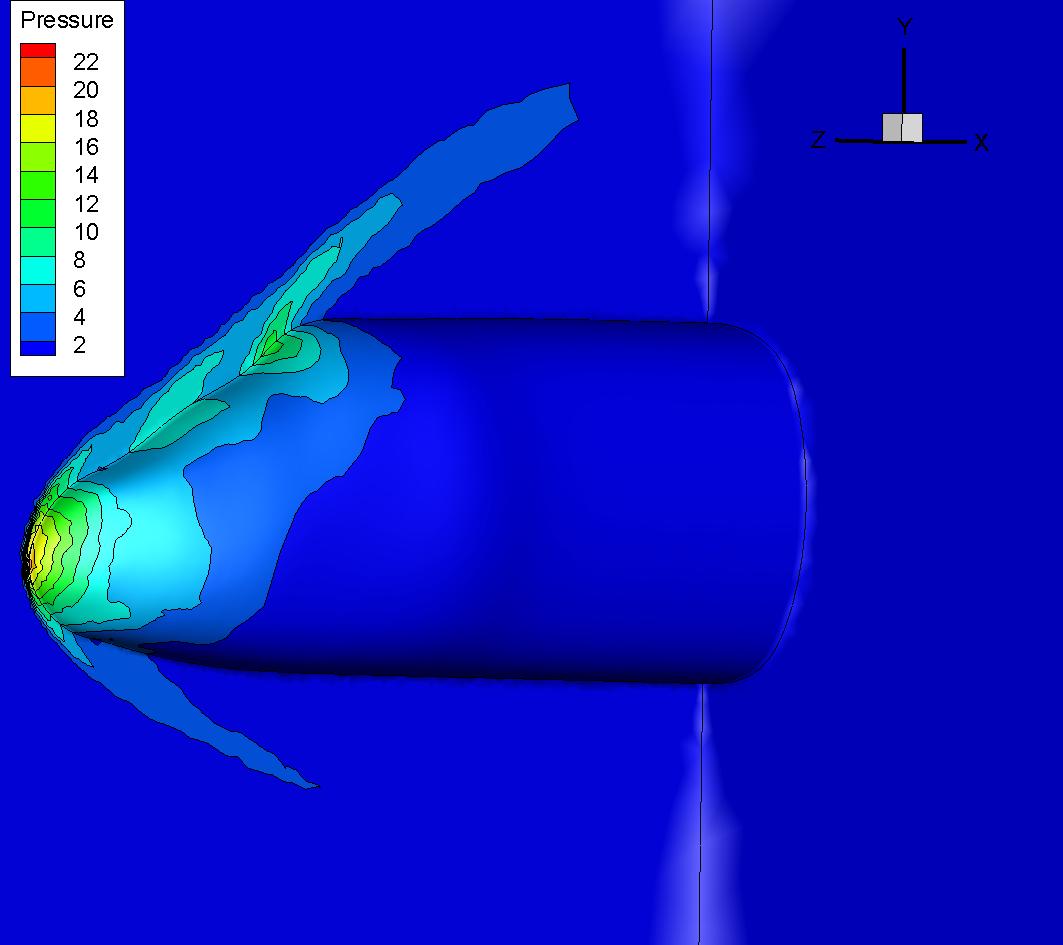}
	\includegraphics[height=0.4\textwidth]
	{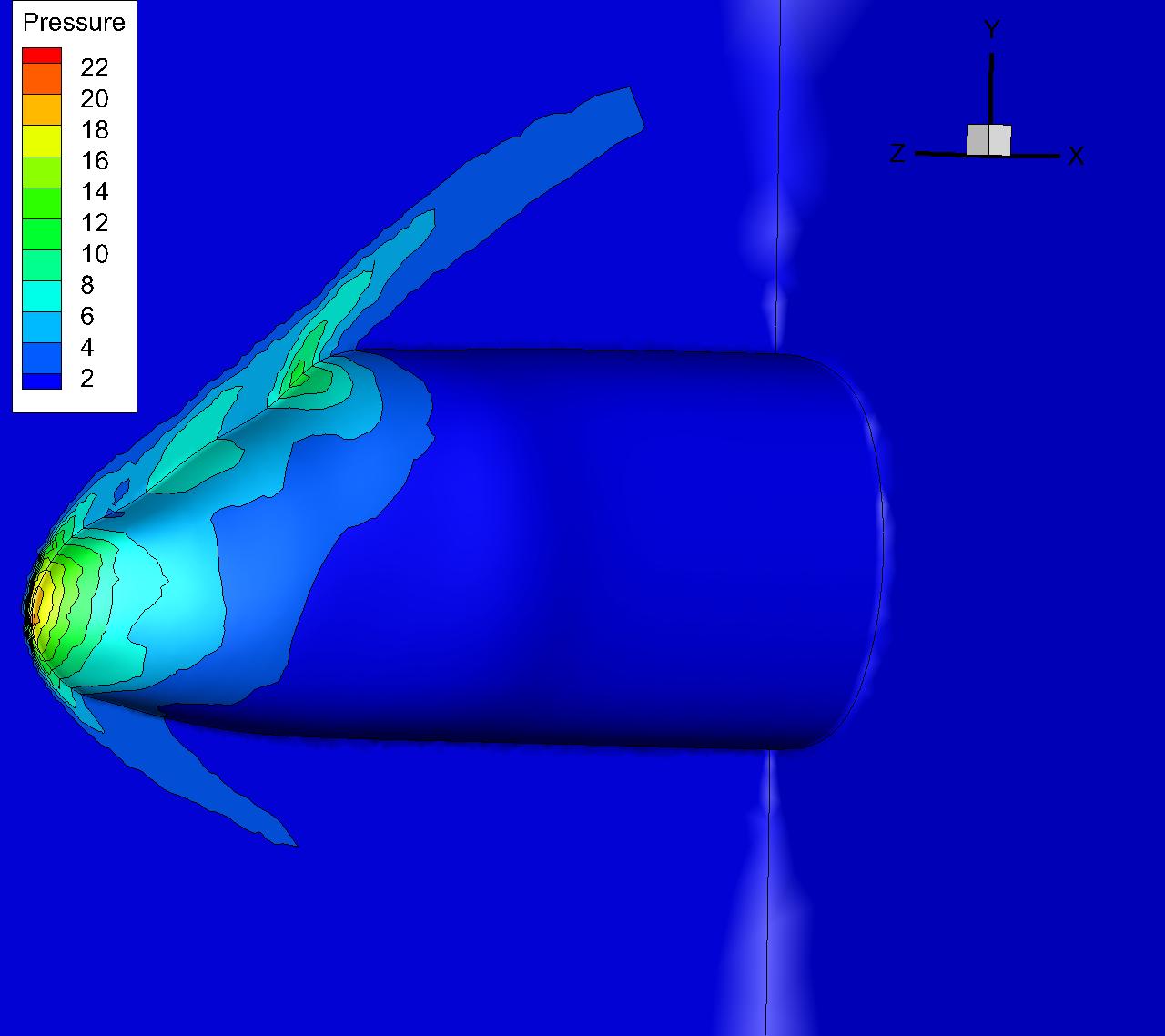}
	 \caption{\label{space-shuttle-pressure}
		Pressure distributions around the blunt body.
		Ma=5.0. AOA=$0.0^{\circ}$. Left: the second-order GKS. Right: the third-order GKS. }
\end{figure}

\begin{figure}[htp]	
	\centering	
	\includegraphics[height=0.4\textwidth]
	{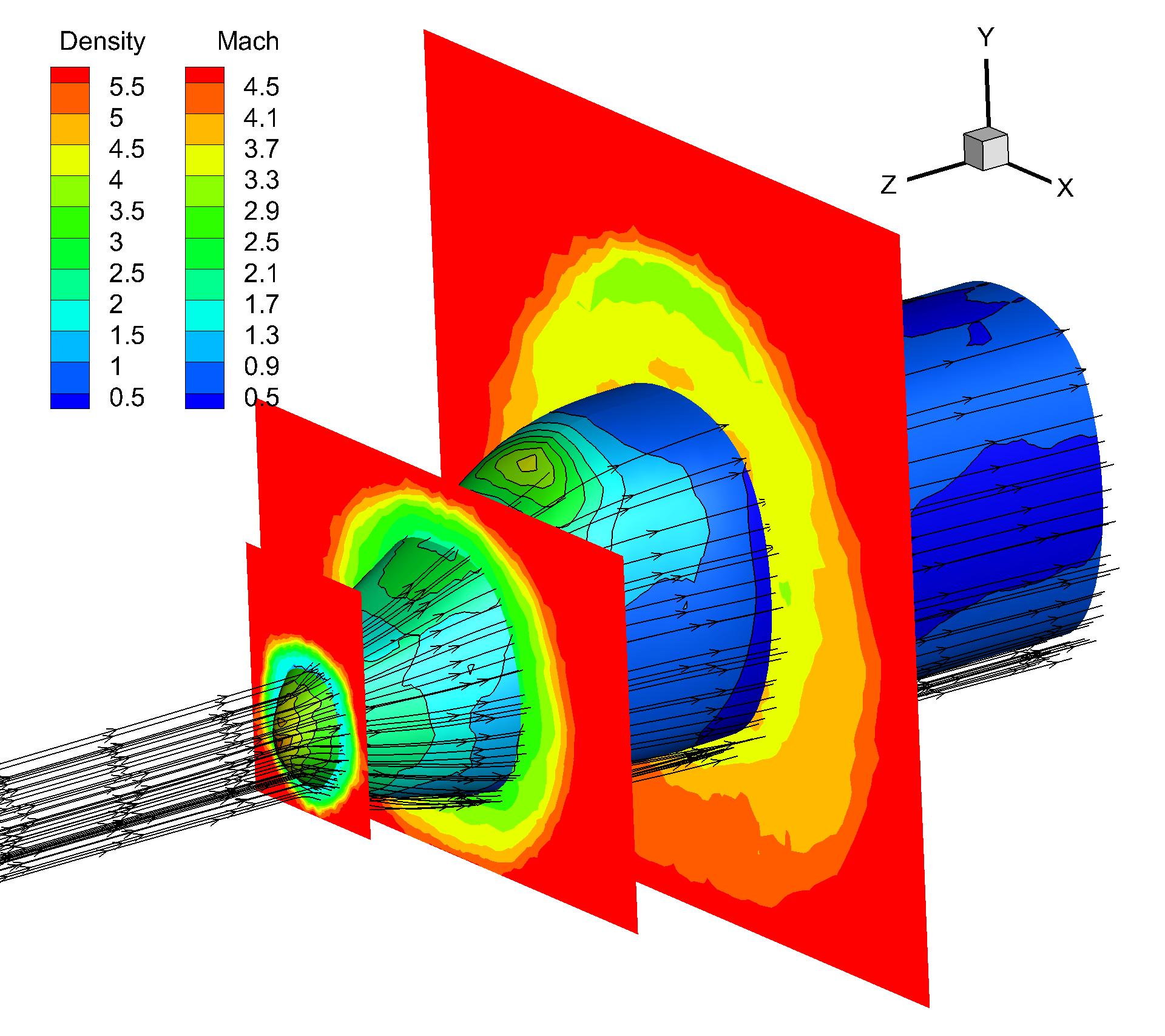}
	\includegraphics[height=0.4\textwidth]
	{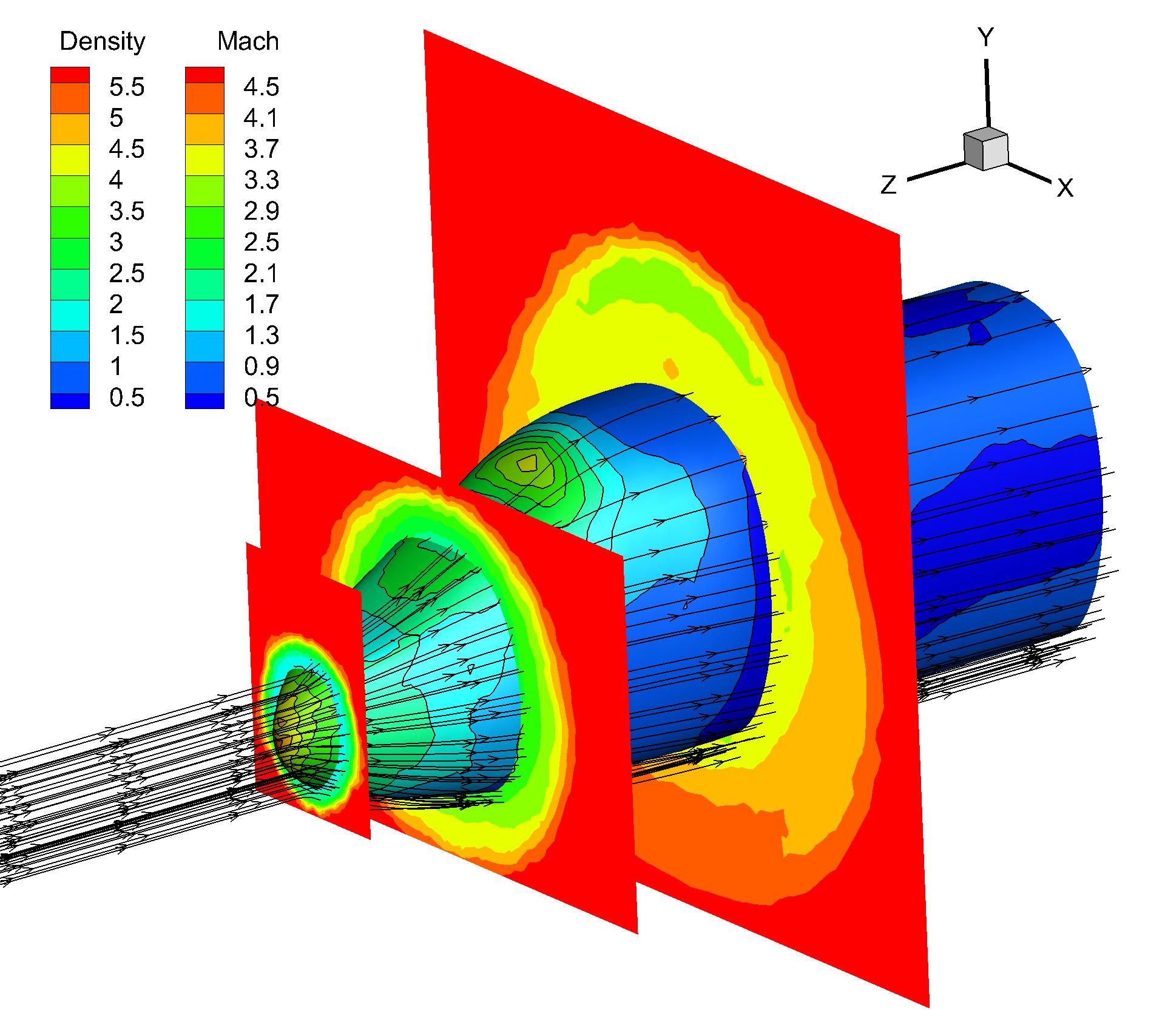}
	\caption{\label{space-shuttle-stream}
		Mach number distribution, stream-lines, and surface density distributions around the blunt body.
		Ma=5.0. AOA=$0.0^{\circ}$.  Left: the second-order GKS. Right: the third-order GKS. }
\end{figure}

A space-shuttle-like blunt-body model is considered to test the robustness of the schemes for the hypersonic inviscid flow.
The initial condition has  Ma=5 and AOA=0$^{\circ}$.
The surface mesh is given in Fig.~\ref{space-shuttle-mesh}, where the controlling points of the quadratic elements are shown.
The pressure distributions are shown in Fig.~\ref{space-shuttle-pressure}, where no significant differences are observed
in the results from the second-order and the third-order GKS. The Mach number distribution and streamlines are also plotted in Fig.~\ref{space-shuttle-stream}.

\section{Conclusion}
A 3rd-order compact GKS is developed for 3-D tetrahedral mesh.
The main difficulty from the structured to tetrahedral mesh is related to the 
linear instability in unstructured mesh. On a compact stencil with von Neumann neighbors only, even a second-order FVM in tetrahedral mesh
can become linearly unstable.
The high-order method based on Riemann-solvers with a compact stencil is also associated with instability. 
Also, the traditional WENO strategy based on the above compact stencil fails in dealing with discontinuities.
However, benefiting from the direct evolution of the cell-averaged first-order spatial derivatives in the compact gas-kinetic scheme,
the linear stability of the compact third-order GKS has been validated on tetrahedral mesh through the smooth inviscid and viscous tests.
To further improve the mesh adaptability and robustness of the scheme,
a new reconstruction based on the two-step and multi-resolution WENO methods is proposed.
At the same time, a new second-order GKS can be naturally obtained as a byproduct.
Both second and third-order schemes keep the compactness.
The reconstruction in this paper is carefully designed with the consideration of possible singularities from the mesh distortion
or the boundary corner, and it becomes suitable for arbitrary mesh.
The compact GKS also uses the two-stage time discretization as a building block for temporal accuracy,
which becomes efficient in comparison with the Runge-Kutta time stepping method for the same third-order temporal accuracy.
Various numerical examples from low-speed smooth flow to hypersonic flow are tested.
The compact GKS shows properties of robustness, high accuracy, and low dissipation.
Reliable mesh adaptability is also validated in the supersonic flow computation over a complete aircraft model.
Moreover, a large explicit time step with a CFL of $1$ can be used for most test cases.
The proposed compact GKS with the two-stage time discretization and the two-step multi-resolution WENO reconstruction
exhibits excellent numerical performance among the current existing compact schemes on tetrahedral mesh.
The compact GKS is currently extended to hybrid mesh with high aspect ratio for the boundary layer flow computation in supersonic and hypersonic viscous flow.

\section*{Acknowledgments}

The authors would like to thank Dr. Jun Zhu for helpful discussion, and be grateful to Mr. Nianhua Wang, Dr. Yangyang Liu, and Dr. Liming Yang for providing computational mesh.
The current research is supported by National Numerical Windtunnel project  and  National Science Foundation of China (11772281, 91852114).

\section*{References}
\bibliographystyle{plain}%
\bibliography{jixingbib}

\end{document}